%% file: main.tex
\documentclass[preprint]{elsarticle}
\usepackage{amsmath,amsfonts,amssymb,amscd,xcolor}
\usepackage{epsfig}
\usepackage{bbm}
\usepackage{bm}
\usepackage{standalone}
\usepackage{expl3}
\usepackage{xspace}
\usepackage{pgfplots}
\usepackage{pgfplotstable}
\usepgfplotslibrary{colorbrewer}
\usepgfplotslibrary{groupplots}
\usepackage{float}
\usepackage{subfig}
\usepackage[title]{appendix}
\usepackage{soul}
\usepackage{hyperref}

\usepackage{tikz}
\usetikzlibrary{calc,arrows}
\usetikzlibrary{matrix}

\biboptions{sort&compress}

\pgfplotsset{compat=1.16}

\newcommand{\AxisRotator}[1][rotate=0]{%
    \tikz [x=0.25cm,y=0.60cm,line width=.2ex,-stealth,#1] \draw (0,0) arc (-150:150:1 and 1);%
}

\begin{document}

\begin{frontmatter}
\title{Single-stage gradient-based stellarator coil design: Optimization for
  near-axis quasi-symmetry}
\author[courant]{Andrew Giuliani}
\author[courant]{Florian Wechsung}
\author[courant]{Antoine Cerfon}
\author[courant]{Georg Stadler}
\author[maryland]{Matt Landreman}
\address[courant]{Courant Institute of Mathematical Sciences, New York University, New York, New York, USA}
\address[maryland]{University of Maryland-College Park, Maryland, USA}

\begin{abstract}
  We present a new coil design paradigm for magnetic confinement in
  stellarators.  Our approach directly optimizes coil shapes and coil currents to
  produce a vacuum quasi-symmetric magnetic field with a target rotational
  transform on the magnetic axis. This approach differs from the
  traditional two-stage approach in which first a magnetic configuration
  with desirable physics properties is found, and then coils to approximately realize this magnetic configuration are designed. The proposed single-stage approach allows us to find a compromise between confinement and engineering requirements, i.e., find easy-to-build coils with good confinement properties.
  Using forward and adjoint sensitivities, we derive derivatives
  of the physical quantities in the
  objective, which is constrained by a nonlinear periodic differential equation. In two numerical
  examples, we compare different gradient-based descent algorithms and
  find that incorporating approximate second-order derivative
  information through a quasi-Newton method is crucial for convergence.  We
  also explore the optimization landscape in the neighborhood of
  a minimizer and find many directions in which the objective is mostly
  flat, indicating ample freedom to find simple and thus easy-to-build
  coils.
\end{abstract}

\begin{keyword}
stellarator optimization \sep quasi-symmetry \sep adjoint/forward sensitivity \sep optimal control \sep magnetic confinement
\end{keyword}

\end{frontmatter}

\section{Introduction}

Stellarators are a promising candidate for
magnetic confinement fusion, in which a high-temperature plasma is confined by a magnetic field that lacks continuous rotation symmetry (axisymmetry) \cite{Boozer1998,Helander2014}. This field is produced by electromagnetic coils, and these coils are designed
using optimization targeting a multitude of desired stellarator
properties. Mathematically, this amounts to complicated optimization
objectives with components that may be in competition with each other,
governed by complex physics
equations \cite{ImbertEtAl20,Reiman2001,Mynick2010}.
At the same time, the optimization must be formulated to avoid coil shapes that are impractical to build \cite{Strykowsky09,Neilson10,Klinger2013}.  Numerical optimization for these problems is challenging
due to the nonlinearity of the governing equations, complicated
minimization landscapes, and the difficulty in obtaining accurate
derivatives. In this article, we propose a new formulation of this
optimization problem in which the coil parameters are the primary optimization unknowns, and which therefore avoids the indirect, two-stage approach currently used by the stellarator community. We use our new formulation to design coils for vacuum stellarator configurations, and illustrate in these optimization problems the advantages of using analytical derivatives of the objective with respect to parameters in the coil representation.

From a physical point of view, the optimization problem may be motivated by highlighting the major weaknesses of stellarators as compared to their currently better performing cousin, the tokamak, which is a toroidally axisymmetric device with generally simpler coils \cite{Helander2012}. 
First, unlike in tokamaks, the orbits of single particles not subject to collisions are not guaranteed to be confined in a generic stellarator. Particle and energy losses can therefore be unacceptably high \cite{Seiwald2008,Helander2012,Helander2014} if the equilibrium magnetic configuration is not designed with care, which can lead to significant deterioration and damage of the first wall, and requiring unacceptably high power to maintain the plasma temperature. Second, planar coils are usually not sufficient to produce the desired equilibrium magnetic field, and stellarators commonly require highly distorted coils, which are expensive and complex to build \cite{Strykowsky09,Neilson10,Klinger2013}. 
In the last four decades, a family of magnetic fields guaranteeing good confinement, named quasi-symmetric magnetic fields \cite{Helander2014}, has been identified and studied extensively \cite{Nuhrenberg1988,Boozer_1995,Canik2007}. It is the design basis for several advanced stellarators that have been partially constructed or proposed for future experiments \cite{Zarnstorff2001,ku_2008,Garabedian_2008,Drevlak_2013,liu_2018,Henneberg_2019,bader2020new}. While other families of magnetic fields may lead to equally good confinement \cite{Helander2014}, quasi-symmetric fields are convenient from a mathematical point of view because there are several equivalent formulations for the quasi-symmetry condition, which are all fairly easy to compute numerically. In this article, we will therefore focus on quasi-symmetric magnetic configurations. 

Regarding coil design, the following optimization strategy has been
favored in the past, which we call the indirect method or the
two-stage approach in this article. In the first stage, coils are
ignored and we optimize the shape of the toroidal boundary surface of the plasma.  The field is assumed
tangent to the boundary, enabling the interior field to be computed
from the boundary shape.  The objective function reflects physics
properties of the plasma inside the surface. In the second optimization stage, the magnetic field determined in the first stage is fixed and we optimize the shapes and currents of electromagnetic coils such that they reproduce this target field
\cite{Merkel_1987,drevlak_1998,strickler_2002,strickler_2004,brown_2015,Zhu2017}. There
are several advantages to this approach. It is efficient, since codes computing the magnetic configuration taking as input a known plasma boundary surface on which the magnetic field is tangent (so-called ``fixed-boundary equilibrium codes") are faster and more robust than codes which instead solve for the magnetic configuration taking as input the geometry of the coils and their currents, and for which the location of the toroidal plasma boundary is one of the outputs of the computation (so-called ``free-boundary equilibrium codes"). 
One also empirically observes that with the two-stage method, the level sets of the magnetic flux inside the plasma often are toroidal surfaces foliating a large fraction of the plasma volume, which is the most basic -- but not sufficient -- requirement for confinement. Furthermore, new methods have been recently proposed to more efficiently optimize coils and reduce their complexity in this two-stage method \cite{Landreman_2017,Paul_2018,Zhu2017}. The two-stage method has been used for the design of many successful stellarator experiments that are optimized for particle confinement and are currently operating. However, this approach has drawbacks. First, when a stellarator is constructed, one builds coils, and not a plasma boundary; the first stage of the two-stage method thus optimizes for aspects of the experiment on which one does not have direct control. Undesired discrepancies in the magnetic field necessarily arising from the coil design process in the second stage may then significantly alter the quality of the equilibrium magnetic configuration. Conversely, the two-stage method cannot capture magnetic configurations that are achievable with relatively simple coils. For example, existing three-dimensional equilibrium codes are not able to compute equilibria for which the plasma boundary has corners, corresponding to separatrices. 
Third, the two-stage approach makes it difficult to balance desired physics properties with coil manufacturing constraints.

A single-stage approach designs optimized stellarators starting directly from coil configurations. It holds the promise to address these drawbacks, reduce the complexity of stellarator coils, improve access to the device, and therefore help to reduce construction costs. However, to the best of our knowledge, a systematic and robust method for obtaining magnetic fields with good confinement properties has never been proposed for that single-stage method. This is particularly challenging since most coil shapes generate magnetic fields whose level sets of the magnetic flux are not closed toroidal surfaces for a large fraction of the plasma volume, i.e.~fields which do not satisfy the most basic requirements for confinement. The purpose of the present article is to demonstrate that a single-stage coil optimization method can be developed that addresses that challenge. Our strategy is to optimize coils so they produce a magnetic field that approximates, on the magnetic axis, the quasi-symmetric field constructed with the Garren-Boozer near-axis expansion \cite{garren1991,landreman_2018,landreman_2019,LandremanSengupta2019} as well as its gradient, and which also has a near-axis rotational transform that is close to a target rotational transform chosen as an input. 
We recall that the magnetic axis is the closed curve corresponding to the innermost level set of the magnetic flux \cite{Helander2014}, see for instance the dotted line in Figure \ref{fig:problem1_surfaces}. 
Additionally, recall that the rotational transform, $\iota$, is a property of the magnetic field and its field lines that plays an important role for confinement \cite{Helander2014}. 
The rotational transform at the axis is given by the number of poloidal transits per toroidal transit for field lines in the neighborhood of the axis.
Mathematically, we solve an optimization problem that is constrained by a nonlinear differential equation. Building on methods from optimization governed by differential equations \cite{BorziSchulz12, Delosreyes15,
  Gunzburger03}, we derive analytical expressions for the gradients
of the optimization objective. Following a discretize-then-optimize
approach, in which we discretize the governing equation and compute the
analytical derivatives, using either forward or adjoint
sensitivities, at the discrete level, we have the guarantee of obtaining
discretely exact gradients, which is advantageous for iterative
descent optimization algorithms \cite{Gunzburger03}.

\subsection{Contributions and limitations}
We make the following contributions.
(1) We present a single-stage approach to stellarator coil design, which
directly targets quasi-symmetry and avoids the two-stage optimization
process that is currently favored by the stellarator community. 
This allows a more thorough exploration of the trade-offs between confinement and engineering constraints, and thus, it has the potential to help design coils that are easier and more cost-effective to manufacture.
(2) We derive analytical derivatives of physical quantities (e.g., the
rotational transform on an axis) in the objective. This is made possible by choosing
suitable characterizations of these quantities, combined with forward
and adjoint sensitivity methods for objectives governed by
differential equations. Accurate derivatives facilitate efficient numerical solution and enable exploration of the optimization objective landscape, but are not commonly used in stellarator optimization codes.
(3) A numerical study of the Hessian at a minimum of the objective
provides some insight into to objective landscape. Namely, there is
only a moderate number of directions in parameter space where the
Hessian has large curvature and thus the objective is very sensitive to
changes in these directions. In many other directions, the objective is
mostly flat, indicating ample freedom for finding coil configurations
with good confinement properties that are simpler and thus easier-to-build.

Our approach also has several limitations.
(1) The presented approach currently focuses on vacuum magnetic
configurations, i.e., it neglects magnetic fields generated by currents inside the
plasma. At the end of this article, we discuss generalizations of the
present approach to include these contributions, which can be important
for stellarator experiments with high plasma pressure.
(2) While our approach targets quasi-symmetry on and in a neighborhood
of the magnetic axis, we do not have explicit control of
quasi-symmetry away from the axis. In the last section of this article,
we discuss how our method can be extended to enforce
quasi-symmetry also on surfaces away from the axis.
(3) Our physics and engineering constraints are simpler than those
which are imposed in detailed stellarator design studies. The purpose
of the present work is to demonstrate the feasibility and the
advantages of the single-stage method for the design of
quasi-symmetric stellarators. Both the objective and constraints
discussed in this article can be expanded depending on the design
needs, in which case additional analytical derivatives would have to be
computed.

\subsection{Structure of article}
The structure of this article is as follows.  In section \ref{sec:qs},
we review the equations describing quasi-symmetric magnetic fields near
the magnetic axis.  In section \ref{sec:optim_prob}, we present our
optimization problem, define our optimization space, and motivate our
cost function physically. In section \ref{sec:gradient}, we explain
how the forward sensitivity method and the adjoint method each allows
us to efficiently and accurately evaluate the gradient with respect
to the design parameters, which we need in order to minimize our cost
function. We present the magnetic configurations obtained with this
new single-stage approach in sections \ref{sec:p1} and \ref{sec:p2}, and summarize our results and suggest future directions for this work in section \ref{sec:conclusion}.

\section{Near-axis quasi-symmetry} \label{sec:qs}
Our optimization formulation aims to find a coil system such that
the magnetic field at the magnetic axis is quasi-symmetric, a highly
desirable property for confinement. While a quasi-symmetric magnetic
field would be desirable throughout the entire plasma region,
published results show that this might not be achievable as the
required system of equations is over-determined \cite{garren1991}.
Thus, we
limit ourselves to finding a quasi-symmetric magnetic field in a
neighborhood of the magnetic axis. Instead of considering the axis
induced by the magnetic coils, we parameterize a so-called expansion
axis independently and include these parameters in the
optimization. The optimization problem detailed in section
\ref{sec:optim_prob} then aims to minimize the difference between
the magnetic field induced by the coils and a
quasi-symmetric magnetic field derived from a near-axis expansion at
that independent axis. This expansion was recently presented in
\cite{landreman_2018,landreman_2019,LandremanSengupta2019} and is
based on the Garren-Boozer expansion \cite{garren1991}. Upon
convergence of the optimization problem detailed in section
\ref{sec:optim_prob}, the magnetic field from the coils and the axis
expansion coincide in a neighborhood of the independent axis. In
particular, the magnetic axis and the expansion axis coincide and the
field induced by the coils is quasi-symmetric close to these
axes. Before presenting the optimization problem, we summarize
elements of this near-axis expansion that are critical to the
understanding of our optimization formulation.

The remainder of this section describes how the choice of an expansion
axis and of a value $\overline \eta\in \mathbb R \setminus \{0\}$
uniquely defines the quasi-symmetric field $\mathbf{B}_{\text{QS}}$
and its gradient $\nabla \mathbf{B}_{\text{QS}}$ on axis. This
nonlinear relationship requires to first solve a nonlinear system
containing an ordinary differential equation (see \eqref{eq:sigma})
for a function $\sigma (\phi)$ and $\iota\in \mathbb R$. 
These
quantities then define the near-axis expansion of the quasi-symmetric
magnetic field in terms of $\sigma$ and $\iota$. 
These technical
expressions, which were derived in \cite{Landreman2020}, will be given
in \eqref{eq:B_QS} and \eqref{eq:gradB_QS} below.  Note that $\sigma$ is a function that
is used in the formula for the gradient of the quasisymmetric magnetic field \eqref{eq:gradB_QS}, and $\iota$ is the rotational transform on axis.

\begin{figure}\centering
\begin{tikzpicture}
    \node (11) at (0,0) {\includegraphics[width=0.4\textwidth]{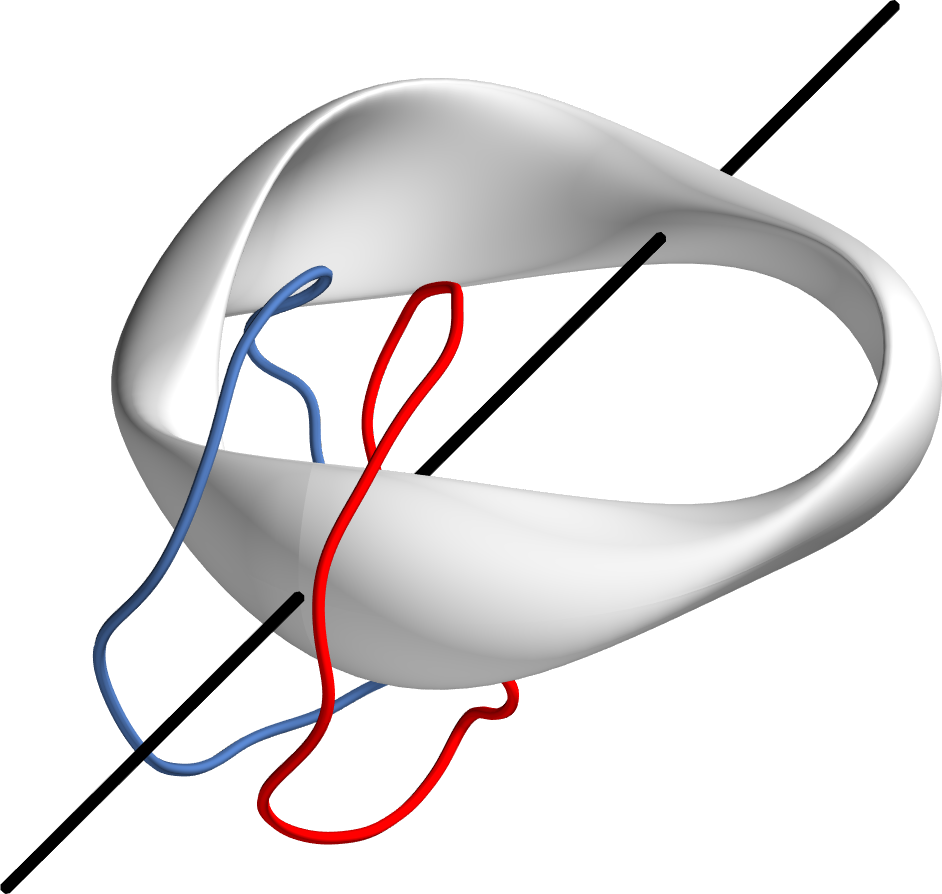}};
    \node (12) at (8,0) {\includegraphics[width=0.4\textwidth]{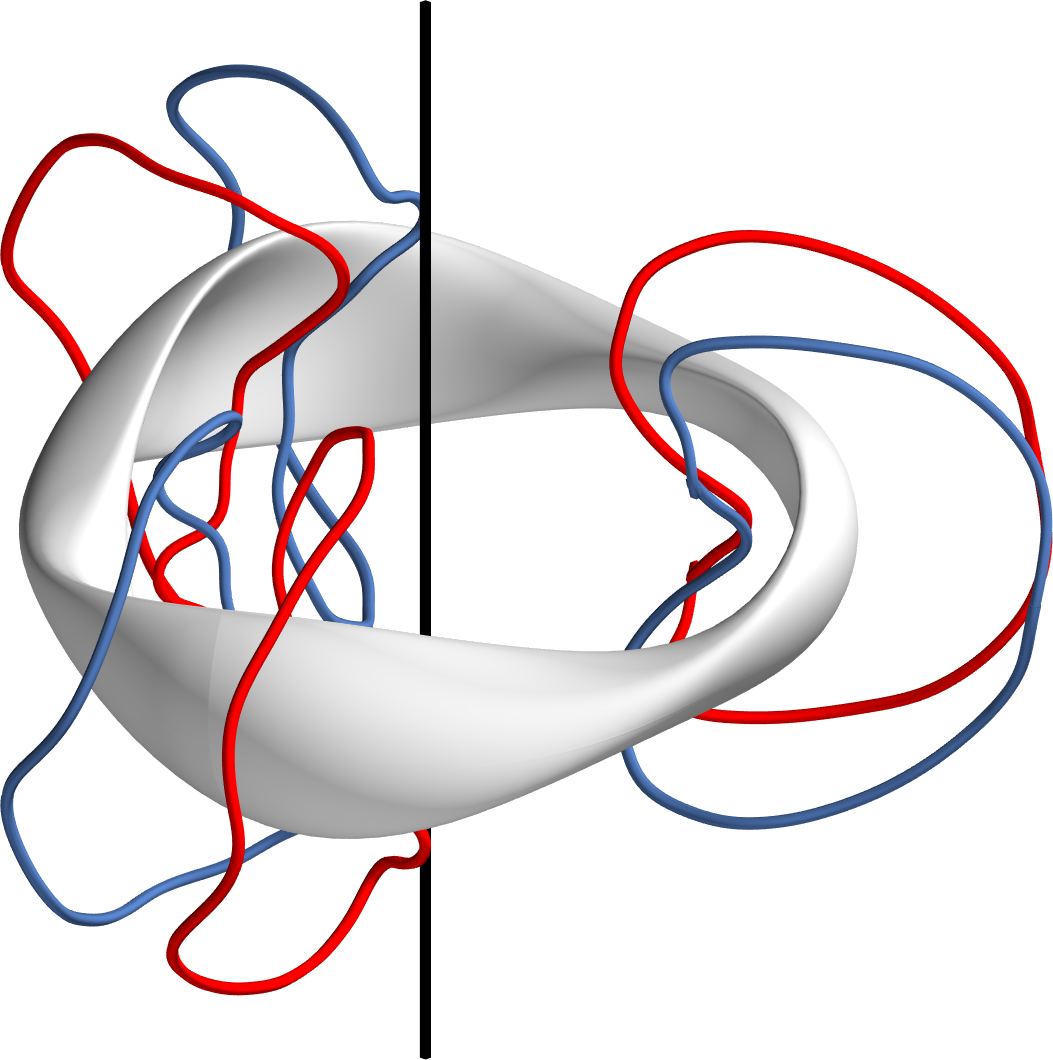}};
    
    \node at (-2,3) {\textcolor{black}{\large (A)}};
    \node at ( 5,3) {\textcolor{black}{\large (B)}};
    \node at (-3.1,-3)  {\AxisRotator[rotate=45]};
    \node at (7.4,3)  {\AxisRotator[rotate=-90]};
\end{tikzpicture}
\caption{\label{fig:symmetries}Illustration of stellarator and rotational symmetry. Figure (a) shows a magnetic surface (grey) and a modular coil in the initial NCSX configuration (red) with its stellarator-symmetric counterpart (blue) obtained by a 180$^{\circ}$ rotation about the $x$-axis (black).  Figure (b) illustrates discrete rotational symmetry ($N_{\text{fp}}=3$) by rotating the stellarator symmetric coils from (a) by 120$^{\circ}$ and 240$^{\circ}$ around the $z$-axis (black). }
\end{figure}
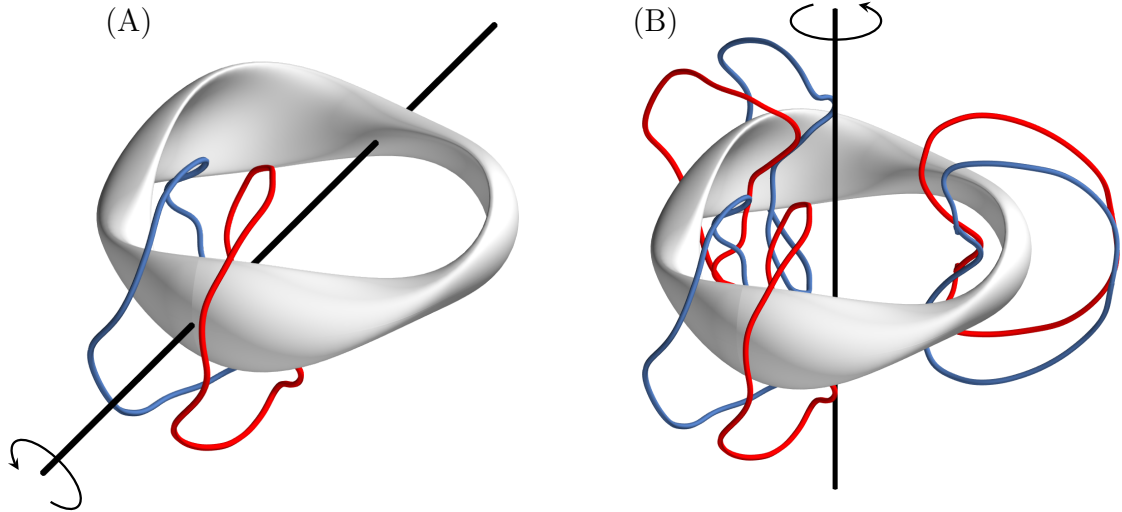

We consider the expansion to the order used in \cite{landreman_2019}, such that
quasi-symmetry is guaranteed through first order in the distance from
the axis, but quasi-symmetry-breaking errors generally arise at second
order.  In what follows, we assume that the current density on axis is
zero, and that the magnetic field is stellarator symmetric \cite{Dewar1998}, and has discrete rotational symmetry. Stellarator symmetry and discrete rotational symmetry imply that in cylindrical coordinates the magnetic field $\mathbf{B} = (B_R, B_\phi, B_Z)$ satisfies respectively
\begin{equation*}
\begin{aligned}
    B_R(R, \phi, Z) &= -B_R(R, -\phi, -Z), \\
    B_\phi(R, \phi, Z) &= B_\phi(R, -\phi, -Z), \\
    B_Z(R, \phi, Z) &= B_Z(R, -\phi, -Z),
\end{aligned}
\quad \text{ and } \quad
\begin{aligned}
    B_R(R, \phi , Z) &= B_R(R, \phi+ 2 \pi / N_{\text{fp}}, Z), \\
    B_\phi(R, \phi, Z) &= B_\phi(R, \phi+ 2 \pi / N_{\text{fp}}, Z), \\
    B_Z(R, \phi, Z) &= B_Z(R, \phi+ 2 \pi / N_{\text{fp}}, Z),
\end{aligned}
\end{equation*}
where $N_{\text{fp}}$ is the number of identical field
periods; see also an illustration of these symmetries for coils in Figure \ref{fig:symmetries}.
The quasi-symmetric expansion field then
depends on two inputs, namely the geometry of the expansion axis, and
a parameter $\bar \eta \in \mathbb R \setminus \{0\}$ that influences how
elongated the magnetic surfaces are.  The expansion axis may be
any closed curve with non-vanishing curvature,
$\Gamma_{\!\mathbf{a}}(\phi)$, where $\phi$ is the standard cylindrical
angle.
The quasi-symmetric field corresponding to a given axis and
$\bar \eta$ is found by first solving the following $2\pi/N_{\text{fp}}$-periodic
first-order ordinary differential equation with quadratic nonlinearity
for $\sigma=\sigma(\phi)$ and $\iota\in \mathbb R$:
\begin{equation}
\begin{aligned}
\frac{|G_0|}{\ell' B_0}
\frac{d \sigma}{d\phi} + (\iota-N)  \left[ \frac{\bar \eta^4}{\kappa^4} + 1 + \sigma^2 \right]
+ \frac{2 G_0 \bar \eta^2 s_\psi \tau}{B_0 \kappa^2} &=0, \\
\sigma(0) &=  0,
\label{eq:sigma}
\end{aligned}
\end{equation}
where $N_{\text{fp}}$ is the number of field periods.  The expansion axis enters \eqref{eq:sigma} through its curvature $\kappa$ and
torsion $\tau$. 
Note that the condition $\sigma(0)=0$ follows from stellarator
symmetry and that the $2\pi/N_{\text{fp}}$-periodicity implies that $\sigma(2\pi/N_{\text{fp}})=0$.  
Additionally, $\sigma$ must be odd for stellarator symmetry, so $\sigma(\pi/N_{\text{fp}})=0$ as well.
Since $\iota$ is part of the solution, 
\eqref{eq:sigma} is not a standard differential
equation boundary value problem. However, it is shown in \cite{landreman_2019} that given a suitable expansion axis with non-vanishing curvature and $\overline \eta$, the solution pair $\sigma(\phi)$, $\iota$ of \eqref{eq:sigma} is unique.  
The constants in \eqref{eq:sigma} are $G_0 = s_G B_0 L / (2\pi)$,
$L$ is the axis length, $s_G = \pm 1$ specifies if the magnetic
field points in the same or opposite direction of increasing $\phi$,
$s_\psi=\pm 1$ is related to the sign of $\iota$, $B_0$ is a constant,
$\ell' := \|\Gamma'_{a}(\phi)\|$ is the incremental axis length, and the integer
$N$ indicates the type of quasi-symmetry.  In this paper we only
consider quasi-axisymmetry, which is characterized by $N=0$.

Solving \eqref{eq:sigma}, we obtain the on-axis rotational transform $\iota$ and a function $\sigma(\phi)$.  
Once they are known, the quasi-symmetric magnetic field
$\mathbf{B}_{\text{QS}}$ and its gradient $\nabla
\mathbf{B}_{\text{QS}}$ are determined by the formulae \cite{Landreman2020}
\begin{equation}\label{eq:B_QS}
\mathbf{B}_{\text{QS}} = B_0 \mathbf{t}
\end{equation}
and
\begin{align}
\label{eq:gradB_QS}
\nabla B_{\text{QS},j}(\phi) = &s_\psi \frac{B_0^2}{|G_0|} \left\{
\left[ s_\psi \frac{G_0}{B_0}\kappa t_j 
+ \left( \widetilde{X'_{1c}} Y_{1s} + \iota X_{1c} Y_{1c} \right) n_j \right.\right.
\\
& \hspace{0.6in} \left.+ \left( \widetilde{Y'_{1c}}Y_{1s} - \widetilde{Y'_{1s}} Y_{1c}
+s_\psi \frac{G_0}{B_0} \tau + \iota \left( Y_{1s}^2 + Y_{1c}^2\right) \right) b_j
\right] \mathbf{n} \nonumber\\
&\hspace{0.5in} \left.+\left[\left(-s_\psi \frac{G_0}{B_0}\tau - \iota X_{1c}^2\right)n_j + \left(X_{1c}\widetilde{Y'_{1s}}-\iota X_{1c}Y_{1c}\right)b_j \right] \mathbf{b} \right\}
+\kappa s_G B_0 n_j \mathbf{t}, \nonumber
\end{align}
where the $j$ subscripts indicate Cartesian components, $\mathbf{t}, \mathbf{n}, \mathbf{b}$ are the Frenet vectors \cite{stewart2012essential}, i.e., the unit tangential, normal, and binormal vectors  associated to the expansion axis $\Gamma_{\!\mathbf{a}}(\phi)$. Moreover,
\begin{align*}
X_{1c}(\phi) = \frac{\bar \eta}{\kappa},
\hspace{0.2in}
Y_{1s}(\phi) = \frac{s_G s_\psi \kappa}{\bar \eta},
\hspace{0.2in}
Y_{1c}(\phi) = \frac{s_G s_\psi \kappa \sigma}{\bar \eta},
\end{align*}
and we use the scaled derivatives
\begin{align} \label{eq:sderiv}
\widetilde{X'_{1c}} = \frac{|G_0|}{\ell' B_0}X'_{1c}(\phi) ,
\hspace{0.2in}
\widetilde{Y'_{1s}} = \frac{|G_0|}{\ell' B_0}Y'_{1s}(\phi) ,
\hspace{0.2in}
\widetilde{Y'_{1c}} = \frac{|G_0|}{\ell' B_0}Y'_{1c}(\phi).
\end{align}
The above expressions for $B_{\text{QS},j}, \nabla B_{\text{QS},j}$ depend on the expansion axis parameters $\mathbf a$, which determine the Frenet frame vectors  $\mathbf{t}, \mathbf{n}, \mathbf{b}$, the incremental axis length $\ell'$, curvature $\kappa$, and the torsion $\tau$.  They additionally depend on the solution pair $\sigma(\cdot),\iota$ computed by solving \eqref{eq:sigma}, which requires a suitable expansion axis characterized by $\mathbf a$. In the next section, we formulate an optimization problem whose objective involves least squares terms for $B_{\text{QS},j}, \nabla B_{\text{QS},j}$, which are considered to be functions of  $\mathbf a$.

\section{Single-stage coil optimization for quasi-symmetry}\label{sec:optim_prob}
We aim to find coils and currents that generate a magnetic field that
is quasi-symmetric on axis. In our formulation, we treat the coils,
their currents, and
an expansion axis as parameters, and formulate an objective that
ensures that the magnetic field generated by the coils coincides (or is very close
to) the quasi-symmetric field on the expansion axis.
As a by-product, this typically also means that the expansion axis
nearly coincides with the magnetic axis of the field generated by the coils.
We emphasize that the quasi-symmetric field is not fixed, i.e., the
unknowns associated to the expansion axis and to the coils are allowed
to vary.

\subsection{Design parameters} \label{sec:os}
The design parameters of our optimization problem are the geometric degrees
of freedom associated to the coils and the expansion axis, the
coil currents and the scalar $\bar \eta$.
We model stellarator coils as current-carrying filaments in
three dimensions.  The Cartesian coordinates of the $i$th coil,
$\Gamma^{(i)}_{c}(\theta) = (x^{(i)}(\theta), y^{(i)}(\theta), z^{(i)}(\theta))$, are given by a Fourier representation,
e.g., the $x$-coordinate of the curve describing the $i$th coil is
\begin{equation}\label{eq:coils}
x^{(i)}(\theta) = c^{(i)}_{x,0} + \sum^{n^{\text{coil}}_p}_{k = 1} s^{(i)}_{x,k} \sin( k \theta) + c^{(i)}_{x,k} \cos( k \theta),
\end{equation}
where $\theta$ varies from 0 to $2\pi$, the real numbers $c^{(i)}_0$ and $s^{(i)}_k,c^{(i)}_k$ for $k
= 1 \hdots n^{\text{coil}}_p$ are the Fourier coefficients, and
$n^{\text{coil}}_p$ is the number of Fourier modes.  The $y$ and $z$-components
of the coil curves are defined analogously.  
In using this
(finite)  Fourier representation, we are imposing some regularity to the coils.  Note that
our approach can be adapted to other, local representations
that imply less regularity, e.g., splines.  The $i$th modular
coil also has an associated current, $I_i$. In total, there are therefore $6n^{\text{coil}}_p + 3$
unknowns per coil corresponding to the Fourier representation and an
additional unknown per coil for the current.  The unknowns
associated to the coils are combined into the vector of optimization
variables $\mathbf{c} = (c^{(i)}_{x,0}, c^{(i)}_{x,1}, s^{(i)}_{x,1}, \cdots, I_0, I_1, \cdots)^T \in \mathbb{R}^{N_c (6n^{\text{coil}}_p + 4)}$, which has $N_c (6n^{\text{coil}}_p + 4)$ entries, where
$N_c$ is the number of modular coils.
Note that given a particular coil geometry, the corresponding Fourier representation is not unique. 
For example, adding a constant angle to $\theta$ does not affect the
coil geometry but produces a different parametrization. 
However, note that all the terms in the objective defined below only depend on the geometry of the coils and expansion axis and not on their parametrizations.
The non-uniqueness of the coil representation may lead to a non-unique representation of minimizers, but the objective value is not affected by this. Additionally, this non-uniqueness is reduced by the fact that the expansion \eqref{eq:coils} is truncated.

Similar to the coils, we represent the expansion axis,
$\Gamma_{\!\mathbf{a}}(\phi) = (R(\phi), \phi, Z(\phi) )$, using a
Fourier representation for the $R(\phi)$, $Z(\phi)$ coordinates of
the curve in a cylindrical coordinate system.  Due to stellarator
symmetry, $R(\phi)$ is only composed of cosines and
$Z(\phi)$ only of sines 
\begin{align*}
R(\phi) = R_{0} + \sum^{n^{\text{axis}}_p}_{k = 1} R_k \cos(k N_{\text{fp}}\phi), \quad
Z(\phi) = \sum^{n^{\text{axis}}_p}_{k = 1} Z_k \sin(k N_{\text{fp}}\phi),
\end{align*}
where $N_{\text{fp}}$ is the number of identical field
periods. 
There is an additional degree of freedom associated to the axis,
$\bar \eta$, that enters in the near-axis expansions. In total,
there are $2n^{\text{axis}}_p + 2$ unknowns associated to the Fourier representation of
the axis curve and the additional unknown $\bar \eta$, both
combined into the vector of design parameters $\mathbf{a} = (R_0, R_1, \cdots, Z_1, \cdots, \bar \eta)^T \in \mathbb{R}^{2n^{\text{axis}}_p+2}$.

\subsection{Optimization objective}
The function that we seek to minimize is 
\begin{align} \label{eq:fhat}
\begin{split}
\hat{J}(\mathbf{c},\mathbf{a}, \sigma, \iota) &= \frac{1}{2}\int_{\Gamma_{\!\mathbf{a}}}  \|\mathbf{B}_{\text{coils}}(\mathbf{c}) - \mathbf{B}_{\text{QS}}(\mathbf{a})\|^2~dl + \frac{1}{2}\int_{\Gamma_{\!\mathbf{a}}} \| \nabla\mathbf{B}_{\text{coils}}(\mathbf{c}) - \nabla\mathbf{B}_{\text{QS}}(\mathbf{a}, \sigma, \iota) \| ^2 ~dl\\
 &+ \frac{1}{2}\left(\frac{\iota-\iota_{0,a}}{\iota_{0,a}} \right)^2 + R(\mathbf{c},\mathbf{a}),
\end{split}
\end{align}
where $\Gamma_{\!\mathbf{a}}$ is the expansion axis, $\mathbf{B}_{\text{coils}}$ and
$\nabla\mathbf{B}_{\text{coils}}$ are the magnetic field and its
gradient generated by the coils computed using the Biot-Savart law \cite{jackson_classical_1999},
$\mathbf{B}_{\text{QS}}$ and $\nabla\mathbf{B}_{\text{QS}}$ are the
quasi-symmetric magnetic field and its gradient computed from near-axis
expansions (section \ref{sec:qs}), and $R(\mathbf{c},\mathbf{a})$
groups together regularization terms which we describe shortly.  The objective is a function of
coil and expansion axis variables $\mathbf{c}$ and $\mathbf{a}$,
respectively, (section \ref{sec:os}) and state variables $\sigma$,
$\iota$ (section \ref{sec:qs}).

Now, let us discuss the terms in \eqref{eq:fhat} in more detail. The
first term forces the on-axis field generated by the coils
$\mathbf{B}_{\text{coils}}$ to coincide with the quasi-symmetric field
$\mathbf{B}_{\text{QS}}$, where $\|\cdot\|$ is the standard Euclidean
norm.  Recall that the quasi-symmetric field on axis and its gradient
are computed from the formulae \eqref{eq:B_QS} and \eqref{eq:gradB_QS}
by solving \eqref{eq:sigma} for $\sigma$ and $\iota$.  The second term
in \eqref{eq:fhat} acts similarly, but for the gradient of the
magnetic field on axis.  Since at every point on the axis, the field
gradients $\nabla \mathbf{B}_{\text{coils}}$ and $\nabla
\mathbf{B}_{\text{QS}}$ are $3\times 3$ matrices, in this term,
$\|\cdot\|$ denotes the Frobenius norm.  The third term forces the
on-axis rotational transform $\iota$ to be close to a given target
value $\iota_{0,a}$.  Similar to $\mathbf{B}_{\text{QS}}$, $\iota$ is
computed by solving \eqref{eq:sigma}, i.e., it is a function of the
expansion axis geometry and $\bar \eta\in \mathbb R$.  Note that in
the formulation of the optimization problem, the expansion axis
defined by $\mathbf{a}$ is treated as an independent variable, i.e.,
during the optimization it is not constrained to being the magnetic
axis corresponding to the field from the coils.  In fact, we do not
require computation of the magnetic axis. However, the first and
second penalty terms in \eqref{eq:fhat} tend to become large if the
magnetic and the expansion axis differ substantially. Thus, upon
convergence, the expansion axis and the magnetic axis (when computed)
typically coincide. Treating the expansion axis as independent
variable and minimizing the difference between
$\mathbf{B}_{\text{coils}}$ and $\mathbf{B}_{\text{QS}}$ has proven
numerically advantageous compared to considering all computed quanties
as functions of the coils.

We now describe multiple regularization terms that we combined in $R(\mathbf{c},
\mathbf{a})$. These terms help to find coils (and a magnetic axis)
that have desirable engineering properties, e.g., avoid points with
large curvature, have a certain length, or are sufficiently far from
other coils.  We use a number of regularization terms such as
\begin{equation*}
\sum_{i = 1}^{N_c}\frac{1}{2}\left(\frac{L^{(i)}_{c}(\mathbf{c})
  -L^{(i)}_{0,c}}{L_{0,c}}\right)^2 \qquad \text{ and } \qquad
\frac{1}{2}\left(\frac{L_a(\mathbf{a})-L_{0,a}}{L_{0,a}} \right)^2,
\end{equation*}
which target coil lengths $L^{(i)}_{c}>0$ and an expansion axis
length $L_{0,a}>0$. These terms prevent unnecessarily complex
coils and impose a target length (and thus a length scale) on the
axis.  Note that to normalize these regularization terms, they are
multiplied by the reciprocal target values. This is also done for the
term involving the rotational transform in the objective
\eqref{eq:fhat}.

One typically wants to avoid coils that contain points with high
curvature, as such coils are difficult and thus costly to manufacture. One
way to prevent this is to define
$$
k_i(\mathbf{c}) = \int_{\Gamma^{(i)}_{\!\mathbf{c}}} \max(0, \kappa_i - \kappa_{i,0})^4 ~dl,
$$
where $\Gamma^{(i)}_{\!\mathbf{c}}$ is the $i$th coil, $\kappa_i$ is the point-wise curvature of coil $i$, and $\kappa_{i,0}:= 2 \pi /
L^{(i)}_c$ is the curvature of a circle with the same target length as
coil $i$.  We use a quartic instead of a quadratic penalty as this
more strongly penalizes extremal values.  The term $k_i$ is then used to
penalize (an approximation of) the maximum curvature by adding
$$
\frac{\delta}{4} \sum_{i = 1} ^{N_c} k_i(\mathbf{c})
$$
to the objective, where $\delta>0$ is a regularization
weight. 

Additionally, we might also want to avoid configurations with coils that are too close to each other.
To this end, we denote a minimal
  distance that we aim for the coils to satisfy by $d_\mathrm{min} > 0$ and for $i\neq j$ define
\begin{equation*}
d_{i,j}(\mathbf{c}) = \int_{\Gamma^{(i)}_{\!\mathbf{c}}}\int_{\Gamma^{(j)}_{\mathbf{c}}} \max(0, d_\mathrm{min} - \| \mathbf{r}_i - \mathbf{r}_j \|)^2 ~ dl_j\, dl_i,
\end{equation*}
where $\mathbf{r}_i$, $\mathbf{r}_j$ is a position on coils $\Gamma^{(i)}_{\!\mathbf{c}}$, $\Gamma^{(j)}_{\mathbf{c}}$, respectively.
Clearly $d_{i,j}(\mathbf{c})\ge 0$, and $d_{i,j}(\mathbf{c}) > 0$ if
and only if the distance between two coils is smaller than
$d_\mathrm{min}$ somewhere.  This function is once differentiable, and
we use $d_{i,j}(\mathbf{c})$ as quadratic penalty for the minimum
distance by adding
\begin{equation}
\frac{\gamma} 2
  \sum_{j<i} d_{i,j}(\mathbf{c}) \label{eqn:min_dist}
\end{equation}
to the objective, where $\gamma>0$ is a regularization
weight. 
It is possible to reduce the computational cost of the double sum in \eqref{eqn:min_dist} by taking into account symmetries and making assumptions on coils that are far away from one another.
For the results presented here, we use the double sum for convenience.
The minimum distance and maximum curvature regularization terms include the arclength as this has been shown to be important to prevent them from acting in a parametrization-dependent manner \cite{Bader2020}.

In summary, we want to find a minimizer of $ \hat{J}(\mathbf{c},\mathbf{a}, \sigma, \iota) $ in equation \eqref{eq:fhat}, constrained by the nonlinear first-order ordinary differential equation \eqref{eq:sigma}.
In practice, we eliminate the dependence of $\hat{J}$ on $\sigma, \iota$ using the constraint and optimize over the smaller space $\mathbf{c},\mathbf{a}$.  

\subsection{Discretization} \label{sec:optim_prob_discrete}
Next, we briefly summarize our discretization of the state equation
\eqref{eq:sigma} and the objective \eqref{eq:fhat}. To distinguish
functions from their finite-dimensional approximations, we use bold
letters to denote finite-dimensional approximation vectors, e.g., the
function $\sigma(\phi)$ corresponds to a vector $\bm \sigma \in \mathbb{R}^{n_\phi}$.

We approximate the solution to \eqref{eq:sigma}, $\sigma$ and $\iota$,
using a Fourier pseudo-spectral collocation discretization \cite{landreman_2019,weideman2000matlab}.  The numerical
solution, $(\bm \sigma,\iota) \in \mathbb{R}^{n_\phi + 1}$, then solves the finite-dimensional
nonlinear system $\mathbf{g}(\mathbf{a},\bm \sigma, \iota) =
\mathbf{0} \in \mathbb{R}^{n_\phi+1}$, the
components of which are 

\begin{equation} \label{eq:sigma_hh}
\begin{aligned}
g_q := \frac{|G_0|}{\ell'_q B_0}
D_q \bm{\sigma} + \iota  \left[ \frac{\bar \eta^4}{\kappa_q^4} + 1 + \sigma_q^2 \right]
+ \frac{2 G_0 \bar \eta^2 s_\psi \tau_q}{B_0 \kappa_q^2} &=0 \text{ for } q = 0 \hdots n_\phi-1, \\
g_{n_\phi} := \sigma_0 &= 0.
\end{aligned}
\end{equation}
Here, $D_q$ is the $q$th row of the spectral differentiation matrix $D$ on
the interval $[0,2\pi/N_{\text{fp}})$ with $n_\phi$ discretization points
  \cite{weideman2000matlab} and $\sigma_{n_\phi}=0$ due to
  periodicity.
  The approximate solution is
  $\bm {\sigma}= (\sigma_0, \hdots, \sigma_{n_\phi-1})^T$, where
  $\sigma_q \approx \sigma(\phi_q)$, $\phi_q = q (2\pi/N_{\text{fp}}) /n_\phi$.
  Additionally, $\ell'_q$, $\tau_q$, and $\kappa_q$ are the
  incremental arc length, torsion and curvature on the expansion axis
  at $\phi_q$.

Integrals are approximated by the trapezoidal rule, using uniformly spaced quadrature points, which is well known to converge spectrally for smooth, periodic functions \cite{trefethen_trapezoidal}.  Thus, after applying
quadrature, the discretized objective function is
\begin{align}\label{eq:fhat_h}
\begin{split}
\hat{J}(\mathbf{c},\mathbf{a}, \bm \sigma, \iota) &= \biggl(\frac{2 \pi}{n_\phi} \biggr)  \sum_{q = 0}^{n_\phi-1} \biggl[ \frac{1}{2}  \|\mathbf{B}_{\text{coils},q}(\mathbf{c}) - \mathbf{B}_{\text{QS},q}(\mathbf{a})\|^2 ~ \| \Gamma'_a(\phi_q) \| \biggr] \\
&+  \biggl(\frac{2 \pi}{n_\phi} \biggr)  \sum_{q = 0}^{n_\phi-1} \biggl[ \frac{1}{2} \| \nabla\mathbf{B}_{\text{coils},q}(\mathbf{c}) - \nabla\mathbf{B}_{\text{QS},q}(\mathbf{a}, \bm \sigma , \iota) \| ^2 ~ \| \Gamma'_a(\phi_q) \| \biggr] 
+ \frac{1}{2}\left(\frac{\iota-\iota_{0,a}}{\iota_{0,a}} \right)^2 + R(\mathbf{c},\mathbf{a}),
\end{split}
\end{align}
where $n_\phi$ is the number of equispaced quadrature points on $[0,2\pi/N_{\text{fp}})$ of the expansion axis and the subscript $q$ indicates that the term is evaluated at the
$q$th quadrature point along the expansion axis. The lengths of the
coils and axis in the regularization terms are
\begin{equation}
    L^{(i)}_{c} = \frac{2 \pi}{n_\theta}\sum_{r = 0}^{n_\theta-1} \| \Gamma^{(i)'}_{\mathbf{c}}(\theta_r) \| ~ \text{ and } ~ L_{a} = \frac{2 \pi}{n_\phi}\sum_{q = 0}^{n_\phi-1} \| \Gamma_{\!\mathbf{a}}'(\phi_q) \|,
\end{equation}
where $n_\theta$ is the number of quadrature points on each coil, and $\theta_r = 2r\pi / n_\theta$ is the $r$th quadrature point on a coil. Note a slight abuse of notation by using the same symbol
$\hat J$ for the continuous objective \eqref{eq:fhat} and its
discretization \eqref{eq:fhat_h}.

The quasisymetric field and its gradient, $\mathbf{B}_{\text{QS}}$,
$\nabla \mathbf{B}_{\text{QS}}$ in the discrete problem, are
straightforward to compute by evaluating \eqref{eq:B_QS} and
\eqref{eq:gradB_QS} at each quadrature point $\Gamma_{\!\mathbf{a}}(\phi_q)$ on the axis.  We use the
spectral differentiation matrix to determine an approximation of
$\sigma'(\phi_q)$ in \eqref{eq:sigma_hh} as required in
\eqref{eq:sderiv}.
The field generated by the coils $\mathbf{B}_{\text{coils}}$ in the
discrete problem are evaluated at each quadrature point on the axis
by using the Biot-Savart law, approximated using the trapezoidal rule
$$
\mathbf{B}_{\text{coils},q} = \frac{\mu_0}{4 \pi} \left(\frac{2\pi}{n_\theta}\right) \sum^{N_c}_{i = 1}\sum^{n_\theta-1}_{r = 0} I_i\frac{ \Gamma^{(i)'}_{\mathbf{c}}(\theta_r)\times \left[\Gamma_{\!\mathbf{a}}(\phi_q) - \Gamma^{(i)}_{\!\mathbf{c}}(\theta_r) \right]}{\|\Gamma_{\! \mathbf{a}}(\phi_q) - \Gamma^{(i)}_{\!\mathbf{c}}(\theta_r)\|^3},
$$
The gradient of the magnetic field generated by the coils, $\nabla
\mathbf{B}_{\text{coils},q}$, is computed in a similar manner.
Thus, given the design parameters
$\mathbf{c}$ and $\mathbf{a}$, the objective can  be evaluated.

\section{Analytical computation of derivatives}\label{sec:gradient}
Next, we summarize available techniques for computing
derivatives of optimization problems constrained by complex physics
equations. The derivations in this sections are formal, i.e., we
assume that all functions are sufficiently smooth and that the Jacobian of the constraint has full rank to guarantee existence of Lagrange multipliers. While
these techniques are well-known, we use this section to discuss
advantages and disadvantages of different approaches for the stellarator
optimization problem.
The discrete optimization problem discussed in section
\ref{sec:optim_prob_discrete} can be written as
\begin{equation} \label{eq:opt}
\begin{aligned}
    &\min_{\mathbf{c},\mathbf{a}, \bm \sigma, \iota} \hat{J}(\mathbf{c},\mathbf{a}, \bm \sigma, \iota), \\
    &\text{subject to }  \mathbf{g}(\mathbf{a},  \bm \sigma, \iota) = 0.
\end{aligned}
\end{equation}
Solving \eqref{eq:opt} requires optimizing $\hat{J}$ over $\mathbf{c}$, $\mathbf{a}$,  $\bm \sigma$ and $\iota$ while
taking into account the nonlinear equality constraint
equation $\mathbf{g}(\mathbf a, \bm \sigma, \iota)=0$.
When this equality constraint admits a unique solution, $(\bm \sigma,
\iota)$ for a given axis parameters $\mathbf{a}$, we can use the
implicit function theorem to transform \eqref{eq:opt} into an
equivalent unconstrained problem over a smaller parameter space
\begin{equation} \label{eq:opt_red}
    \begin{aligned}
    &\min_{\mathbf{c}, \mathbf{a}} J(\mathbf{c},\mathbf{a}),
\end{aligned}
\end{equation}
where $J(\mathbf{c},\mathbf{a}) = \hat{J}(\mathbf{c},\mathbf{a}, \bm
\sigma(\mathbf{a}), \iota(\mathbf{a}) )$, and $\bm \sigma(\mathbf{a}),
\iota(\mathbf{a})$ are considered as implicitly defined functions of $\mathbf{a}$
through the solution of $\mathbf{g}(\mathbf{a}, \bm \sigma, \iota) = 0$.
The main advantage of \eqref{eq:opt_red} compared to \eqref{eq:opt} is
that we have eliminated the equality constraint, allowing us to
use methods from unconstrained optimization. However, the implicit
dependence of $(\bm \sigma,\iota)$ on $\mathbf a$ makes the
computation of derivatives of $J$ in \eqref{eq:opt_red}
challenging.
In the following, we summarize the \emph{forward} and \emph{adjoint
  sensitivity} approaches to compute gradients of $J$ and illustrate
their application to the stellarator optimization problem \eqref{eq:opt_red}.
Our derivations are for the discretized equation and
objective, which avoids inconsistencies compared to when derivatives are computed in
infinite dimensions, followed by discretization of the resulting
equations.  Standard references for computation of derivatives for
problems constrained by infinite-dimensional ordinary of partial differential equations are
\cite{BorziSchulz12, Delosreyes15, Gunzburger03}.

In both, forward and adjoint sensitivity methods, for the optimization
problem \eqref{eq:opt_red} we have
\begin{equation} \label{eq:dfdc}
\frac{\partial J}{\partial \mathbf{c}} = \frac{\partial \hat J}{\partial \mathbf{c}}.
\end{equation}
since the coil design parameters ($\mathbf{c}$) do not affect the state variables ($\bm \sigma$, $\iota$).  In
contrast, the axis design parameters ($\mathbf{a}$) directly affect the state
variables, thus we use either a forward or an adjoint sensitivity
approach to determine the gradient of $J$ with respect to
$\mathbf{a}$.

\subsection{Gradient from forward sensitivities}
One approach to computing the gradient of the objective is to use forward sensitivities.
Direct application of the chain rule to \eqref{eq:opt_red} gives
\begin{equation} \label{eq:fs}
   \frac{\partial J}{\partial \mathbf{a}} = \frac{\partial \hat J}{\partial
     \mathbf{a}} + \frac{\partial \hat J}{\partial (\bm \sigma,
     \iota)}  \frac{d (\bm \sigma, \iota)}{d\mathbf{a}},
\end{equation}
where ${\partial (\bm \sigma, \iota)}$ denotes partial differentiation
with respect to the components of $\bm \sigma$ and with respect to
$\iota$.
Next, we differentiate the equality constraint \eqref{eq:sigma_hh} to
obtain
\begin{equation}\label{eq:diff}
   \frac{\partial \mathbf g}{\partial (\bm \sigma, \iota)}  \frac{d (\bm \sigma, \iota)}{d\mathbf a} = -  \frac{\partial \mathbf g}{\partial \mathbf{a}}.
\end{equation}
This identity can be used to compute the sensitivity matrix $d (\bm
\sigma, \iota) / d \mathbf a$, where the computation of each column requires
a linear solve with the matrix ${\partial \mathbf g}/{\partial (\bm
  \sigma, \iota)}$ and a different right hand side, namely the columns
of $\partial \mathbf g / \partial \mathbf{a}$.  The sensitivity
matrix is then used in \eqref{eq:fs} to compute the gradient
${\partial J}/{\partial \mathbf{a}}$.

\subsection{Gradient using adjoint sensitivities}
An alternative approach to computing the analytical gradient of $J$ is
using the adjoint method. The approach  can
be  derived in different ways. Here, we derive it from \eqref{eq:fs}
and \eqref{eq:diff} by introducing a Langrange multiplier vector $\bm\lambda$
as solution of the (linear) adjoint system
\begin{equation}\label{eq:adjoint}
\biggl[\frac{\partial \mathbf{g}}{\partial (\bm \sigma,\iota)} \biggr]^T \bm \lambda = - \frac{\partial \hat{J}}{\partial (\bm \sigma,\iota)}.
\end{equation}
Using the definition of $\bm\lambda$ in \eqref{eq:fs} shows that the
gradient of $J$ can also be computed as
\begin{equation} \label{eq:adjoint2}
\frac{\partial J}{\partial \mathbf{a}} = \frac{\partial \hat{J}}{\partial \mathbf{a}} + \bm \lambda^T \frac{\partial \mathbf{g}}{\partial \mathbf{a}}.
\end{equation}
Using the adjoint approach only requires solving the nonlinear state
equation \eqref{eq:sigma_hh} for $\bm \sigma$, $\iota$ and one linear
equation \eqref{eq:adjoint} for $\bm \lambda$.  In particular, the number
of equation solves is independent of the number of axis design
variables.

The expressions for the partial derivatives used in \eqref{eq:dfdc},
\eqref{eq:adjoint} and \eqref{eq:adjoint2} are complex and
details are provided in \ref{app:state} and \ref{app:objective}.
However, as an illustration of the adjoint approach, we differentiate here the
simple objective function
\begin{equation}
\begin{aligned}
    \hat{J}(\mathbf{a}, \iota) =  \frac{1}{2}\biggl( \frac{\iota - \iota_{0,a}}{\iota_{0,a}} \biggr)^2,
\end{aligned}
\end{equation}
with respect to $\bar \eta \in \mathbf{a}$.
Substituting the expressions for $\hat{J}$ and $\mathbf{g}$ in \eqref{eq:adjoint2} and simplifying, the derivative becomes 
\begin{align*}
\frac{\partial J}{\partial \bar \eta} = \frac{\partial \hat{J}}{\partial \bar \eta } + \bm \lambda^T \frac{\partial \mathbf{g}}{\partial \bar \eta} 
=\sum^{n_\phi-1}_{i = 0} \lambda_i \biggl( \frac{4\iota\bar \eta^3}{\kappa^4_i} + \frac{4 G_0 \bar \eta s_{\psi}\tau_i}{B_0 \kappa^2_i} \biggr),
\end{align*}
where $\bm \lambda = (\lambda_0, \lambda_1, \cdots,
\lambda_{n_\phi})^T$ solves the adjoint equation \eqref{eq:adjoint}.

\subsection{Comparison between forward and adjoint approaches}
Next we compare the above approaches to compute gradients in terms of
computational efficiency. The forward sensitivity approach requires as
many solves of a linearized state equation as there are parameters,
where each problem uses the same system matrix but a different right
hand side. When the number of parameters is moderate (as for the
problem considered in this paper), this linearized system matrix can
be assembled and stored.  One can thus compute a matrix factorization
and reuse it for each linear solve.
In contrast, the adjoint method requires only a single linear solve
independently of the number of parameters. The adjoint system matrix
is the transpose of the linearized state matrix. Thus, if that matrix
can be assembled (as in the case here), the adjoint system can be
found straightforwardly. When matrices are too large to be assembled
and stored, formulation and solution of the adjoint system can be a
more substantive challenge, e.g., \cite{BorziSchulz12, Gunzburger03}.
For the problems we consider in this article, both methods perform
similarly well due to the moderate parameter dimension (usually, a
few hundred). Additionally, a dominant cost is the construction of the
matrix $\partial \mathbf{g} / \partial (\bm \sigma,\iota)$ and the
vector $\partial \hat{J}/\partial (\bm \sigma,\iota)$, which are
required for both the forward and adjoint sensitivity methods.

Note that optimization with gradients computed using
the adjoint and forward sensitivity approaches vastly 
outperforms derivative-free optimization methods or gradient-based
optimization with finite difference gradients. The performance of
derivative-free methods \cite{ConnScheinbergVicente09} typically
degrades with increasing parameter dimension and thus requires many
objective evaluations. For the problem considered here, each objective
evaluation requires the solution of the nonlinear state equation and is
thus expensive.  Similarly, computation of gradients with finite
differences requires as many objective evaluations, and thus solutions
of the nonlinear state equation, as there are design
parameters. One cannot reuse matrix factorizations as in the
forward sensitivity approach, and the method therefore becomes vastly more
expensive.  However, finite difference (directional) gradients are a
useful tool to verify gradients computed with the forward or adjoint
sensitivity approach, as used in the next section.

\subsection{Implementation and gradient verification} \label{sec:gradver}
We have implemented the optimization objective and its gradient using
the forward and adjoint sensitivity approaches in Python and MATLAB.
In both implementations, the evaluation of the magnetic field and its derivatives with the Biot-Savart law is a dominating part of the objective function and gradient evaluation run time.  
Therefore, we call an optimized C++ implementation of the Biot-Savart law.

\begin{figure}[tb]
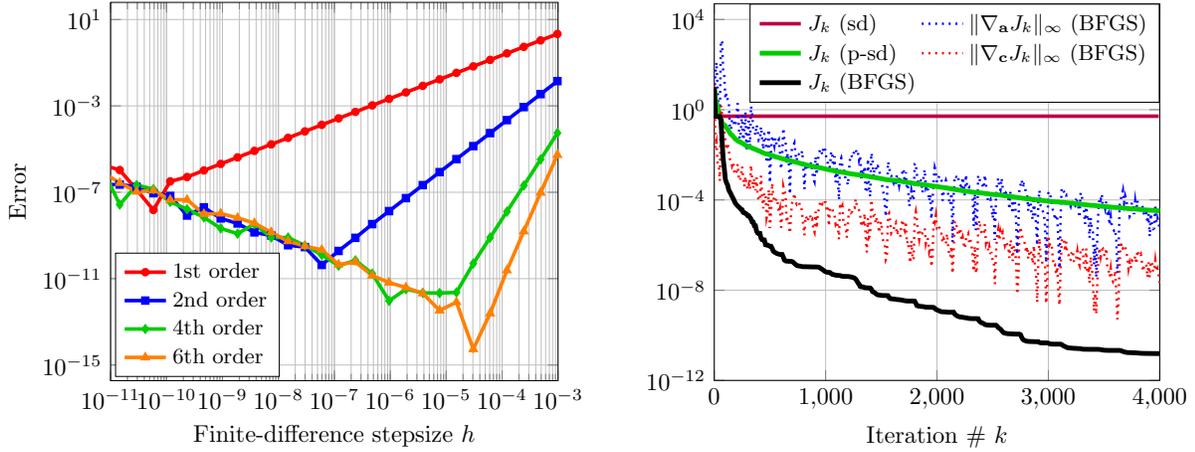

   \begin{center}
     \includestandalone[mode=buildnew, height=.38\textwidth]{taylortest/taylortest}
     \hfill
     \includestandalone[mode=buildnew, height=.38\textwidth]{problem1/convergence_history/convergence_history}
   \end{center}
   \caption{Shown on the left is a comparison between analytical derivatives and
     finite-difference approximation using different orders. Plotted
     are the absolute errors of the directional derivatives in a
     random direction. The right figure shows the convergence behavior for the steepest descent, a
     preconditioned steepest descent, and the BFGS quasi Newton
     algorithms for Problem I. Besides the evolution of the objective function, we
     show the max-norm of the part of the gradient with
     respect to the expansion axis ($\|\nabla_{\mathbf{a}} J
     \|_{\infty}$) and the coil design parameters ($\|
     \nabla_{\mathbf{c}} J \|_{\infty}$) during the BFGS iteration.
     \label{fig:taylor_convergence}}
\end{figure}

Before presenting two stellarator coil optimization problems in
Sections \ref{sec:p1} and \ref{sec:p2}, we numerically verify our
derivations and implementation of the analytical gradients by
comparing with finite-difference gradients.
Namely, we compare the directional derivative in a random direction
with standard first, second, fourth and sixth-order finite difference
approximations.  These derivatives are computed at the initial
configuration of Example \ref{sec:p2}. Our results are shown in the
left plot of Figure \ref{fig:taylor_convergence}. These results are based
on forward sensitivities, but we have verified that the adjoint
approach gives identical results.
As can be seen, the finite
difference step size $h$ is reduced, the difference between the finite
difference approximation and the analytical directional gradient is
reduced at the expected rate, i.e., at a linear rate for first-order
finite differences, a quadratic rate for second-order finite
differences, and so on. As expected, we also see that all finite
difference approximations are numerically unstable and floating point
error eventually leads to an increase of the error again.  These
results indicate the correctness of our analytical expressions and
their implementation for the gradient of the discretized problem
\eqref{eq:fhat_h}.

We also observe that standard first-order finite differencing at most
gives seven digits of accuracy when the ideal step size $h$ is chosen.
The fourth and sixth order approximations level off at an error
of approximately $10^{-12}$ and $10^{-14}$ respectively.  However, we emphasize that first-order
finite differences only require a single additional objective
evaluation, whereas the fourth order finite difference approximation
requires four evaluations for each directional derivative.

\section{Problem I: A quasi-axisymmetric vacuum field with low
  rotational transform for electron-positron confinement}\label{sec:p1}

In this and the next section, we study the performance of
gradient-based optimization algorithms for minimization of
\eqref{eq:fhat_h}, reformulated as an unconstrained optimization of
the form \eqref{eq:opt_red}.
We also study the optimization landscape, the
influence of regularization, and properties of the obtained coil
designs.
In these sections, we set the constants $s_{\psi} = s_G = 1$.

As a first test, we solve \eqref{eq:opt_red} with regularization
\begin{equation}\label{eq:reg1}
R(\mathbf{c},\mathbf{a}) := \frac{1}{2}\sum_{i = 1}^{N_c}\left(\frac{L^{(i)}_{c}(\mathbf{c}) -L_{0,c}}{L_{0,c}}\right)^2 + \frac{1}{2}\left(\frac{L_a(\mathbf{a}) -L_{0,a}}{L_{0,a}}\right)^2,
\end{equation}
and the target quantities
\begin{equation}
    \begin{aligned}
    L_{c,0} = 0.7 \times 2\pi,\quad
    L_{a,0} = 2\pi,\quad
    \iota_{0,a} = 0.103,
    \end{aligned}
\end{equation}
where units of meters are assumed throughout.
We use $n^{\text{coil}}_p=n^{\text{axis}}_p = 9$, and $N_c = 4$, initially flat, modular coils and
apply $2$-fold toroidal symmetry as well as stellarator symmetry. The
coils are arranged uniformly on the initially flat axis as shown in
Figures \ref{fig:problem1_1234}a, \ref{fig:problem1_1234}b, where the initial currents in the coils are zero, we fix the constant $B_0 = 1$, and use the initialization $\bar \eta = 1$.
The target length of the coils and axis correspond to those in the
initial configuration. Therefore, the regularization terms in
\eqref{eq:reg1} are initially zero.  
The target rotational transform is chosen as relatively small.
Thus the desired $\iota$ and quasisymmetry should be achievable to a good approximation with coils of moderate curvature.
Since simple planar coil configurations easily yield near-perfect axisymmetry with $\iota=0$, physical intuition suggests that achieving a small iota and quasiaxisymmetry may not require a large curvature of the coils.
High-performing stellarators typically have larger
rotational transforms \cite{Zarnstorff2001,Drevlak_2013,Geiger_2014,Henneberg_2019}, but the magnetic configuration we will obtain with this set-up of the optimization problem is relevant for experiments designed to confine and study electron-positron pair plasmas \cite{stenson2019,stoneking2020}. For a low temperature pair plasma, with temperatures of the order of 1 electron-volt, confined in a relatively strong magnetic field, of the order of 1 or 2 Tesla, the drift of charged particles in a purely toroidal magnetic field is slow, because the particles are strongly magnetized. Therefore, a small amount of poloidal field, and a correspondingly small rotational transform is sufficient to provide the desired quality of confinement.

\subsection{Gradient-based optimization performance}\label{subsec:p1:algos}
We solve the stellarator optimization problem with various
gradient-based methods, namely steepest descent with and without
preconditioning, and a quasi-Newton method, which uses consecutive
gradients to approximate the Hessian using the BFGS algorithm
\cite{NocedalWright06}.  All methods use a cubic line search procedure
to find an appropriate step length with sufficient decrease of the
objective.

First, in the right plot of Figure \ref{fig:taylor_convergence} we compare the convergence of
the steepest descent algorithms and the BFGS quasi-Newton method. For clarity of the figure, we only show the convergence for the norm of the gradient for the BFGS quasi-Newton methods. For
the steepest descent method, we find that the objective function plateaus at $(J, \| \nabla J \|_{\infty}) = (5.15 \times 10^{-1}, 4.47 \times
10^{-1})$. Little progress is made over 4,000 iterations, which require
6,250 function and gradient evaluations.  The excess function
evaluations are due to the line search procedure. This slow
convergence is due to the vastly different sensitivities of the
objective with respect to the parameters, which we study in more
detail in the next subsection.  Improved convergence is obtained by
multiplying the steepest descent direction with a preconditioning
matrix that helps to balance these sensitivities. The preconditioning
matrix we use is motivated by the sensitivity study in the next
subsection and detailed there. The convergence history of the
preconditioned gradient descent method seen on the right in Figure
\ref{fig:taylor_convergence} shows that after 4,000 iterations and 6,199
function and gradient evaluations, we obtain $(J, \| \nabla J \|_{\infty}) =
(3.26\times 10^{-5},1.12)$. As can be seen, the norm of the
gradient does not decrease substantially. However, the norm of the
preconditioned gradient decreases by about 4 orders of magnitude,
resulting in substantial decrease of the objective.

After 4,000 iterations of the quasi-Newton method, we have $(J, \| \nabla J \|_{\infty}) = (1.54 \times 10^{-11}, 4.58 \times 10^{-6})$.
The resulting coil design is shown in Figures \ref{fig:problem1_1234}c,
\ref{fig:problem1_1234}d and are discussed in section \ref{subsec:p1:physics}.
These 4,000 iterations required 4,629 evaluations of $J$ and its gradient,
where the excess evaluations are again due to the line search
procedure.  From Figure \ref{fig:taylor_convergence} it can also be seen
that the norm of the part of the gradient corresponding to the
expansion axis, $\nabla_\mathbf{a} J$, dominates the norm of the part
of the gradient corresponding to the coils $\nabla_\mathbf{c} J$.
The much improved performance of  the quasi-Newton algorithms as opposed to the steepest-descent algorithms shows
that it is crucial for fast convergence to use second-order
information.

Note that using a Fourier expansion with $n_{p}^{\text{coil}} = 9$ modes is a modeling decision and has the advantage that the coils will be easier to manufacture than those with more modes.
The Fourier coefficients of the first modular coil in the optimized configuration decay with increasing mode number (Table \ref{tab:fourier_coeffs_ex1}), typically by several orders of magnitude even though the objective in this example does not explicitly include regularization on coil curvature and torsion.
Finally, in contrast to finite difference methods, enriching the coil parameter space does not directly increase the computational cost associated with calculating the gradient when using an adjoint approach, as done here.

\begin{table}[]
\small
    \centering
    \begin{tabular}{|c|c|c|c|c|c|c|}
    \hline
    Mode \#& sin $x$  & cos $x$  & sin $y$  & cos $y$  & sin $z$  & cos $z$  \\
    \hline
0 & - & 1.0884e+00 & - & 2.2511e-01 & - & 2.6995e-02 \\
\hline
1& -5.8491e-03 & 6.5503e-01 & -2.8165e-02 & 1.2335e-01 & 6.2858e-01 & -5.3515e-03 \\
\hline
2 & -8.0674e-03 & 3.9426e-02 & 9.8105e-02 & 3.0468e-02 & 3.5265e-02 & -2.1969e-02 \\
\hline
3 & -2.3760e-03 & 4.1630e-02 & -5.7865e-02 & -7.1681e-03 & 1.6093e-02 & 8.6299e-04 \\
\hline
4 & 1.9382e-03 & -1.1647e-02 & 5.9301e-03 & 1.0502e-02 & 2.7819e-03 & 1.3789e-02 \\
\hline
5 & 9.1377e-04 & -5.2997e-03 & 1.2073e-02 & 1.2459e-03 & 8.0925e-03 & 5.7870e-03 \\
\hline
6 & -5.9186e-03 & 1.4024e-02 & 2.8668e-03 & -2.0511e-03 & 3.1622e-03 & -5.1570e-03 \\
\hline
7 & 5.9704e-03 & -1.0150e-02 & -4.8427e-03 & -8.0999e-03 & -6.4598e-03 & -7.1050e-03 \\
\hline
8 & -1.4711e-03 & 3.4307e-03 & -3.0153e-03 & 7.3435e-03 & 2.1565e-03 & 3.0483e-03 \\
\hline
9 & -8.2496e-04 & -7.6827e-04 & -4.1981e-04 & 4.4477e-05 & -9.8108e-04 & 2.3497e-03 \\
\hline
    \end{tabular}
    \caption{Fourier coefficients of the $x$, $y$, and $z$ expansions \eqref{eq:coils} associated to the first modular coil in the optimized configuration computed in Problem I.}
    \label{tab:fourier_coeffs_ex1}
\end{table}

\subsection{Local study of optimization landscape}\label{subsec:p1:landscape}
Next, we study the optimization landscape locally around the solution
we found with the BFGS method.  For convenience of notation, we denote
the parameters for the coils and the expansion axis corresponding to
this (approximate) minimum by $\mathbf{x}^\star := (\mathbf{c}^\star,
\mathbf{a}^\star)$.  Since the gradient vanishes at
$\mathbf{x}^\star$, the objective function in the neighborhood of
$\mathbf{x}^\star$ can be written as
\begin{equation}\label{eq:f_approx}
J(\mathbf{x}^\star + \mathbf{h}) = J(\mathbf{x}^\star) + \frac 12\mathbf{h}^T
H \mathbf{h} + \mathcal{O}(\|\mathbf{h}\|^3),
\end{equation}
where $H=\nabla^2 J(\mathbf{x}^\star)$ is the Hessian of $J$ at
$\mathbf{x}^\star$. This Taylor expansion shows that the local
landscape of the objective around the minimizer is governed by
properties of the Hessian matrix $H$. While we do not have direct
access to second derivatives, we can approximate the application of
$H$ to a vector using finite differences of gradients. Applying $H$ to
unit vectors, we can thus construct an approximation to the Hessian
matrix. Due to finite-difference errors, the approximate Hessian matrix constructed in this way would not be exactly symmetric. Thus,
in the following we use the symmetrization $\bar H:=(\tilde H+\tilde
H^T)/2$, where $\tilde H$ is obtained using finite differences of
gradients.

\begin{figure}[tb]
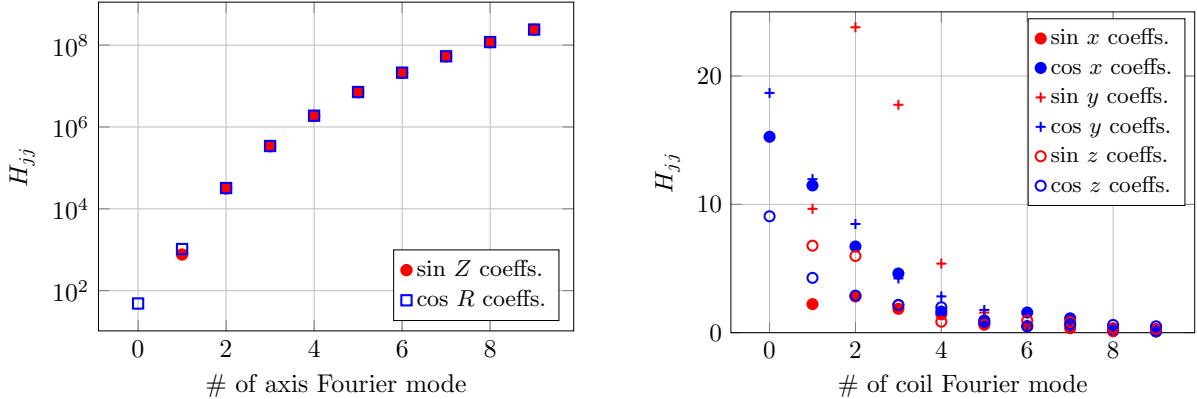

   \begin{center}
    \includestandalone[mode=buildnew, width=0.48\textwidth]{problem1/eigenvalues/diag_axis}
    \hfill
    \includestandalone[mode=buildnew, width=0.48\textwidth]{problem1/eigenvalues/diag_coil}
   \end{center}
   \caption{Hessian diagonal entries for Problem I: Shown on the left are the diagonal entries
     corresponding to the Fourier coefficients of the $R$ and $Z$
     coordinates of the expansion axis. Shown on the right are the
     diagonal entries corresponding to the Fourier
     coefficients of the $x$, $y$ and $z$-coordinates of the first
     coil. For the expansion axis, the sensitivity of the objective
     grows with the mode number. For the coil, the sensitivity
     decreases with the mode number.} \label{fig:diag_axis_coils}
\end{figure}

We first focus on the diagonal entries $\bar H_{jj}$ of the Hessian
$\bar H$. These entries describe how sensitive the objective is to the
$j$th design parameter in a neighborhood of $\mathbf{x}^\star$, which
follows by choosing $\mathbf{h} = \varepsilon \mathbf{e}_j$ in
\eqref{eq:f_approx}, where $\mathbf{e}_j$ is the $j$th unit vector,
and $\varepsilon>0$ is small.
If $\bar H_{jj}$ is large, then the objective function is very
sensitive to variation of the $j$th design parameter.  Alternatively,
if $\bar H_{jj}$ is small or even zero, the objective function is
insensitive to variation in the $j$th design parameter.  On the left
in Figure \ref{fig:diag_axis_coils}, we show the diagonal entries of $\bar
H$ that correspond to the Fourier coefficients of the $R$ and
$Z$-coordinates of the expansion axis.  Clearly, the objective is
highly sensitive to the magnetic axis as the corresponding diagonal
entries of $\bar H$ range from $\mathcal{O}(10^2)$ to
$\mathcal{O}(10^8)$.  Moreover, it can be seen that the objective is
several orders of magnitude more sensitive to higher Fourier mode
perturbations of the expansion axis than to low Fourier mode perturbations.

On the right in Figure \ref{fig:diag_axis_coils}, we show the diagonal
entries of $\bar H$ that correspond to the sine and cosine Fourier
coefficients of the first coil's coordinates.  Note that the objective
function is not as sensitive to the coil parameters as it is to the
expansion axis.  Moreover, in contrast to what we found for the
expansion axis, $J$ is more sensitive to lower than to higher Fourier
mode coil perturbations.  This is due to the fact that only the evaluation
of the magnetic field (and its gradient) at the expansion axis enter
in the objective. Since the expansion axis is relatively far away from
the coils and magnetic fields may cancel each other, higher Fourier
modes in the coil representations have a smaller influence on the
objective.

These different sensitivities result in extremely stretched contour
lines of the objective around $\mathbf{x}^\star$. This explains the
slow convergence of the (un-preconditioned) steepest descent method
\cite{NocedalWright06} we observed in subsection
\ref{subsec:p1:algos}. Namely, the narrow valleys in the objective
landscape result in small steps due to a behavior sometimes referred
to as ``zig-zagging'' \cite{NocedalWright06}. As discussed above,
using a preconditioner improves the slow convergence of the steepest
descent method.  To illustrate this we used the inverse of a modified
approximation of the Hessian as a preconditioning matrix for the
results shown on the right in Figure \ref{fig:taylor_convergence}.
Specifically, we use the previously computed symmetrized finite
difference approximation of the Hessian and modify it to be positive
definite by adding the identity matrix $(\bar{H}+I)^{-1}$. Clearly,
this preconditioner is not available in practice as the Hessian at the
minimizer, and the minimizer itself, are not available. However, this
experiment highlights that preconditioning is crucial for problems
with extremely different parameter sensitivities.  The quasi-Newton
BFGS method that we use iteratively builds a Hessian approximation for
preconditioning and thus significantly improves the convergence.

\begin{figure}[H]
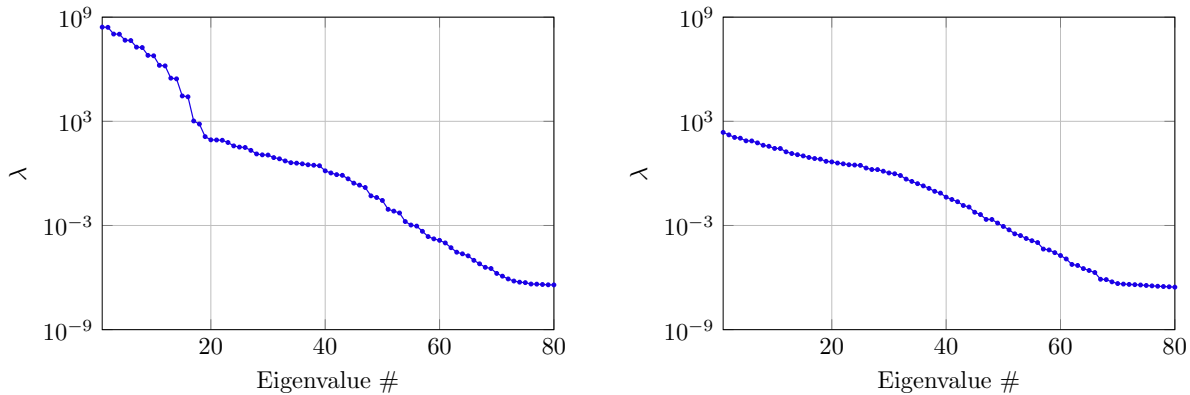

   \begin{center}
       \includestandalone[mode=buildnew, width=0.48\textwidth]{problem1/eigenvalues/hessianeig}
           \hfill
       \includestandalone[mode=buildnew, width=0.48\textwidth]{problem1/eigenvalues/reduced_hessian_coils}
   \end{center}
   \caption{Hessian spectra for Problem I: Shown on the left are the largest (out of 252) eigenvalues
     of the Hessian approximation $\bar H$. The strong
     variation in eigenvalues results in extremely stretched contours
     of the objective and shows that the objective is mostly sensitive
     to perturbations in $\sim40$ parameter space directions, but
     mostly insensitive in other directions. Shown on the right are the
     largest (out of 228) eigenvalues of $\bar H$ when the
     magnetic axis and coil currents are fixed.  The large number of small eigenvalues
     indicate that we can introduce additional design constraints on
     the coils.} \label{fig:hessian}
\end{figure}

After studying sensitivities with respect to individual parameters, we
now study the spectrum of the (approximate) Hessian $\bar H$ at
$\mathbf{x}^\star$. At a minimizer, the Hessian must be positive
semi-definite and the eigenvalues provide insight into the local
geometry of the objective. In Figure \ref{fig:hessian} (left), we show
the rapid decay of eigenvalues taking into account all 252
design parameters. In the right figure, we show the eigenvalues
obtained when only considering the subspace of dimension 228  corresponding to the Fourier representation of the coil
parameters. The large eigenvalues correspond to eigenvectors, i.e.,
directions in the parameter space, in which the objective has large
curvature and thus is very sensitive to perturbations in these
directions.  Clearly, there are many directions in which the minimizer
can be perturbed that do not substantially affect the value of the
objective function.  Some of these directions are due to the coil
parametrization we use (section \ref{sec:os}), while others indicate that there is
substantial freedom to introduce additional design constraints on the coils or include coil manufacturing uncertainty, without changing the objective substantially.  We examine this freedom
in the next numerical example.

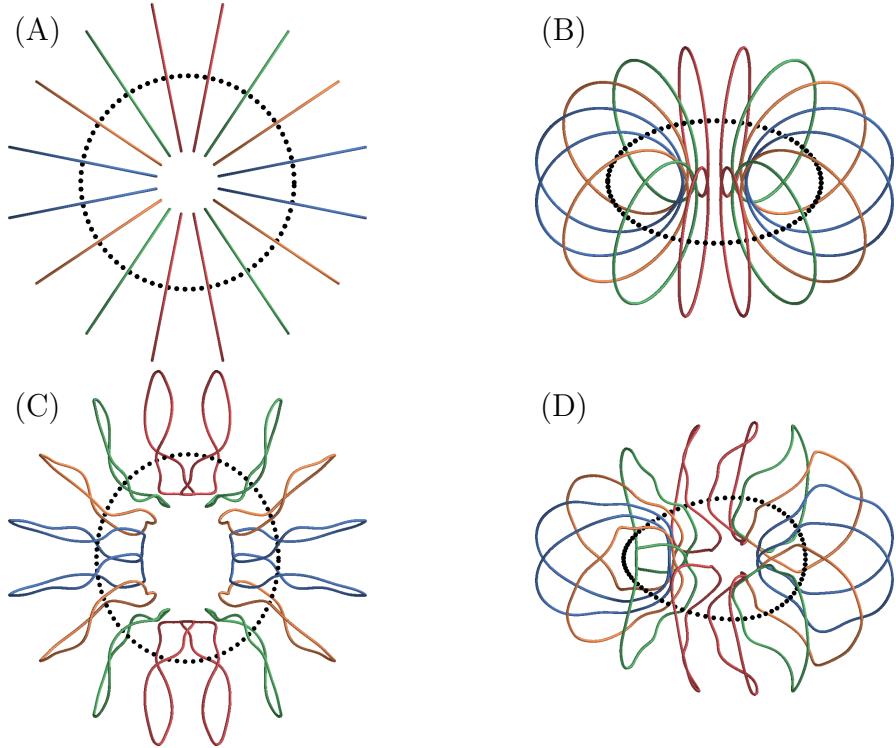
\begin{figure}[tb]
\centering
    \begin{tikzpicture}
    \node (11) at (0,0) 
    {\includegraphics[width=0.3\textwidth]{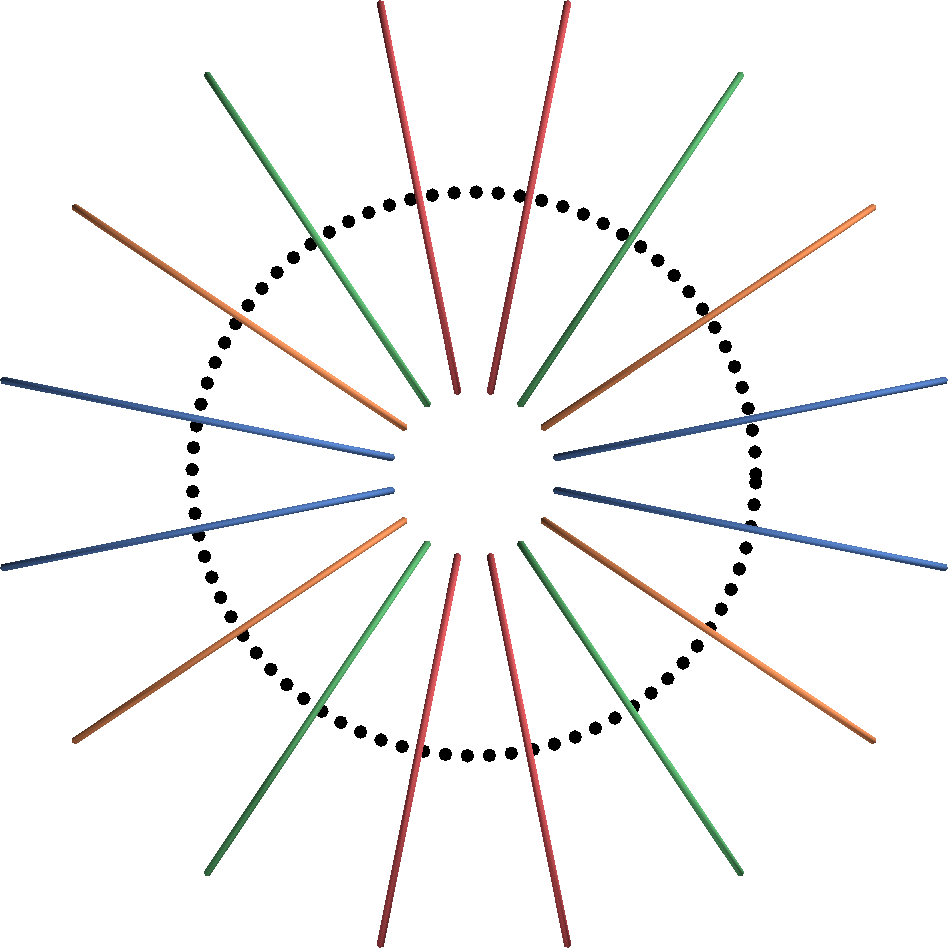}};
    \node (12) at (7,0) 
    {\includegraphics[width=0.3\textwidth]{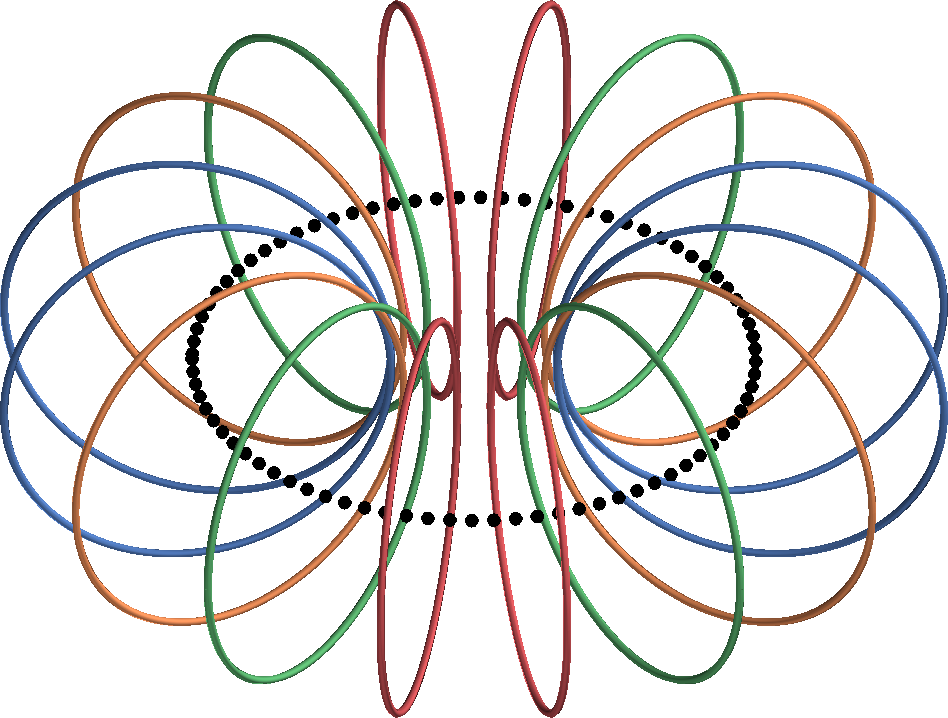}};
    \node (21) at (0,-5) 
    {\includegraphics[width=0.3\textwidth]{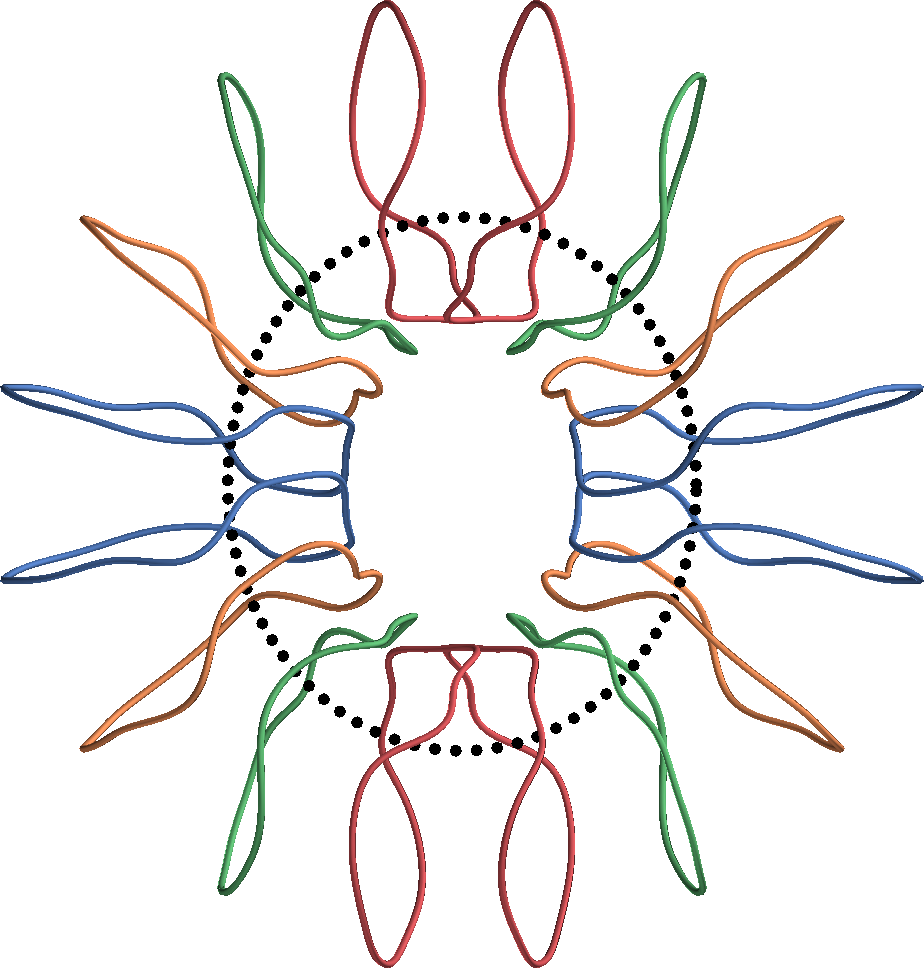}};
    \node (22) at (7,-5) 
    {\includegraphics[width=0.3\textwidth]{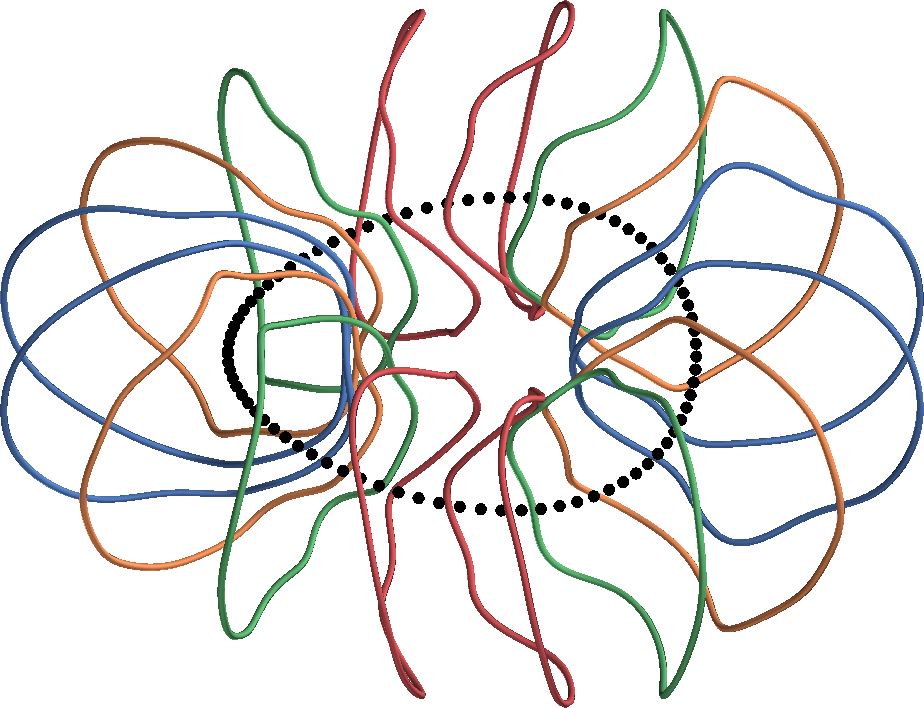}};
    
    \node at (-2,2) {\textcolor{black}{\large (A)}};
    \node at (5,2) {\textcolor{black}{\large (B)}};
    \node at (-2,-3) {\textcolor{black}{\large (C)}};
    \node at (5,-3) {\textcolor{black}{\large (D)}};
    \end{tikzpicture}
    \caption{Initialization and optimized designs for Problem I: Shown are two views of the initial (upper row) and
      optimized (lower row) coils (in red, blue, green, magenta) and
      expansion axis (dotted black). Coils that correspond to the same modular coil after application of the symmetries are drawn in the same color.  For the optimized setup, the expansion
      axis and the magnetic axis corresponding to the coils
      coincide. }
    \label{fig:problem1_1234}
\end{figure}

\subsection{Discussion of coil designs}\label{subsec:p1:physics}
We now discuss the physical properties of the optimal solution obtained by the BFGS algorithm.
In Figure \ref{fig:problem1_1234}, we show the initial and final
coil configurations.  The initially flat coils are uniformly arranged around
the initially flat expansion axis (Figures \ref{fig:problem1_1234}a,
\ref{fig:problem1_1234}b).  From this initially axisymmetric
configuration, the coils and axis reach a non-axisymmetric
configuration where each individual term in the objective is nearly zero (Figures \ref{fig:problem1_1234}c,
\ref{fig:problem1_1234}d). 

\begin{figure}
      \begin{tikzpicture}
            \node (11) at (-4,4)
            { \includegraphics[width=0.32\textwidth]{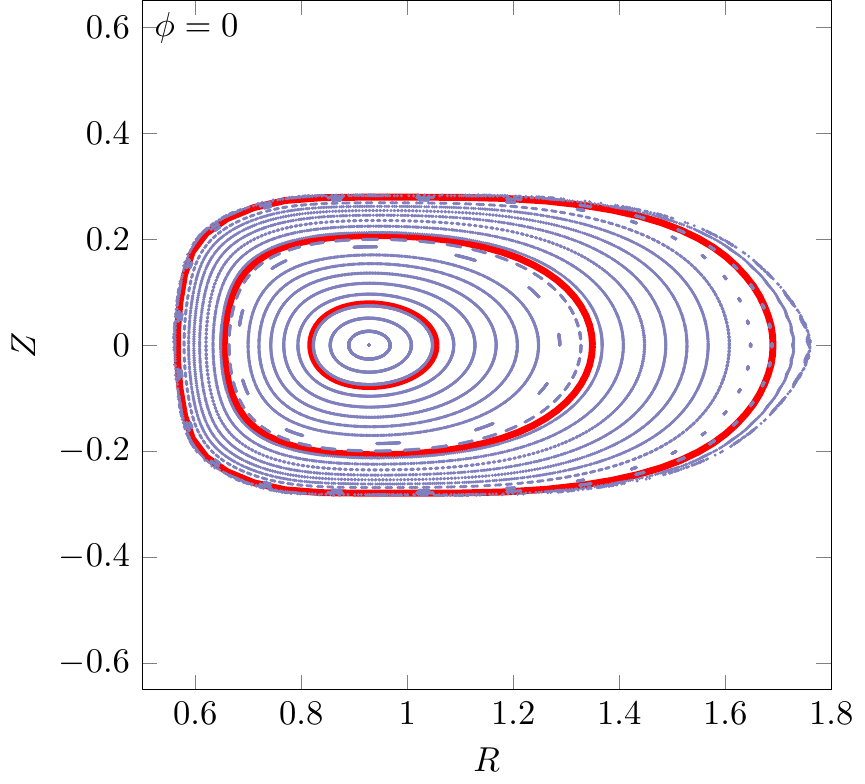} };
            \node (12) at (1,4)
            { \includegraphics[width=0.305\textwidth]{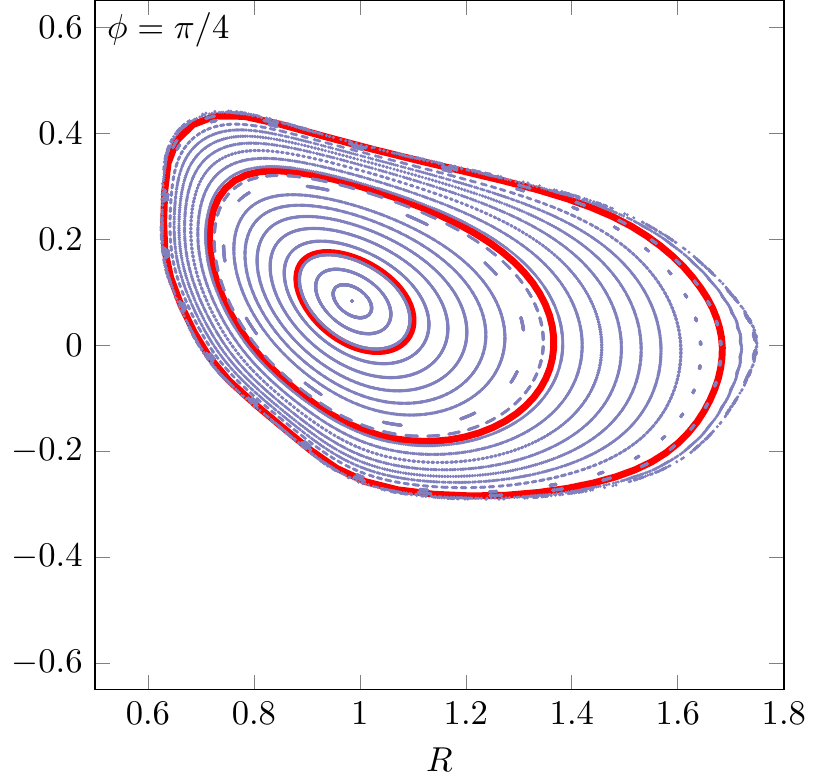} };
            \node (13) at (6,4)
            { \includegraphics[width=0.305\textwidth]{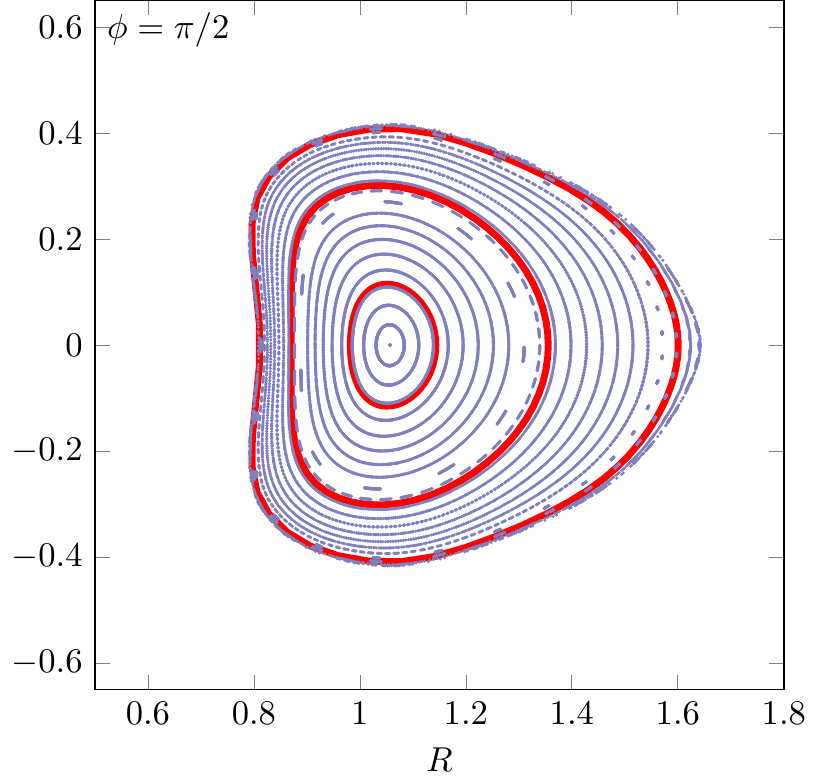} };
    \end{tikzpicture}
     \caption{Poincare plots for final design in Problem I.  These plots are generated by following field lines around the origin and computing their intersection with the cylindrical angles $\phi = 0$, $\pi/4$, and $\pi/2$, shown as blue dots.  The red lines correspond to the three surfaces plotted in Figure \ref{fig:problem1_surfaces}.}\label{fig:p1poincare}
\end{figure}

We study the magnetic field generated by the optimized coils in two different ways. The first method is through Poincar\'e plots: starting at a large number of radial initial positions along the midplane $Z=0$ at the toroidal angle $\phi=0$, we numerically integrate the coupled ordinary differential equations to trace field lines, and compute their intersections with the plane $\phi=0$, as well as the planes $\phi=\pi/4$ and $\phi=\pi/2$. Poincar\'e plots give a visual indication of the existence of flux surfaces, which are closed toroidal surfaces on which the magnetic field is everywhere tangent to the surface, as well as the existence of magnetic islands and regions in which the magnetic field is chaotic. Large islands and chaotic magnetic field lines must be avoided in the plasma core to guarantee good confinement properties. Figure \ref{fig:p1poincare} demonstrates that our optimized coil configuration produces a vacuum magnetic field with nested toroidal flux surfaces over a large volume, and with no significant internal island chains. This is remarkable, since in our formulation of the optimization problem, we do not have terms which explicitly drive the optimum to a configuration with nested surfaces. We obtain a vacuum field with such a desirable property for at least two reasons: quasi-symmetric fields approximately constructed with the Garren-Boozer expansion are empirically found to have a large set of nested flux surfaces \cite{landreman_2019}, and by focusing on a low value of the rotational transform, we avoid the large islands and chaotic magnetic fields which would be present at radii where the rotational transform is a low-order rational number \cite{Wobig1987}. 

\begin{figure}[H]
   \begin{tikzpicture}
   \node (11) at (-4,4.1)
           {\includegraphics[width=0.31\textwidth]{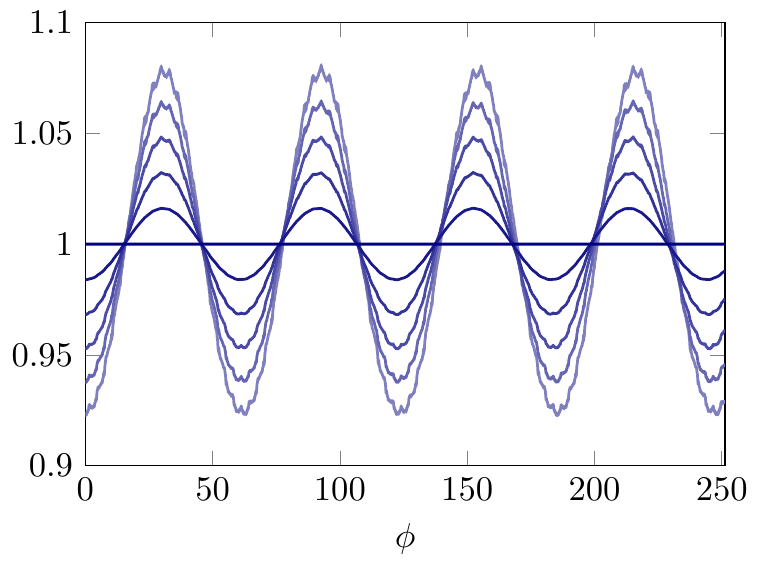} }; 
   \node (12) at (1.1,4)
           {\includestandalone[mode=buildnew, width=0.32\textwidth]{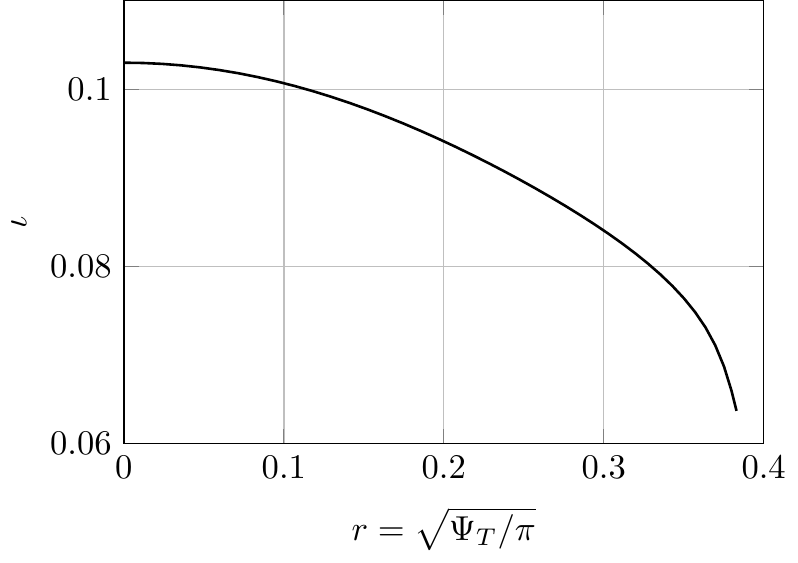} };
    \node (13) at (6.25,4.1)
            {\includestandalone[mode=buildnew, width=0.32\textwidth]{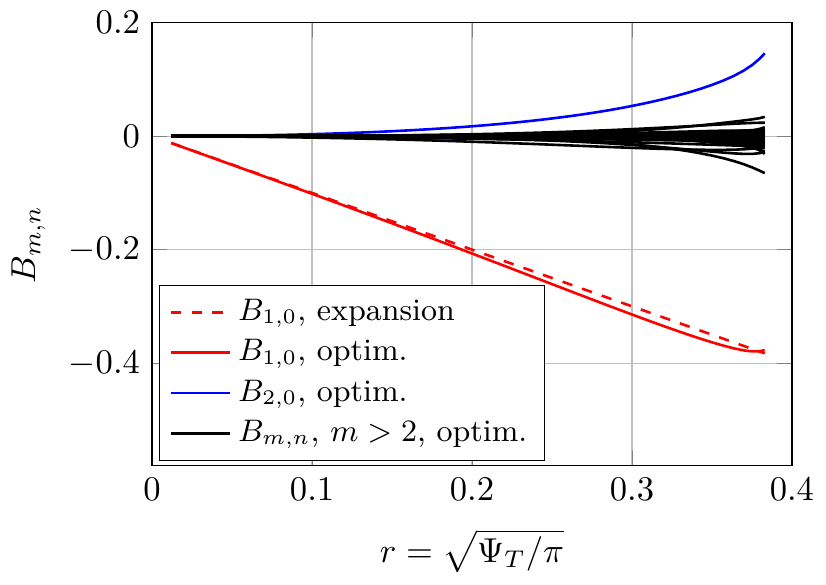} };
    \end{tikzpicture}
   \caption{Results for the optimized stellarator in Problem I: Shown on the left is the field magnitude along fieldlines in the neighborhood of the magnetic axis. Shown
     in the middle is the rotational transform $\iota$ as function of effective minor radius
     $r$. Shown on the right is the spectrum of the field magnitude
     $B$ as function of $r = \sqrt{\Phi_T/\pi}$, where $\Phi_T$ is the toroidal flux.  
     Subscripts $m$ and $n$ on $B$ refer to Fourier modes with respect to the Boozer poloidal and toroidal angle respectively.} \label{fig:iota_B}
\end{figure}

Once the magnetic field lines have been constructed numerically, one can compute the field strength along the magnetic field lines, and obtain a first measure of the quality of the quasi-symmetry of the magnetic field produced by the coils obtained in our optimization. Indeed, the magnitude of a perfectly quasi-symmetric magnetic field is a periodic function along a magnetic field line \cite{Helander2014}, and only has a global maximum and a global minimum along a field line, and no local extrema. In the left of Figure \ref{fig:iota_B}, we plot the magnetic field strength along the fieldlines as a function of the toroidal angle for fieldlines close to the magnetic axis. This plot confirms that our optimization procedure led to coils corresponding to a good approximation of a vacuum quasi-axisymmetric field close to the axis. The magnetic field strength along the field lines can be viewed as the sum of a periodic function with only global maxima and minima which line up for all the field lines, and a much smaller ripple with local minima and maxima. The amplitude of the ripple increases as one moves away from the magnetic axis, as one would expect from our formulation of the optimization problem, in which quasi-symmetry is only imposed on the expansion axis. We return to this observation shortly, from a slightly different viewpoint. To do so, we consider our second method of analyzing the vacuum field resulting from the coil optimization. It provides an alternative quantitative assessments of the quality of quasi-symmetry achieved. 

When flux surfaces do exist, we can construct a numerical parameterization for them, as well as Boozer coordinates \cite{boozer1981A,boozer1981B} on each surface. Since the purpose of the present article is to focus on the formulation of the optimization problem, and the properties of the obtained optima, we will present our numerical scheme for doing so, as well as our method for computing the rotational transform on the surface, in a forthcoming article. The rotational transform profile is shown in the middle of Figure \ref{fig:iota_B}. As expected given our low target value $\iota_{0,a}=0.103$, the rotational transform is small throughout the plasma volume. Furthermore, the radial variation of $\iota$ is small. The coils we obtained thus generate what is called a ``low-shear" vacuum magnetic field \cite{Wobig1987}, where the $\iota$ profile does not cross resonant values corresponding to low order rational numbers. As mentioned previously, this is one of the explanations for the fact that we have a large volume with nested flux surfaces, and do not have any island chains with large islands.

With the Boozer coordinates numerically constructed on each surface, we can compute the Fourier spectrum of the field strength. In the right plot of Figure 6, we show the radial profile of the Fourier modes $B_{mn}$ of the field strength. The plot confirms that quasi-axisymmetry is achieved with very high accuracy in the neighborhood of the axis, and degrades as one moves away from the axis, which is consistent with theoretical results in the Garren-Boozer asymptotic analysis \cite{landreman_2019}. Throughout the plasma, the quasi-axisymmetric modes, with $n=0$, have a much larger amplitude than the symmetry-breaking modes with $n\neq 0$. 

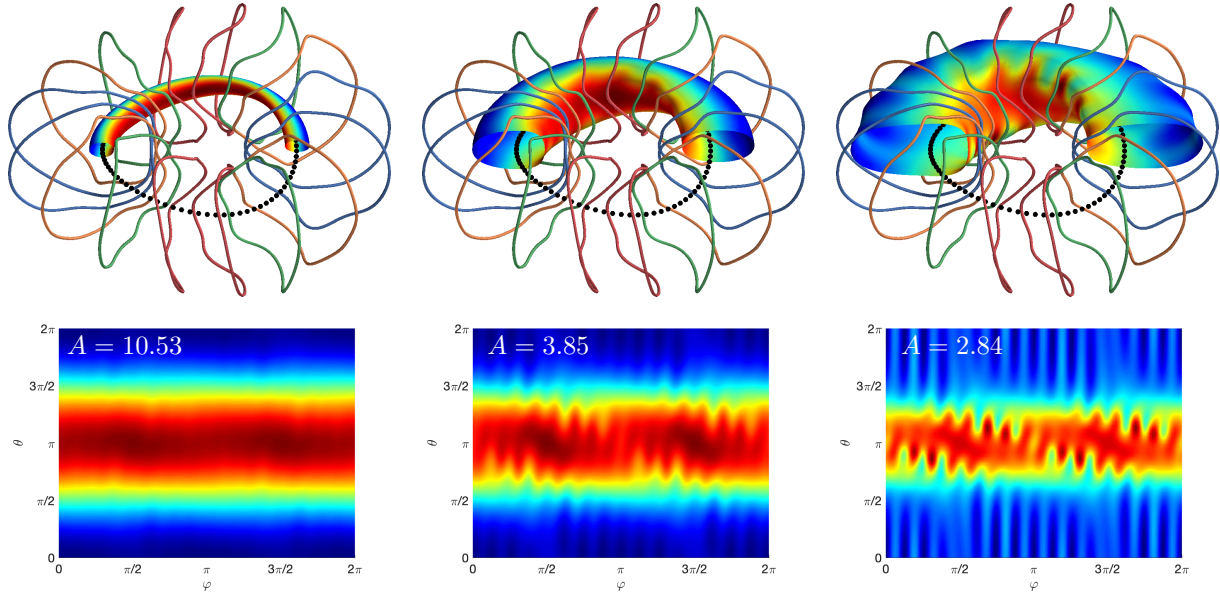
\begin{figure}
  \begin{tikzpicture}
    \node (11) at (-4,4)
          {\includegraphics[width=0.32\textwidth]{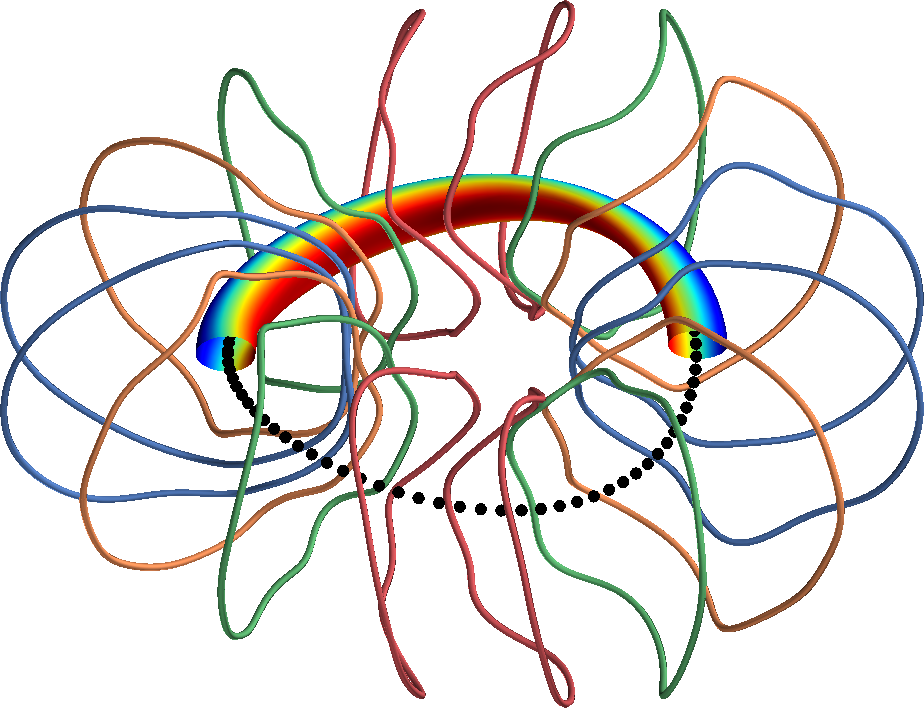}};
    \node (12) at (1.5,4)
          {\includegraphics[width=0.32\textwidth]{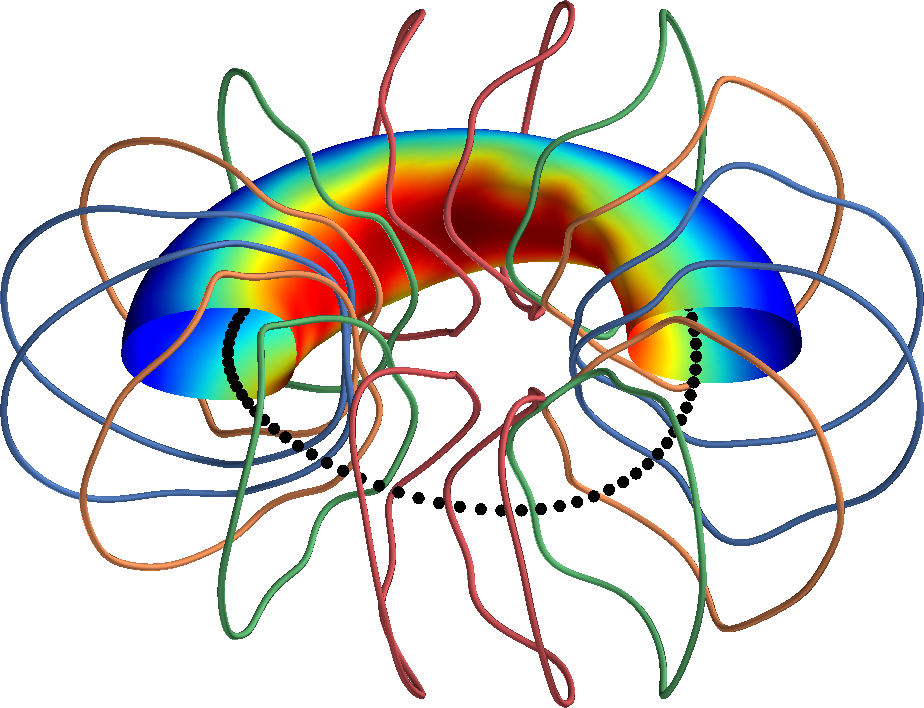}};
    \node (13) at (7,4)
          {\includegraphics[width=0.32\textwidth]{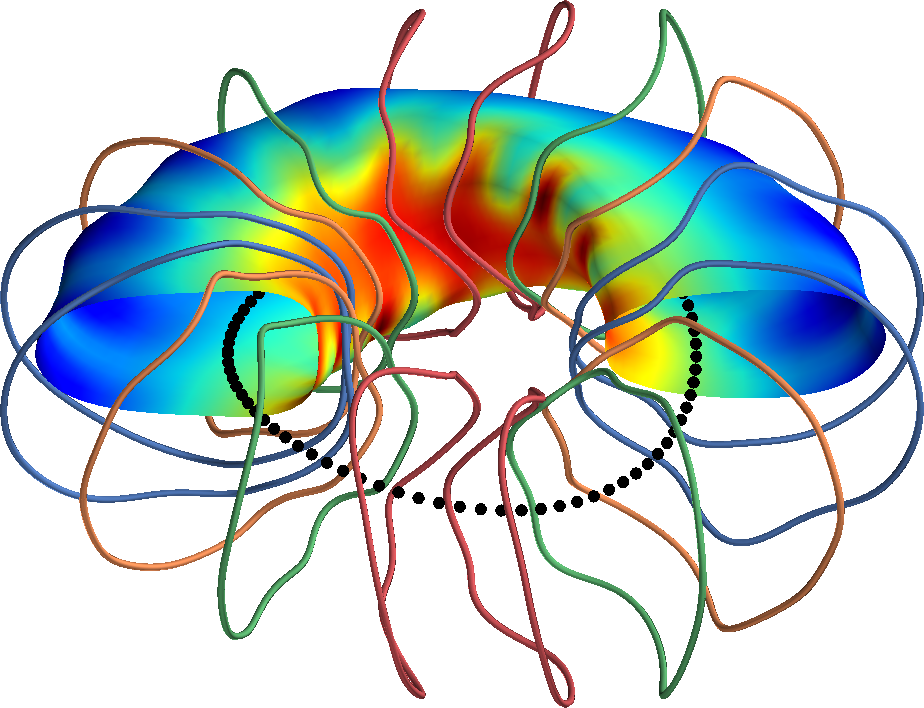}};
    \node (21) at (-4,0)
          {\includegraphics[width=0.32\textwidth]{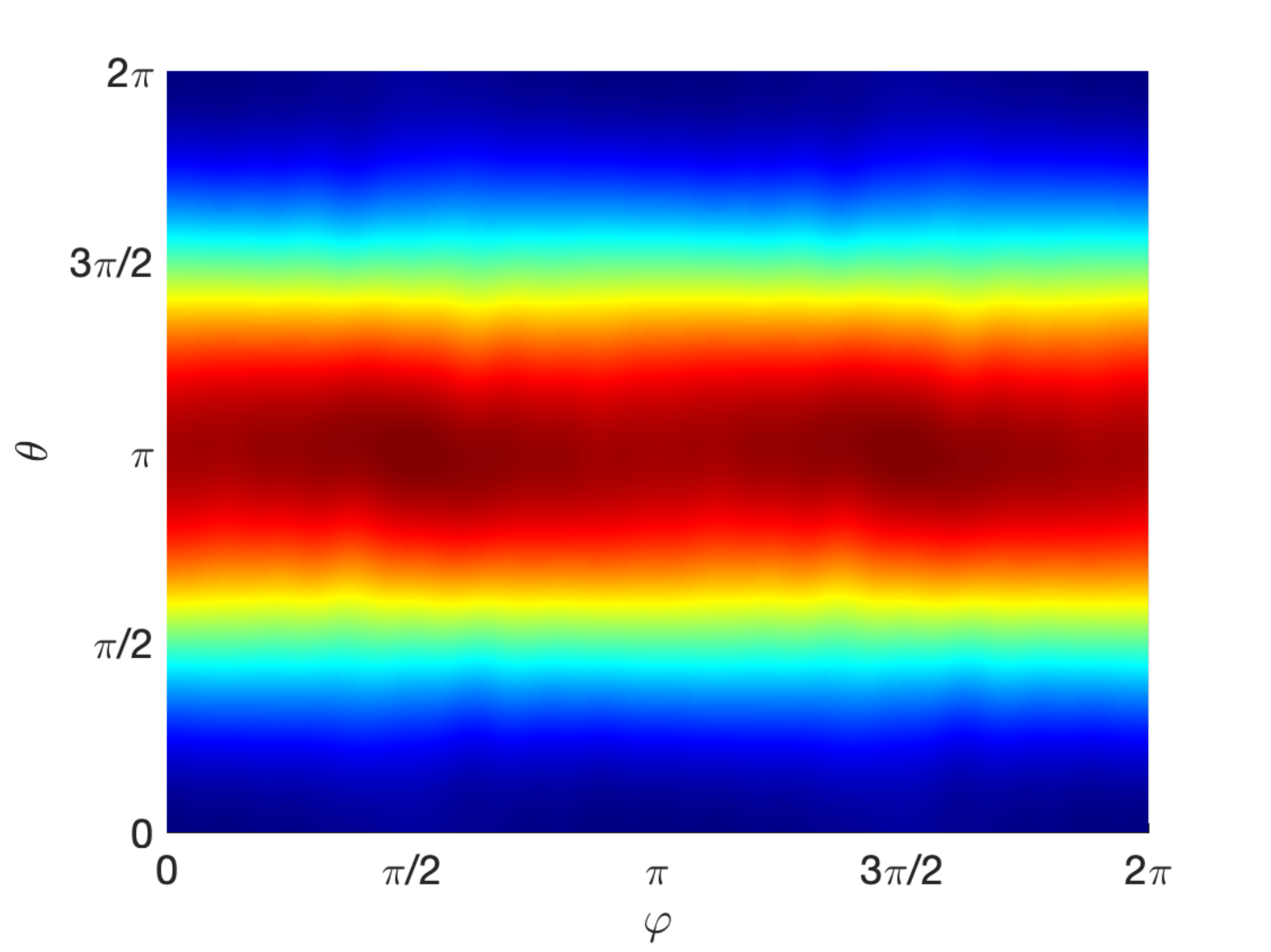}};
    \node (22) at (1.5,0)
          {\includegraphics[width=0.32\textwidth]{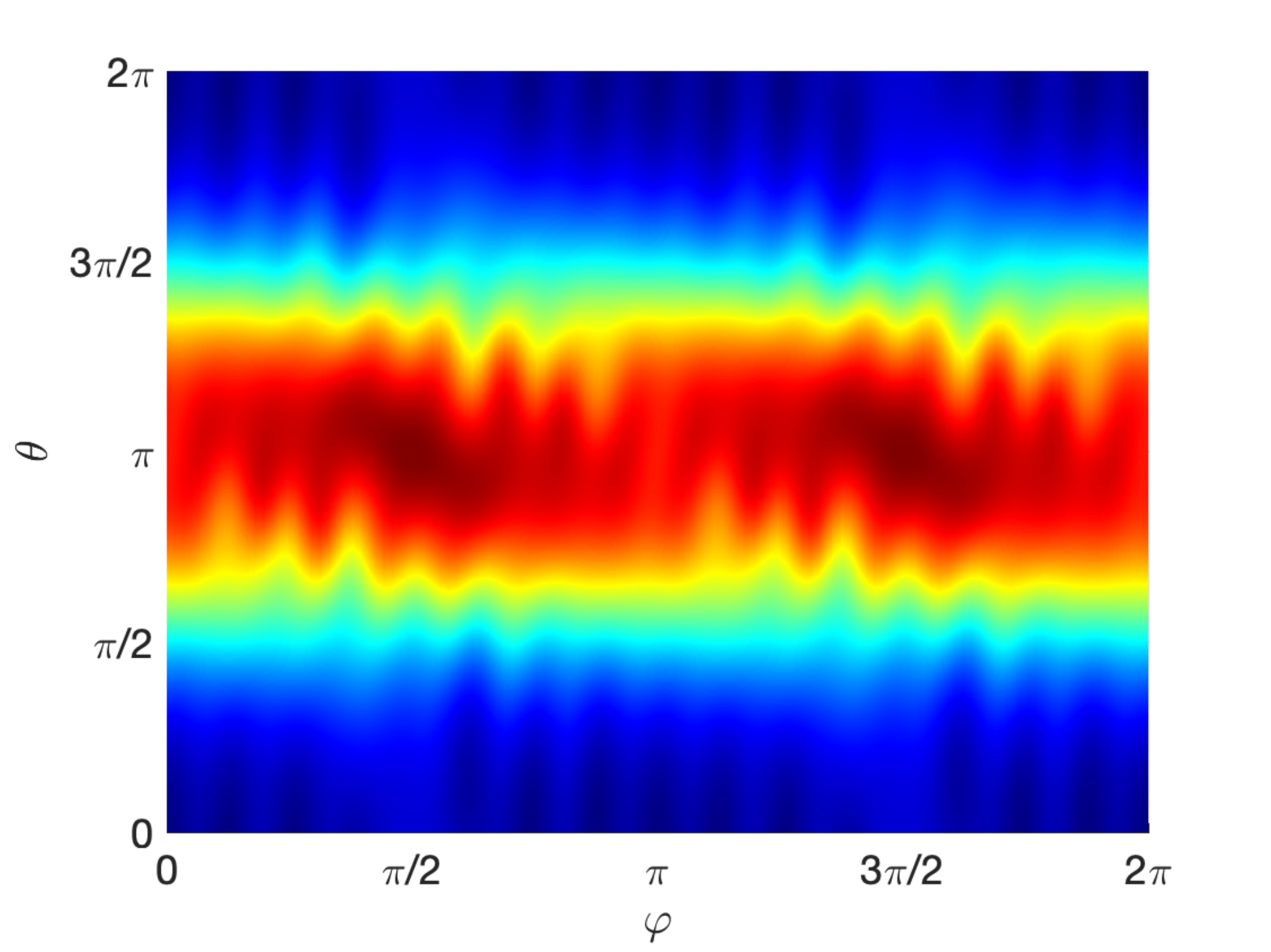}};
    \node (23) at (7,0)
          {\includegraphics[width=0.32\textwidth]{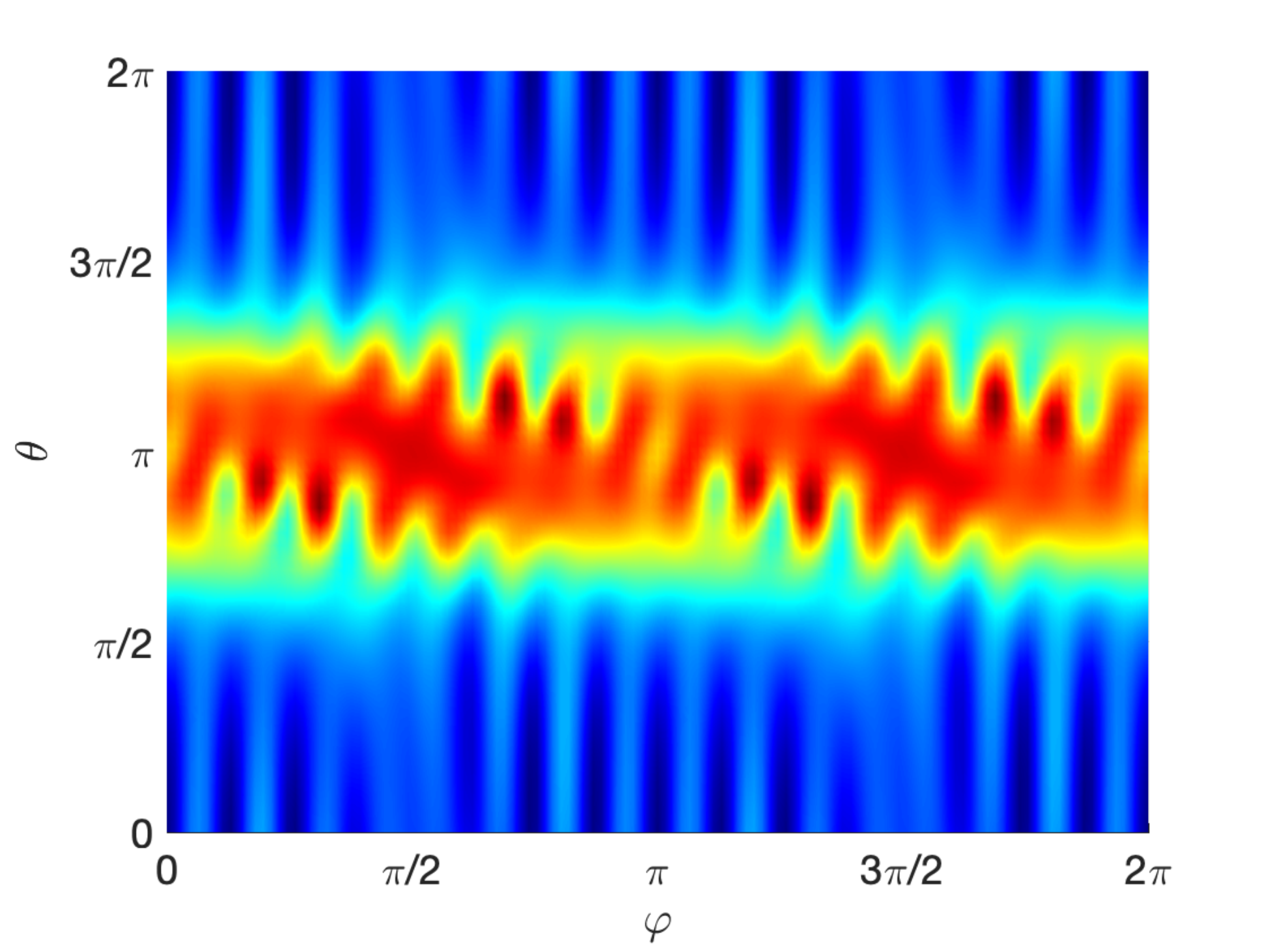}};
    \node at (-5,1.4) {\textcolor{white}{$A=10.53$}};
    \node at (0.5,1.4) {\textcolor{white}{$A=3.85$}};
    \node at (6,1.4) {\textcolor{white}{$A=2.84$}};
  \end{tikzpicture}
    \caption{Surfaces corresponding to different aspect ratios $A$ (upper row) and field strength on these surfaces (lower row) for the optimized stellarator of Problem I.  In the lower row, the axes $\theta$ and $\varphi$ are the Boozer poloidal and toroidal angles.
    Hot colors on
      the surface correspond to high field strength and cold colors
      correspond to low field strength. Color scales are different between columns. Perfect quasi-axisymmetry on the  surfaces corresponds to field strengths that only vary with $\theta$. On surfaces close to the coils, we observe the field strength variations caused by the coils, indicating a degraded approximation of quasi-axisymmetry. }
    \label{fig:problem1_surfaces}
\end{figure}

Finally, for a perfectly quasi-axisymmetric field, the contours of constant magnitude of the magnetic field on a flux surface should be horizontal lines in Boozer coordinates \cite{Helander2014}. We look at the degree to which our vacuum magnetic field satisfies this property by considering three different flux surfaces increasingly far from the magnetic axis. To visualize the radial location of the three surfaces, we highlighted them in the Poincar\'e plots in Figure \ref{fig:p1poincare}. The first surface is in the neighborhood of the magnetic axis, the second surface is approximately at mid-radius, and the last surface is close to the last closed flux surface. In order to more quantitatively describe the distance of the three surfaces from the magnetic axis and the plasma volume enclosed by the flux surfaces, we define the aspect ratio of a flux surface as $A := R_{\text{major}}/R_{\text{minor}}$, where $R_{\text{minor}}$ is the effective minor radius $R_{\text{minor}} = \sqrt{\bar A /
  \pi}$, with $\bar A$ the average toroidal cross-sectional area of
the surface, and $R_{\text{major}}$ is the effective major radius $R_{\text{major}} = V/ (2\pi^2 R_{\text{minor}}^2)$, where $V$ the volume enclosed by the surface. The same definition of $A$ is employed in the popular stellarator equilibrium code VMEC \cite{Hirshman1986}. If the surface was a toroidally axisymmetric torus with a circular cross section, our definition of the aspect ratio would correspond to the intuitive definition of the ratio of the major radius to the minor radius.  
In Figure \ref{fig:problem1_surfaces}, we show the three magnetic surfaces
mentioned before which have decreasing aspect ratios of $10.43$, $3.85$, and $2.84$, together with the optimized coil system. An aspect ratio of $2.84$ is quite low as compared to current stellarator experiments and recent designs of new machines, and corresponds to a compact magnetic configuration, which can be desirable regarding cost of construction.
In agreement with Figure \ref{fig:iota_B}, on the surfaces close to the axis, we observe a good approximation to quasi-axisymmetry, with field strengths which almost exclusively vary  with the Boozer poloidal angle $\theta$.
For surfaces further away from the
axis, the discrete nature of the coils becomes apparent and the
quality of the quasi-axisymmetry degrades.  Since we are only
optimizing for near-axis quasi-symmetry, this motivates an off-axis
optimization of quasi-symmetry, which will be the subject of future research.

\section{Problem II: Optimizing the nonplanar coils of the National Compact Stellarator Experiment (NCSX)}\label{sec:p2}
We consider a second example that is inspired by the coil configuration of
the National Compact Stellarator Experiment (NCSX). NCSX is a compact high performance stellarator with quasi-axisymmetry which was designed in the 1990s and early 2000s \cite{Zarnstorff2001,Reiman2001}.
NCSX was partially built but the project was cancelled in 2008, because the estimated cost and schedule for completing the project grew as the technical requirements and risks became better understood \cite{Neilson10}. 
The NCSX stellarator is composed of three unique modular coil shapes to
which stellarator symmetry and three-fold toroidal symmetry are
applied.  
The NCSX design also included planar toroidal field coils and poloidal field coils, which are not included in our study for simplicity.
Thus, we consider a total of 18 coils in our optimization for near-axis quasi-symmetry. 
Since we ignore NCSX's poloidal and planar toroidal field coils, and we consider the case of no plasma current, our configuration is not directly comparable to the complete NCSX design on which the construction of the experiment was based. However we can still expect the NCSX coil shapes to provide an initial condition in an interesting region of parameter space, as an alternative to the planar circular shapes used to initialize Problem I.

The initial coil geometries use the first $n^{\text{coil}}_p = 6$ Fourier modes of the NCSX modular coils, and the expansion axis geometry uses $n^{\text{axis}}_p = 4$ Fourier modes.
Our target
rotational transform, axis length, and coil lengths are set
to coincide with the corresponding values for the original NCSX non-planar coil shapes with their original currents, i.e.,
\begin{equation*}
    \iota_{0,a} =        0.395939, \quad
     L^{(1)}_{0,c} = 6.851923,\quad
     L^{(2)}_{0,c} = 6.480790,\quad
     L^{(3)}_{0,c} = 5.816940,\quad
     L_{0,a} =       9.421513.
\end{equation*}
We divide the NCSX coil currents by the mean field strength on the
magnetic axis of the initial configuration, $1.474$, since we fix $B_0
= 1$. Moreover, we set  $s_{\psi} = s_G = 1$ 
and choose the initial guess $\bar \eta = 0.685$ since
this results in an initial quasi-symmetric rotational transform close
to the target value $\iota_{0,a}$.

We present solutions of \eqref{eq:opt} with different coil
regularization parameters. Namely, we use regularizations of the
following form
\begin{equation}\label{eq:reg4}
R(\mathbf{c},\mathbf{a}) :=  \sum_{i = 1}^{N_c}\left[\frac  12 \left(\frac{L_{i,c}(\mathbf{c}) -L_{0,c}}{L_{0,c}}\right)^2 + \frac{\delta}{4} k_i(\mathbf{c}) + \frac{\gamma}{2} \sum_{j=1}^{i-1} d_{i,j}(\mathbf{c}) \right] + \frac{1}{2}\left(\frac{L_a(\mathbf{a}) -L_{0,a}}{L_{0,a}}\right)^2,
\end{equation}
and consider three cases to study the effect of different
regularization choices. 
First, for finding design (II.A), we only use length penalties on the coils and
the expansion axis, i.e., we set $\delta = \gamma = 0$. For the design
(II.B), we add the minimum
distance penalty, i.e., $\delta = 0$, $\gamma = 2000$, $d_{\text{min}} =
0.2$. Finally, to compute the design (II.C), we additionally  add a
curvature penalty by setting $\delta
= 4\times 10^{-6}$, $\gamma = 2000$, $d_{\text{min}} = 0.2$.  The
weights multiplying the penalty terms are chosen by trial-and-error,
to ensure that the physics penalty terms are not overpowered by the
regularization terms. 
\begin{figure}\centering
  \begin{tikzpicture}
    \node (11) at (0.5,0)
          {\includegraphics[width=0.3\textwidth]{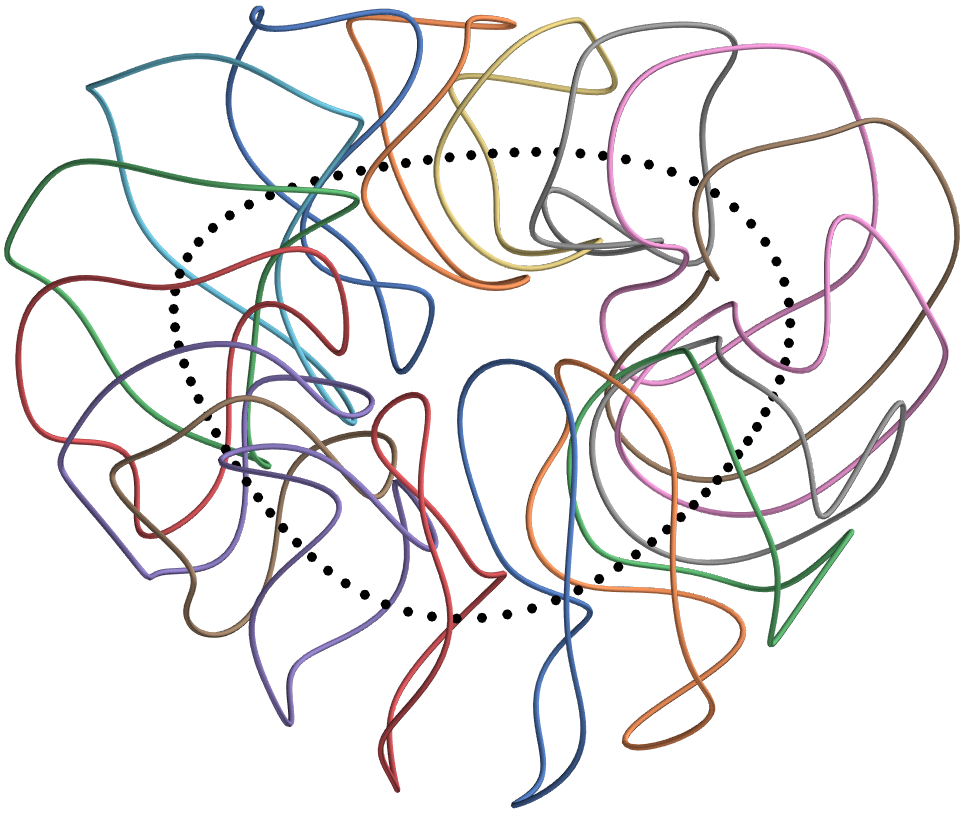}};
    \node (12) at (7.5,0)
          {\includegraphics[width=0.3\textwidth]{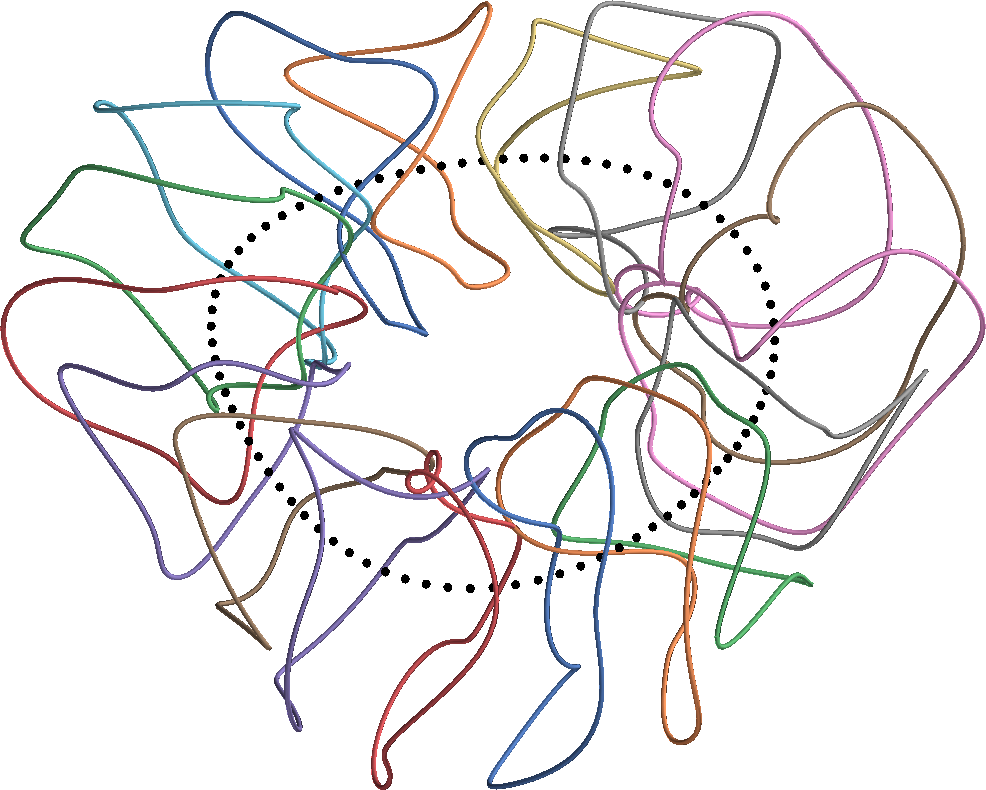}};
    \node (13) at (0.5,-9.)
          {\includegraphics[width=0.3\textwidth]{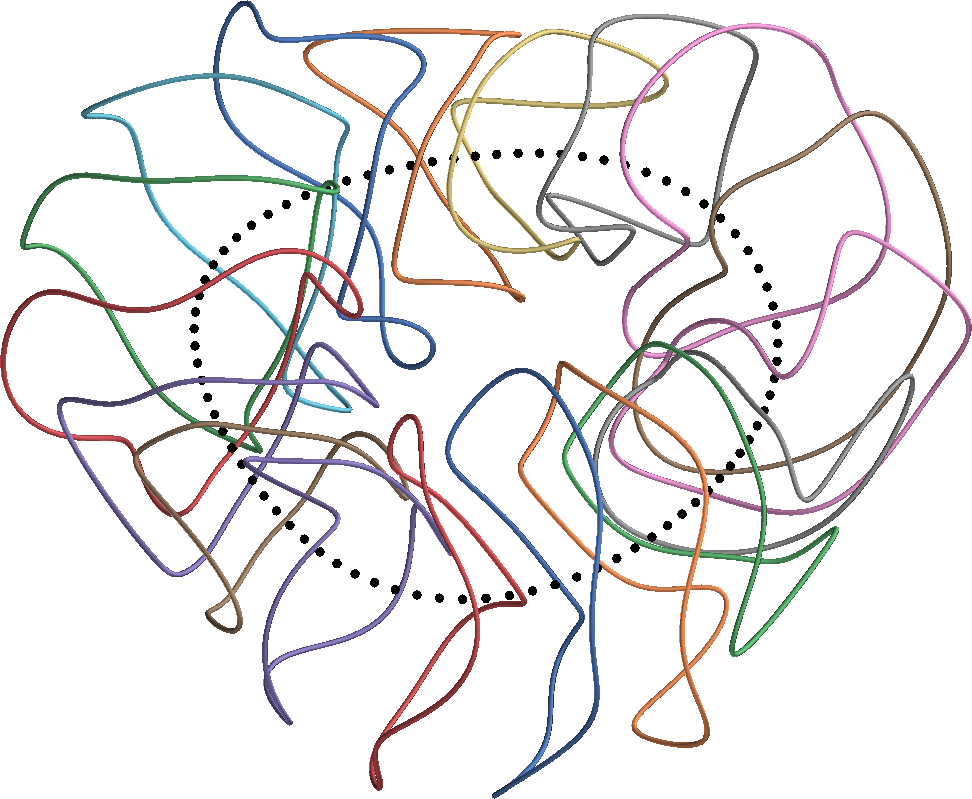}};
    \node (14) at (7.5,-9.)
          {\includegraphics[width=0.3\textwidth]{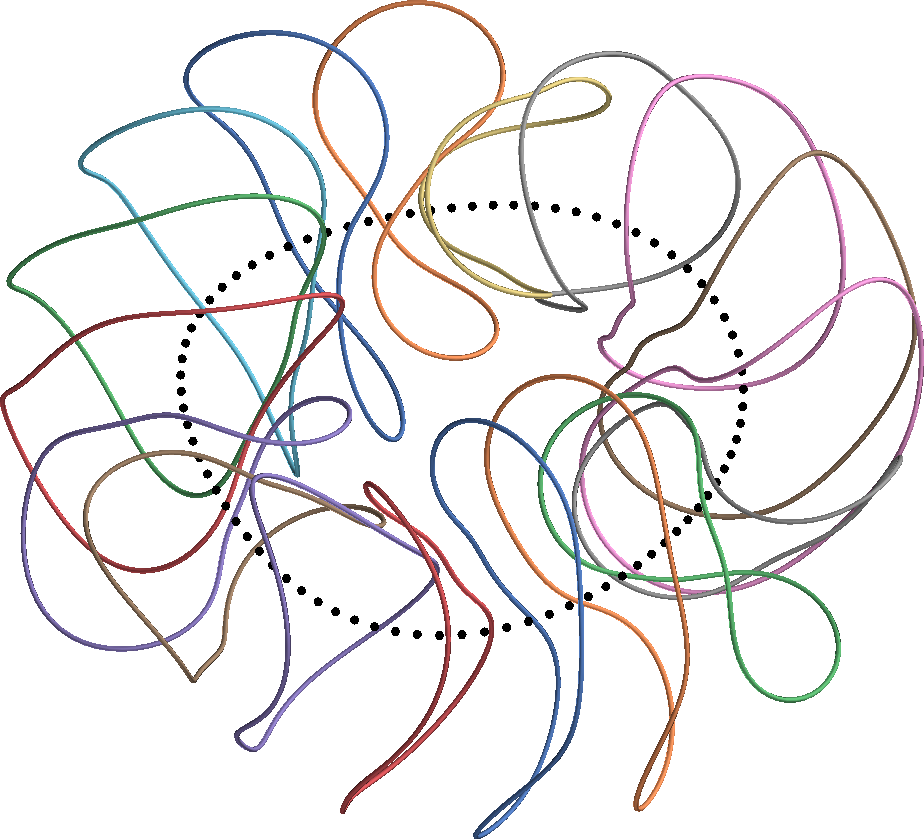}};
    \node (21) at (0,-4.5)
          {\includestandalone[width=0.35\textwidth]{problem2/modB/absB_initial}};
    \node (22) at (7.,-4.5)
          {\includestandalone[width=0.35\textwidth]{problem2/modB/absB_final_noreg}};
    \node (23) at (0.0,-13.5)
          {\includestandalone[width=0.35\textwidth]{problem2/modB/absB_final_withdist}};
    \node (24) at (7.,-13.5)
          {\includestandalone[width=0.35\textwidth]{problem2/modB/absB_final_withdistandreg}};
    \node at (-2.,-2) {\textcolor{black}{\large (II.Init)}};
    \node at (5.1,-2) {\textcolor{black}{\large (II.A)}};
    \node at (-2.,-11.) {\textcolor{black}{\large (II.B)}};
    \node at (5.1,-11.) {\textcolor{black}{\large (II.C)}};
  \end{tikzpicture}
  \caption{Designs for Problem II: Shown are the coils and field strength along field lines for intial
    configuration and from optimization with different coil
    regularizations.  The initial coils are denoted by (II.Init),
    coils (II.A) are obtained with only coil and axis length
    regularization, coils (II.B) are obtained with additional minimum
    distance regularization, and coils (II.C) are obtained with
    additional curvature regularization.}
    \label{fig:coils-problem2}
\end{figure}
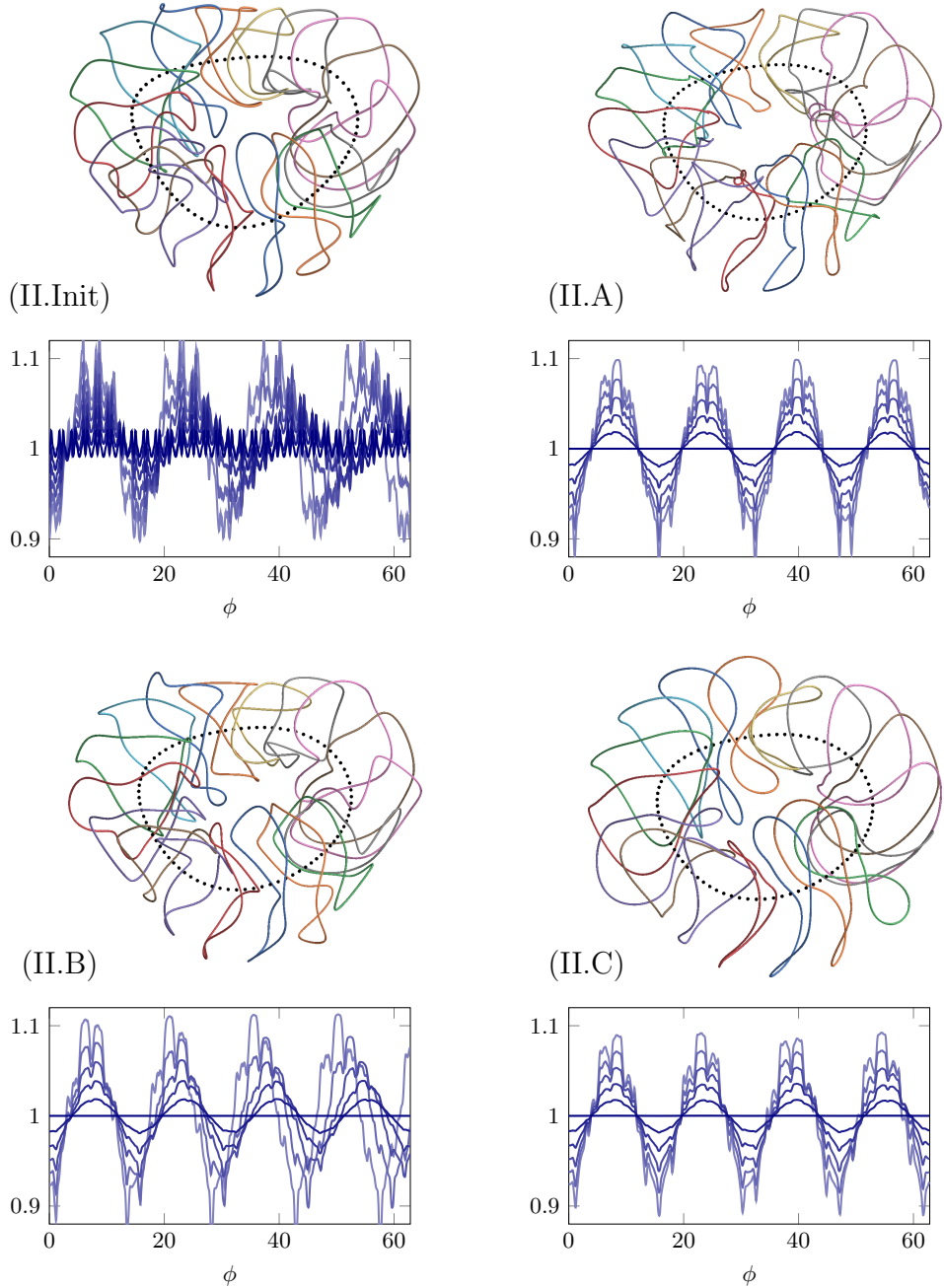

As can be seen in Figure
\ref{fig:coils-problem2}, compared to the initial coils (II.Init), the
design (II.A) yields a much better approximation of 
quasi-axisymmetry near the expansion axis.  However, the coils are irregularly shaped and quite close to one another---they even interweave with each
other.  In fact, the coil distance is less than 0.022 in some
locations. This
motivates introducing the minimum distance term in the regularization to compute design (II.B).
Doing so successfully prevents the coils
from becoming too close too one another, with the minimum distance now
approximately $0.2$, as we show in the lower left of Figure \ref{fig:coils-problem2}. However, the
maximum curvature of the coils in design (II.A) is 22.79 and is 26.76 with the additional penalty term in design (II.B). In order to address this issue,
we introduce the maximum curvature term in the regularization, which produces design (II.C). The lower right of Figure \ref{fig:coils-problem2} shows that design (II.C) has smooth and
well-behaved coils, with maximum curvature of just 2.29.  We note
that introducing the additional penalty terms for designs (II.B) and (II.C) does not substantially degrade the
quasi-axisymmetry in the neighborhood of the expansion axis, as can be seen in the bottom of Figure
\ref{fig:coils-problem2}.
This is a consequence of the many
directions in parameter space in which the quasi-symmetry remains nearly unchanged as observed in Problem I, resulting in
substantial freedom for the coils while maintaining good confinement
properties.  

Poincare plots are provided for design (II.C). 
They show the existence of magnetic surfaces with low aspect ratio,
and the existence of a few wider island chains compared to Problem 1.
\begin{figure}
    \begin{center}
    \includegraphics[width=0.99\textwidth]{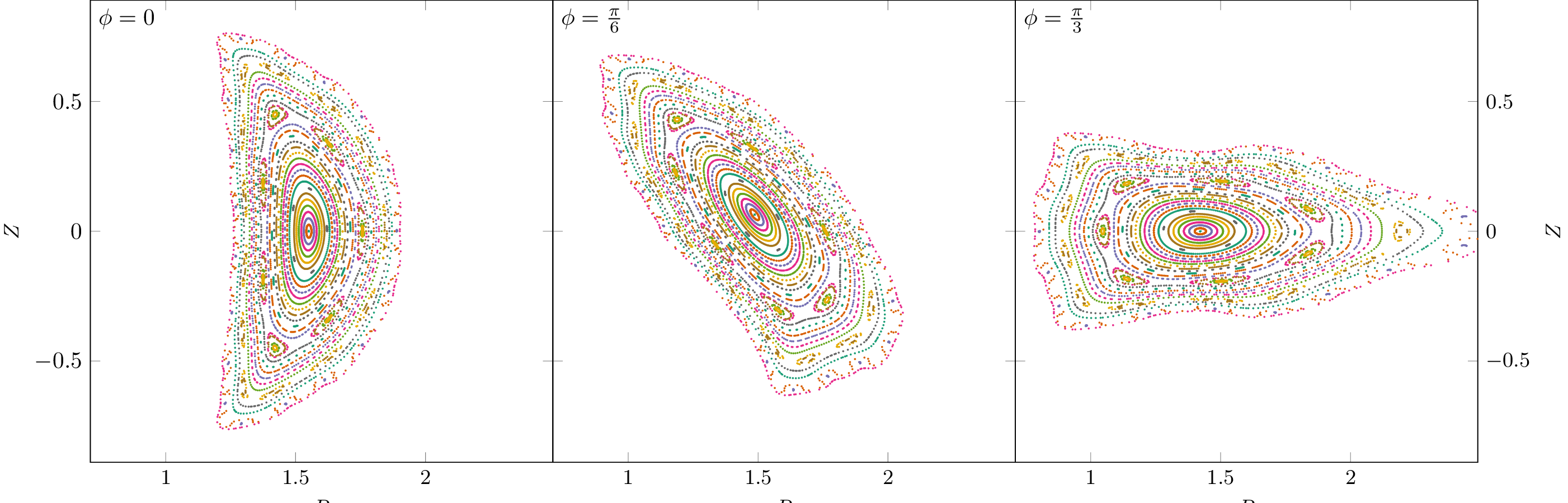}
     \end{center}
     \caption{Poincare plots for Problem II.C with minimum
       distance and curvature regularization
       terms.  These plots are generated by following field lines around the origin and computing their intersection with the cylindrical angles $\phi = 0$, $\pi/6$, and $\pi/3$, shown as dots.
       Points of intersection belonging to the same field line are
       plotted with the same color.}\label{fig:p2poincare}
\end{figure}
We again plot the eigenvalues of the Hessian at minimizers in Figure
\ref{fig:p2eig}. We show the spectrum for all design parameters (axis,
coils and current) and the spectrum for the part of the Hessian
corresponding to the geometric coil parameters only.  For (II.A), the overall
spectrum varies over more than 12 orders of magnitude, with a large
number of small eigenvalues. These small eigenvalues mean that the optimizer can
be perturbed in many directions with little impact on the objective to
second order.  
For (II.B), obtained with the minimum distance constraint,
the optimal coils have a minimum pairwise distance of $0.199999$.
This value is quite close to the target minimal distance
$d_{\min}=0.2$ since we chose a large penalty parameter $\gamma$.
This distance
constraint increases most of the small Hessian eigenvalues to at
least $10^{-4}$. Finally, for (II.C), obtained with an additional curvature
regularization, most eigenvalues are increased to $10^{-3}$.  This shows
that adding constraints increases the curvature of the Hessian at the
minimizer and thus has a (local) convexification effect on the
minimization landscape, in particular in the mostly flat
directions. Such regularization does not lead to considerable
degradation of confinement properties of the stellarator but helps to
identify coils that are simpler and thus more cost-effective to
manufacture.

\pgfplotstableread[col sep = comma, header=false]{problem2/data/e/noreg.txt}\nodist
\pgfplotstableread[col sep = comma, header=false]{problem2/data/e/mindist.txt}\distnoreg
\pgfplotstableread[col sep = comma, header=false]{problem2/data/e/mindist_curvature.txt}\distcurvreg
\pgfplotstableread[col sep = comma, header=false]{problem2/data/ec/noreg.txt}\nodistc
\pgfplotstableread[col sep = comma, header=false]{problem2/data/ec/mindist.txt}\distnoregc
\pgfplotstableread[col sep = comma, header=false]{problem2/data/ec/mindist_curvature.txt}\distcurvregc
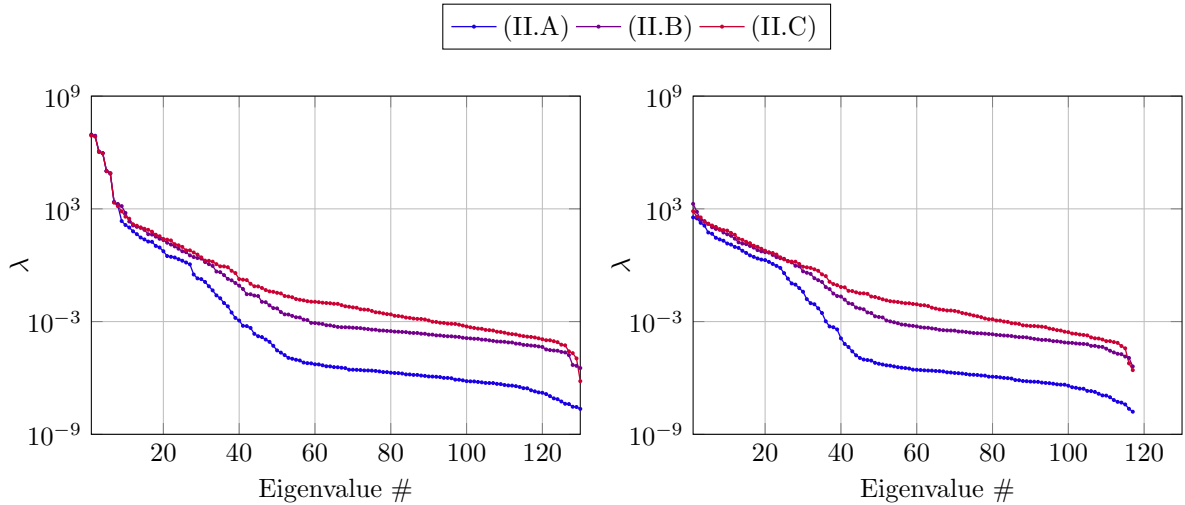
\begin{figure}
\begin{tikzpicture}
    \begin{groupplot}[%
      group style={%
        group name={my plots},
        group size=2 by 1,
        horizontal sep=1.5cm
      },
      width=6.5cm,
      height=4.5cm,
      ylabel={$\lambda$}, 
      ylabel near ticks,
	  log basis x=10,
      scale only axis,
            grid=both,
      xmin=1, xmax = 130,
      ymin=1e-9,ymax=1e9,
      xlabel={Eigenvalue $\#$}
    ]
     \nextgroupplot[ymode=log, y label style={at={(axis description cs:-0.11,.5)},rotate=0,anchor=south},legend to name=plotA,legend columns=-1]
      \addplot+[solid, color=blue!90!red, mark=*, mark options={fill=none}, mark size=0.5pt, line width=0.5pt] table[x expr=\coordindex+1, y index=0] {\nodist};\addlegendentry{(II.A)}
      \coordinate (top) at (rel axis cs:0,1);%
     \addplot+[solid, mark=*, color=blue!55!red, mark options={fill=none}, mark size=0.5pt, line width=0.5pt] table[x expr=\coordindex+1, y index=0] {\distnoreg};\addlegendentry{(II.B)}
      \addplot+[solid, mark=*, color=blue!20!red, mark options={fill=none}, mark size=0.5pt, line width=0.5pt] table[x expr=\coordindex+1, y index=0] {\distcurvreg};\addlegendentry{(II.C)}
      \nextgroupplot[ymode=log,y label style={at={(axis description cs:-0.11,.5)},rotate=0,anchor=south}]
    \addplot+[solid, mark=*, color=blue!90!red, mark options={fill=none}, mark size=0.5pt, line width=0.5pt] table[x expr=\coordindex+1, y index=0] {\nodistc};
      \addplot+[solid,  color=blue!55!red, mark=*, mark options={fill=none}, mark size=0.5pt, line width=0.5pt] table[x expr=\coordindex+1, y index=0] {\distnoregc};
      \addplot+[solid,  color=blue!20!red, mark=*, mark options={fill=none}, mark size=0.5pt, line width=0.5pt] table[x expr=\coordindex+1, y index=0] {\distcurvregc};
      \coordinate (bot) at (rel axis cs:1,0);%
    \end{groupplot}

\path (top|-current bounding box.north)--
      coordinate(legendpos)
      (bot|-current bounding box.north);
\node[right=0em, above = 1em,inner sep=0pt] at (legendpos) {\pgfplotslegendfromname{plotA}};
\end{tikzpicture}
     \caption{Hessian spectra for Problem II: Shown on the left are the eigenvalues of the Hessian at the minimizers for designs (II.A), (II.B) and (II.C) characterized by different regularization parameters for Problem II as explained in section \ref{sec:p2}.  Shown on the right are the eigenvalues of the Hessian when the magnetic axis and coil currents are fixed.} \label{fig:p2eig}
\end{figure}

\section{Conclusions and perspective}\label{sec:conclusion}

We have presented a new approach to coil design, where we directly optimize stellarator coils for quasi-symmetry on an expansion axis. We relied on gradient-based optimization algorithms, for which we computed exact derivatives of the objective function using either the forward sensitivity or the adjoint method. Both methods yield more accurate derivatives and are substantially less computationally expensive than evaluating derivatives through numerical finite differences.  
We have applied our new formulation to design two configurations of physical interest. The first application is a compact vacuum quasi-axisymmetric magnetic field which could be suitable for the confinement of a low-density electron-positron plasma. The second application is a simplification of the non-planar coils of the NCSX experiment, targeting good vacuum quasi-axisymmetry. In both problems, we showed that our method worked in a satisfactory manner, leading to a good approximation of quasi-axisymmetry, and the existence of nested flux surfaces for a large plasma volume. By analyzing the optimization landscape locally around the optimal solutions, we observed that there is substantial freedom in imposing design constraints on the coils, corresponding to the many directions in which the minimizer can be perturbed that do not substantially affect the value of the objective. In our numerical tests, we also find that using an optimization method that incorporates approximate second-order derivative information, such as the BFGS method, is crucial for numerical convergence.

Our results motivate the future use of single-stage optimization for
the detailed design of new stellarator experiments. Our method is
directly applicable to other quasi-symmetries, particularly the also
attractive quasi-helical symmetry, which also fit in the Garren-Boozer
near-axis expansion framework. However, for complete design studies,
our method needs to be extended in several directions before it can be
widely used. We need to develop a robust method to ensure good
quasi-symmetry throughout the plasma for confinement, and not just near the magnetic
axis. For that purpose, we are
currently exploring the idea of including a term driving off-axis
quasi-symmetry in the objective. Our single-stage approach also lends itself toward incorporating coil manufacturing errors into the design process, resulting in stochastic stellarator optimization. Recently results in this direction show that incorporating such stochasticity reduces the occurrence of local minimizers and near-flatness of the objective in some directions \cite{WechsungGiulianiLandremanEtAl21}.
In addition, in stellarators designed
for magnetic confinement fusion research, plasma pressure is
sufficiently high that the magnetic field is not well approximated by
a vacuum field. We thus need to generalize our approach to apply it to
finite-pressure magnetic configurations. This can be done in several
ways. One way is to consider higher-order terms in the Garren-Boozer
asymptotic expansion, which contain the modifications to the magnetic
field due to the plasma pressure. Another way, which has the advantage
of not being restricted to quasi-symmetric configurations, is to couple our method with a solver that computes the
magnetic field accounting for the plasma
pressure starting from the geometry of the coils and their currents, i.e. a ``free-boundary" equilibrium solver, as it is called in the magnetic confinement fusion community. Finally, for detailed stellarator designs, the objective
must include many more terms corresponding to additional physics and
engineering goals, such as the minimization of turbulent transport, or
the loss of energetic particles. Such terms typically involve complex
models and equations, and analytical derivatives would likely be
complicated to obtain. Automatic differentiation could then be a
promising approach to compute exact derivatives and avoid the
limitations of finite differences for the evaluation of the gradients,
as was shown recently for stellarator coil design with the two-stage
method \cite{mcgreivy2020}. All these questions are the subject of
ongoing research, with progress to be reported in the future.

\section*{Acknowledgements}
This research was supported by the Simons Foundation/SFARI (560651, AB).  AC was additionally supported by the U.S.\ Department of Energy, Office of Science, Fusion Energy Sciences under Awards No.\ DE-FG02-86ER53223 and DE-SC0012398. GS was additionally supported by the US National Science Foundation under Award No.\ DMS-1723211.
AG was additionally supported by the Natural Sciences and Engineering Research Council of Canada (NSERC) postdoctoral fellowship.

\section*{Code availability}
For reproducibility, we provide the code used to generate the examples presented in this paper. The code for the first example can be found at
\begin{center}
    \url{https://github.com/andrewgiuliani/matplasmaopt.git}
\end{center}
The code for the second example can be found under 
\begin{center}
\url{https://github.com/florianwechsung/PyPlasmaOpt/releases/tag/v1.0.0}
\end{center}
and was archived on Zenodo at~\cite{pyplasmaopt-zenodo}.
\appendix
\section{Jacobian of the discrete state equation}\label{app:state}
We recall that the discrete state equation is \eqref{eq:sigma_hh}.  The components of $\partial \mathbf{g}/\partial \bm \sigma$ are
\begin{equation*}
\begin{aligned}
    \frac{\partial g_i}{\partial \sigma_j } &= \frac{|G_0|}{\ell'_i B_0}
D_{i,j} +   2\iota \delta_{i-j}  \sigma_i  \quad & ~ 0 \leq i,j \leq n_\phi-1,\\
\frac{\partial g_{n_\phi}}{\partial \sigma_j } &= 0 &\quad 1 \leq j \leq n_\phi-1, \\
\frac{\partial g_{n_\phi}}{\partial \sigma_0 } &= 1, \\
\end{aligned}
\end{equation*}
where $\delta_{i-j}$ is the Kronecker delta.  The components of $\partial \mathbf{g}/\partial \iota$ are
\begin{equation*}
    \begin{aligned}
        \frac{\partial g_i}{\partial \iota} &=  \frac{\bar \eta^4}{\kappa_i^4} + 1 + \sigma_i^2 \quad &  ~ 0 \leq i \leq n_\phi-1,\\
        \frac{\partial g_{n_{\phi}}}{\partial \iota } &= 0   .
    \end{aligned}
\end{equation*}

\section{Derivatives of discrete objective}\label{app:objective}
We recall that the discretized objective function is \eqref{eq:fhat_h}.
The components of $\partial \hat{J}/\partial c_i$ and $\partial \hat{J}/\partial a_i$ are
\begin{align*}
    \frac{\partial \hat{J}}{\partial c_i} = \biggl(\frac{2 \pi}{n_\phi} \biggr) \biggl \{ \sum_{q = 0}^{n_\phi-1} &\biggl[ \sum_{j=1,2,3} (B^j_{\text{coils},q} - B^j_{\text{QS},q}) \frac{\partial}{\partial c_i} B^j_{\text{coils},q}  \\
    &+  \sum_{j=1,2,3}\sum_{k=1,2,3} (\nabla_k B^j_{\text{coils},q} - \nabla_k B^j_{\text{QS},q})\frac{\partial}{\partial c_i} \nabla_k B^j_{\text{coils},q}    \biggr] \| \Gamma'(\phi_q) \| \biggr \}
    + \frac{\partial }{\partial c_i} R(\mathbf{c}, \mathbf{a}), \\ 
     \frac{\partial \hat{J}}{\partial a_i} = \biggl(\frac{2 \pi}{n_\phi} \biggr) \biggl \{ \sum_{q = 0}^{n_\phi-1} &\biggl[   \sum_{j=1,2,3} (B^j_{\text{coils},q}-B^j_{\text{QS},q})  \frac{\partial}{\partial a_i} (B^j_{\text{coils},q}-B^j_{\text{QS},q})    \\
     &+   \sum_{j=1,2,3}\sum_{k=1,2,3} (\nabla_k B^j_{\text{coils},q} - \nabla_k B^j_{\text{QS},q} )\frac{\partial}{\partial a_i} (\nabla_k B^j_{\text{coils},q} - \nabla_k B^j_{\text{QS},q} )   \biggr] \| \Gamma'(\phi_q) \| \\ 
    +&\biggl[ \frac{1}{2}  \|\mathbf{B}_{\text{coils},q}(\mathbf{c}) - \mathbf{B}_{\text{QS},q}(\mathbf{a})\|^2 + \frac{1}{2} \| \nabla\mathbf{B}_{\text{coils},q}(\mathbf{c}) - \nabla\mathbf{B}_{\text{QS},q}(\mathbf{a}, \bm \sigma, \iota) \| ^2 \biggr]  \frac{\partial }{\partial a_i}\| \Gamma'(\phi_q) \| \biggr \}\\
    &+ \frac{\partial }{\partial a_i} R(\mathbf{c}, \mathbf{a}),
\end{align*}
where $\nabla_k B^j_{\text{QS}}$, $\nabla_k B^j_{\text{coils}}$ are the $k$th component of the gradient in the $j$th Cartesian component of the magnetic field from the quasi-symmetric expansions and the one generated by the coils, respectively.
The components of $\partial \hat{J}/\partial \bm \sigma$, and $\partial \hat{J}/\partial \iota$ are
\begin{align*}
    \frac{\partial \hat{J}}{\partial \sigma_i} &= \biggl(\frac{2 \pi}{n_\phi} \biggr)\sum_{q = 0}^{n_\phi-1}  \biggl \{  \sum_{j = 1,2,3} \sum_{k=1,2,3} (\nabla_k B^j_{\text{QS},q} - \nabla_k B^j_{\text{coils},q} )  \frac{\partial\nabla_k B^j_{\text{QS},q}}{\partial \sigma_i} ~\|\Gamma_{\!\mathbf{a}}(\phi_q)\| \biggr\}, \\
    \frac{\partial \hat{J}}{\partial \iota} &= \frac{1}{\iota_{0,a}^2}(\iota - \iota_{0,a}) + \biggl(\frac{2 \pi}{n_\phi} \biggr)\sum_{q = 0}^{n_\phi-1}  \biggl \{  \sum_{j = 1,2,3} \sum_{k=1,2,3} ( \nabla_k B^j_{\text{QS},q} - \nabla_k B^j_{\text{coils},q} )  \frac{\partial\nabla_k B^j_{\text{QS},q}}{\partial  \iota} ~\|\Gamma_{\!\mathbf{a}}(\phi_q)\| \biggr\} .
\end{align*}

\bibliography{refs}

\end{document}

%% file: taylortest/taylortest.tex
\pgfplotstableread[col sep=comma, row sep=\\]{%
eps,first_order, second_order, fourth_order, sixth_order\\
9.765625E-04, 2.149074E+00, 1.403309E-02, 5.556612E-05, 5.355209E-06\\
4.882812E-04, 1.079021E+00, 3.510759E-03, 3.316736E-06, 9.582713E-08\\
2.441406E-04, 5.405108E-01, 8.778433E-04, 2.047281E-07, 1.549595E-09\\
1.220703E-04, 2.704902E-01, 2.194704E-04, 1.275492E-08, 2.430077E-11\\
6.103516E-05, 1.353019E-01, 5.486819E-05, 7.967352E-10, 2.465424E-13\\
3.051758E-05, 6.766490E-02, 1.371709E-05, 5.010333E-11, 5.416595E-15\\
1.525879E-05, 3.383591E-02, 3.429274E-06, 2.303101E-12, 8.144113E-13\\
7.629395E-06, 1.691882E-02, 8.573193E-07, 2.146544E-12, 3.360036E-13\\
3.814697E-06, 8.459623E-03, 2.143286E-07, 2.192323E-12, 2.074905E-12\\
1.907349E-06, 4.229865E-03, 5.358353E-08, 3.264809E-12, 3.892086E-12\\
9.536743E-07, 2.114946E-03, 1.339327E-08, 9.388182E-13, 6.553556E-12\\
4.768372E-07, 1.057476E-03, 3.353577E-09, 1.704183E-11, 1.386701E-11\\
2.384186E-07, 5.287390E-04, 7.756639E-10, 7.027228E-11, 5.917805E-11\\
1.192093E-07, 2.643698E-04, 1.916613E-10, 3.994389E-11, 4.495477E-11\\
5.960464E-08, 1.321850E-04, 4.308482E-11, 1.260961E-10, 2.159140E-10\\
2.980232E-08, 6.609264E-05, 2.889951E-10, 3.405335E-10, 3.197795E-10\\
1.490116E-08, 3.304730E-05, 3.522627E-10, 7.988442E-10, 5.426263E-10\\
7.450581E-09, 1.652456E-05, 9.935204E-10, 8.102952E-10, 1.402885E-09\\
3.725290E-09, 8.260210E-06, 1.388294E-09, 3.403684E-09, 3.707138E-09\\
1.862645E-09, 4.129411E-06, 3.586892E-09, 1.205086E-09, 6.541268E-09\\
9.313226E-10, 2.056866E-06, 6.151923E-09, 2.121169E-09, 1.034301E-08\\
4.656613E-10, 1.026456E-06, 2.007638E-08, 6.884798E-09, 9.770459E-09\\
2.328306E-10, 5.075871E-07, 8.350521E-09, 1.641206E-08, 4.490222E-08\\
1.164153E-10, 3.199734E-07, 6.786754E-08, 3.562143E-08, 4.499383E-08\\
5.820766E-11, 1.510118E-08, 9.043151E-08, 1.402664E-07, 1.208455E-07\\
2.910383E-11, 1.323597E-07, 1.792632E-07, 2.144407E-07, 1.067094E-07\\
1.455192E-11, 1.070428E-06, 2.261666E-07, 2.682704E-08, 2.701386E-07\\
7.275958E-12, 1.837584E-06, 2.261666E-07, 1.134033E-06, 8.408863E-07\\
3.637979E-12, 2.102304E-06, 1.649970E-06, 4.304814E-07, 9.297180E-07\\
1.818989E-12, 8.481169E-06, 6.013940E-07, 6.417419E-06, 6.129247E-06\\
}\taylortesttable
\begin{tikzpicture}
    \begin{loglogaxis}[
      width=6.5cm,
      height=5.5cm,
      xlabel={Finite-difference stepsize $h$},
      ylabel near ticks,
      ylabel={Error},
	  log basis x=10,
      legend cell align=left,
      legend style={at={(0.01, 0.01)}, anchor=south west},
      legend columns=1,
      legend style={font=\small},
      scale only axis,
      grid=both,
      xmin=1e-11,
      xmax=1e-3
      ]

      \addplot+[thick, solid, red, mark=*, mark options={fill=none}, mark size=1pt, line width=1.5pt] table[x=eps, y=first_order] {\taylortesttable};
      \addlegendentry{1st order}
      \addplot+[thick, solid, blue, mark=square*, mark options={fill=none}, mark size=1pt, line width=1.5pt] table[x=eps, y=second_order] {\taylortesttable};
      \addlegendentry{2nd order}
      \addplot+[thick, solid, green!80!black, mark=diamond*, mark options={fill=none}, mark size=1pt, line width=1.5pt] table[x=eps, y=fourth_order] {\taylortesttable};
      \addlegendentry{4th order}
      \addplot+[thick, solid, orange, mark=triangle*, mark options={fill=none}, mark size=1pt, line width=1.5pt] table[x=eps, y=sixth_order] {\taylortesttable};
      \addlegendentry{6th order}

    \end{loglogaxis}
  \end{tikzpicture}

%% file: problem1/convergence_history/convergence_history.tex
\pgfplotstableread[col sep = comma]{problem1/convergence_history/convergence_history.txt}\convergencehistorytable
\pgfplotstableread[col sep = comma]{problem1/convergence_history/convergence_history_gd.txt}\convergencehistorytablegd
\pgfplotstableread[col sep = comma]{problem1/convergence_history/convergence_history_pgd.txt}\convergencehistorytablewgd

\begin{tikzpicture}
    \begin{semilogyaxis}[
      width=6.5cm,
      height=5.5cm,
      xlabel={Iteration $\#$ $k$},
      ylabel near ticks,
	  log basis x=10,
      legend cell align=left,
      legend columns=1,
      legend style={font=\small},
      legend style={at={(1,1)},anchor=north east},
      legend columns=2, 
      scale only axis,
      grid=both,
      ymin=1e-12,
      ymax=5e4,
      xmin=0,
      xmax = 4000
      ]
        \addplot+[very thick, solid, purple, mark=none, mark options={fill=none}, mark size=0.5pt, line width=1.5pt] table[x=iter, y=fgd] {\convergencehistorytablegd};
      \addlegendentry{$J_k$ (sd)}
         \addplot+[thick, blue, mark=none, dotted, mark options={fill=none}, mark size=0.5pt, line width=1.0pt] table[x=idx,y=grad_a] {\convergencehistorytable};
      \addlegendentry{$\| \nabla_{\mathbf{a}} J_k \|_{\infty}$ (BFGS)}
       \addplot+[solid, green!80!black, mark=none, mark options={fill=none}, mark size=0.5pt, line width=2pt]
       table[x=iter, y=f] {\convergencehistorytablewgd};
      \addlegendentry{$J_k$ (p-sd)}
      \addplot+[thick, red, mark=none, dotted, mark options={fill=none}, mark size=0.5pt, line width=1.0pt] table[x=idx,y=grad_c] {\convergencehistorytable};
      \addlegendentry{$\| \nabla_{\mathbf{c}} J_k \|_{\infty}$ (BFGS)}
    \addplot+[solid, black, mark=none, mark options={fill=none}, mark size=0.5pt, line width=2pt] table[x=idx, y=f] {\convergencehistorytable};
      \addlegendentry{$J_k$ (BFGS)}
    \end{semilogyaxis}
  \end{tikzpicture}

%% file: problem1/eigenvalues/diag_axis.tex
\pgfplotstableread[col sep = comma]{problem1/eigenvalues/hessian_diag_axis_p1.txt}\axisstable
\begin{tikzpicture}
  \begin{semilogyaxis}[
      width=6.5cm,
      height=4.5cm,
      xlabel={$\#$ of axis Fourier mode},
      ylabel near ticks,
      ylabel={$H_{jj}$},
	  log basis x=10,
      legend cell align=left,
      legend pos=south east,
      legend columns=1,
      legend style={font=\small},
      scale only axis,
      grid=both
      ]
      \addplot+[thick, solid, red, mark=*, mark options={fill=none}, mark size=2pt, only marks] table[x=idx, y=sZ] {\axisstable};
      \addlegendentry{sin $Z$ coeffs.}
      \addplot+[thick, solid, blue, mark=square, mark options={fill=none}, mark size=2pt, only marks] table[x=idx, y=cR] {\axisstable};
      \addlegendentry{cos $R$ coeffs.}
    \end{semilogyaxis}
\end{tikzpicture}

%% file: problem1/eigenvalues/diag_coil.tex
\pgfplotstableread[col sep = comma]{problem1/eigenvalues/hessian_diag_cc_coil1x_p1.txt}\axisstablex
\pgfplotstableread[col sep = comma]{problem1/eigenvalues/hessian_diag_cc_coil1y_p1.txt}\axisstabley
\pgfplotstableread[col sep = comma]{problem1/eigenvalues/hessian_diag_cc_coil1z_p1.txt}\axisstablez

\pgfplotstablesave[skip rows between index={0}{1}]{\axisstablex}{testx.dat}
\pgfplotstablesave[skip rows between index={0}{1}]{\axisstabley}{testy.dat}
\pgfplotstablesave[skip rows between index={0}{1}]{\axisstablez}{testz.dat}

\begin{tikzpicture}
    \begin{axis}[
      width=6.5cm,
      height=4.5cm,
      xlabel={$\#$ of coil Fourier mode},
      ylabel near ticks,
      ylabel={$H_{jj}$},
      legend cell align=left,
      legend columns=1,
      legend style={font=\small},
      scale only axis,
            grid=both,
        ymin = 0,
        ymax = 25
      ]
      \addplot+[thick, solid, red, mark=*, mark options={fill=none}, mark size=2pt, only marks] table[x=idx, y=sin] {testx.dat};
      \addlegendentry{sin $x$ coeffs.}
      \addplot+[thick, solid, blue, mark=*, mark options={fill=none}, mark size=2pt, only marks] table[x=idx, y=cos] {\axisstablex};
      \addlegendentry{cos $x$ coeffs.}
      
      \addplot+[thick, solid, red, mark=+, mark options={fill=none}, mark size=2pt, only marks] table[x=idx, y=sin] {testy.dat};
      \addlegendentry{sin $y$ coeffs.}
      \addplot+[thick, solid, blue, mark=+, mark options={fill=none}, mark size=2pt, only marks] table[x=idx, y=cos] {\axisstabley};
      \addlegendentry{cos $y$ coeffs.}
      
      \addplot+[thick, solid, red, mark=o, mark options={fill=none}, mark size=2pt, only marks] table[x=idx, y=sin] {testz.dat};
      \addlegendentry{sin $z$ coeffs.}
      \addplot+[thick, solid, blue, mark=o, mark options={fill=none}, mark size=2pt, only marks] table[x=idx, y=cos] {\axisstablez};
      \addlegendentry{cos $z$ coeffs.}
    \end{axis}
\end{tikzpicture}

%% file: problem1/eigenvalues/hessianeig.tex
\pgfplotstableread[col sep = comma]{problem1/eigenvalues/all_eigenvalues_pb1.txt}\eigstable

\begin{tikzpicture}
    \begin{semilogyaxis}[
      width=6.5cm,
      height=4.5cm,
      ylabel={$\lambda$},
      xlabel={Eigenvalue \#},
      ylabel near ticks,
	  log basis x=10,
      scale only axis,
            grid=both,
      ymin=1e-9,ymax=1e9,
      xmin=1, xmax = 80
      ]
      \addplot+[solid, color=blue!90!red, mark=*, mark options={fill=none}, mark size=0.75pt, line width=0.5pt] table[x=idx, y=eig] {\eigstable};
    \end{semilogyaxis}
  \end{tikzpicture}

%% file: problem1/eigenvalues/reduced_hessian_coils.tex
\pgfplotstableread[col sep = comma]{problem1/eigenvalues/reduced_H_evals_coils_p1.txt}\eigstable
\begin{tikzpicture}
    \begin{semilogyaxis}[
      width=6.5cm,
      height=4.5cm,
      ylabel={$\lambda$}, 
      xlabel={Eigenvalue \#},
      ylabel near ticks,
	  log basis x=10,
      scale only axis,
            grid=both,
      ymin=1e-9,ymax=1e9,
      xmin=1, xmax = 80
      ]
      
      \addplot+[solid, blue!90!red, mark=*, mark options={fill=none}, mark size=.75pt, line width=0.5pt] table[x=idx, y=eig] {\eigstable};
    \end{semilogyaxis}
  \end{tikzpicture}

%% file: problem2/modB/absB_initial.tex
\pgfplotstableread[format=file, col sep=comma, header=false]{problem2/data/absB/output-ncsx_atopt_ppp-20_Nt_ma-4_Nt_coils-6_curvature-0p0_torsion-0p0_tikhonov-0p0_sobolev-0p0_arclength-0p0_min_dist-0p2_dist_weight-0p0-init.txt}\absb
\begin{tikzpicture}
    \begin{axis}[
            width=5cm,
            height=3cm,
            xlabel near ticks,
            xlabel={$\phi$},
            ylabel near ticks,
            scale only axis,
            cycle list/Dark2,
            xmin=0,
            xmax=62.83,
            ymin=0.88,
            ymax=1.12,
            ylabel style={font=\small}, 
            xlabel style={font=\small}, 
            yticklabel style={font=\small},
            xticklabel style={font=\small},
        ]
        \addplot+[thick, blue!50!black!50!white] table[col sep=comma,header=false,x index=0,y index=11]  {\absb};
        \addplot+[thick, blue!50!black!60!white] table[col sep=comma,header=false,x index=0,y index=9]  {\absb};
        \addplot+[thick, blue!50!black!70!white] table[col sep=comma,header=false,x index=0,y index=7]  {\absb};
        \addplot+[thick, blue!50!black!80!white] table[col sep=comma,header=false,x index=0,y index=5]  {\absb};
        \addplot+[thick, blue!50!black!90!white] table[col sep=comma,header=false,x index=0,y index=3]  {\absb};
        \addplot+[thick, blue!50!black!100!white] table[col sep=comma,header=false,x index=0,y index=1]  {\absb};

    \end{axis}
\end{tikzpicture}

%% file: problem2/modB/absB_final_noreg.tex
\pgfplotstableread[format=file, col sep=comma, header=false]{problem2/data/absB/noreg.txt}\absb
\begin{tikzpicture}
    \begin{axis}[
            width=5cm,
            height=3cm,
            xlabel near ticks,
            xlabel={$\phi$},
            ylabel near ticks,
            scale only axis,
            cycle list/Dark2,
            xmin=0,
            xmax=62.83,
            ymin=0.88,
            ymax=1.12,
            ylabel style={font=\small}, 
            xlabel style={font=\small}, 
            yticklabel style={font=\small},
            xticklabel style={font=\small},
        ]
        \addplot+[thick, blue!50!black!50!white] table[col sep=comma,header=false,x index=0,y index=11]  {\absb};
        \addplot+[thick, blue!50!black!60!white] table[col sep=comma,header=false,x index=0,y index=9]  {\absb};
        \addplot+[thick, blue!50!black!70!white] table[col sep=comma,header=false,x index=0,y index=7]  {\absb};
        \addplot+[thick, blue!50!black!80!white] table[col sep=comma,header=false,x index=0,y index=5]  {\absb};
        \addplot+[thick, blue!50!black!90!white] table[col sep=comma,header=false,x index=0,y index=3]  {\absb};
        \addplot+[thick, blue!50!black!100!white] table[col sep=comma,header=false,x index=0,y index=1]  {\absb};

    \end{axis}

\end{tikzpicture}

%% file: problem2/modB/absB_final_withdist.tex
\pgfplotstableread[format=file, col sep=comma, header=false]{problem2/data/absB/mindist.txt}\absb
\begin{tikzpicture}
    \begin{axis}[
            width=5cm,
            height=3cm,
            xlabel near ticks,
            xlabel={$\phi$},
            ylabel near ticks,
            scale only axis,
            cycle list/Dark2,
            xmin=0,
            xmax=62.83,
            ymin=0.88,
            ymax=1.12,
            ylabel style={font=\small}, 
            xlabel style={font=\small}, 
            yticklabel style={font=\small},
            xticklabel style={font=\small},
        ]
        \addplot+[thick, blue!50!black!50!white] table[col sep=comma,header=false,x index=0,y index=11]  {\absb};
        \addplot+[thick, blue!50!black!60!white] table[col sep=comma,header=false,x index=0,y index=9]  {\absb};
        \addplot+[thick, blue!50!black!70!white] table[col sep=comma,header=false,x index=0,y index=7]  {\absb};
        \addplot+[thick, blue!50!black!80!white] table[col sep=comma,header=false,x index=0,y index=5]  {\absb};
        \addplot+[thick, blue!50!black!90!white] table[col sep=comma,header=false,x index=0,y index=3]  {\absb};
        \addplot+[thick, blue!50!black!100!white] table[col sep=comma,header=false,x index=0,y index=1]  {\absb};
    \end{axis}
\end{tikzpicture}

%% file: problem2/modB/absB_final_withdistandreg.tex
\pgfplotstableread[format=file, col sep=comma, header=false]{problem2/data/absB/mindistcurvature.txt}\absb
\begin{tikzpicture}
    \begin{axis}[
            width=5cm,
            height=3cm,
            xlabel near ticks,
            xlabel={$\phi$},
            ylabel near ticks,
            scale only axis,
            cycle list/Dark2,
            xmin=0,
            xmax=62.83,
            ymin=0.88,
            ymax=1.12,
            ylabel style={font=\small}, 
            xlabel style={font=\small}, 
            yticklabel style={font=\small},
            xticklabel style={font=\small},
        ]
        \addplot+[thick, blue!50!black!50!white] table[col sep=comma,header=false,x index=0,y index=11]  {\absb};
        \addplot+[thick, blue!50!black!60!white] table[col sep=comma,header=false,x index=0,y index=9]  {\absb};
        \addplot+[thick, blue!50!black!70!white] table[col sep=comma,header=false,x index=0,y index=7]  {\absb};
        \addplot+[thick, blue!50!black!80!white] table[col sep=comma,header=false,x index=0,y index=5]  {\absb};
        \addplot+[thick, blue!50!black!90!white] table[col sep=comma,header=false,x index=0,y index=3]  {\absb};
        \addplot+[thick, blue!50!black!100!white] table[col sep=comma,header=false,x index=0,y index=1]  {\absb};
    \end{axis}
\end{tikzpicture}

%% file: main.bbl
\begin{thebibliography}{54}
\expandafter\ifx\csname natexlab\endcsname\relax\def\natexlab#1{#1}\fi
\providecommand{\url}[1]{\texttt{#1}}
\providecommand{\href}[2]{#2}
\providecommand{\path}[1]{#1}
\providecommand{\DOIprefix}{doi:}
\providecommand{\ArXivprefix}{arXiv:}
\providecommand{\URLprefix}{URL: }
\providecommand{\Pubmedprefix}{pmid:}
\providecommand{\doi}[1]{\href{http://dx.doi.org/#1}{\path{#1}}}
\providecommand{\Pubmed}[1]{\href{pmid:#1}{\path{#1}}}
\providecommand{\bibinfo}[2]{#2}
\ifx\xfnm\relax \def\xfnm[#1]{\unskip,\space#1}\fi
\bibitem[{Bader et~al.(2020)Bader, Faber, Schmitt, Anderson, Drevlak, Duff,
  Frerichs, Hegna, Kruger, Landreman, McKinney, Singh, Schroeder, Terry and
  Ware}]{bader2020new}
\bibinfo{author}{Bader, A.}, \bibinfo{author}{Faber, B.J.},
  \bibinfo{author}{Schmitt, J.C.}, \bibinfo{author}{Anderson, D.T.},
  \bibinfo{author}{Drevlak, M.}, \bibinfo{author}{Duff, J.M.},
  \bibinfo{author}{Frerichs, H.}, \bibinfo{author}{Hegna, C.C.},
  \bibinfo{author}{Kruger, T.G.}, \bibinfo{author}{Landreman, M.},
  \bibinfo{author}{McKinney, I.J.}, \bibinfo{author}{Singh, L.},
  \bibinfo{author}{Schroeder, J.M.}, \bibinfo{author}{Terry, P.W.},
  \bibinfo{author}{Ware, A.S.}, \bibinfo{year}{2020}.
\newblock \bibinfo{title}{A new optimized quasihelically symmetricstellarator}.
\newblock \href{http://arxiv.org/abs/2004.11426}{{\tt arXiv:2004.11426}}.
\bibitem[{Bader and Kruger(2020)}]{Bader2020}
\bibinfo{author}{Bader, A.}, \bibinfo{author}{Kruger, T.},
  \bibinfo{year}{2020}.
\newblock \bibinfo{title}{Ensuring equilibrium evaluations do not depend on
  parametrization}.
\newblock \bibinfo{note}{Presentation in Wisconsin Optimized Stellarator
  Project Group Meetings}.
\bibitem[{Boozer(1981a)}]{boozer1981A}
\bibinfo{author}{Boozer, A.H.}, \bibinfo{year}{1981}a.
\newblock \bibinfo{title}{Establishment of magnetic coordinates for a given
  magnetic field}.
\newblock \bibinfo{type}{Technical Report}. Princeton Univ., NJ (USA). Plasma
  Physics Lab.
\bibitem[{Boozer(1981b)}]{boozer1981B}
\bibinfo{author}{Boozer, A.H.}, \bibinfo{year}{1981}b.
\newblock \bibinfo{title}{Plasma equilibrium with rational magnetic surfaces}.
\newblock \bibinfo{journal}{The Physics of Fluids} \bibinfo{volume}{24},
  \bibinfo{pages}{1999--2003}.
\newblock \URLprefix \url{https://aip.scitation.org/doi/abs/10.1063/1.863297},
  \DOIprefix\doi{10.1063/1.863297},
  \href{http://arxiv.org/abs/https://aip.scitation.org/doi/pdf/10.1063/1.863297}{{\tt
  arXiv:https://aip.scitation.org/doi/pdf/10.1063/1.863297}}.
\bibitem[{Boozer(1995)}]{Boozer_1995}
\bibinfo{author}{Boozer, A.H.}, \bibinfo{year}{1995}.
\newblock \bibinfo{title}{Quasi-helical symmetry in stellarators}.
\newblock \bibinfo{journal}{Plasma Physics and Controlled Fusion}
  \bibinfo{volume}{37}, \bibinfo{pages}{A103--A117}.
\newblock \DOIprefix\doi{10.1088/0741-3335/37/11a/007}.
\bibitem[{Boozer(1998)}]{Boozer1998}
\bibinfo{author}{Boozer, A.H.}, \bibinfo{year}{1998}.
\newblock \bibinfo{title}{What is a stellarator?}
\newblock \bibinfo{journal}{Physics of Plasmas} \bibinfo{volume}{5},
  \bibinfo{pages}{1647--1655}.
\newblock \DOIprefix\doi{10.1063/1.872833}.
\bibitem[{Borz{\`\i} and Schulz(2012)}]{BorziSchulz12}
\bibinfo{author}{Borz{\`\i}, A.}, \bibinfo{author}{Schulz, V.},
  \bibinfo{year}{2012}.
\newblock \bibinfo{title}{Computational Optimization of Systems Governed by
  Partial Differential Equations}.
\newblock \bibinfo{publisher}{SIAM}.
\bibitem[{{Brown} et~al.(2015){Brown}, {Breslau}, {Gates}, {Pomphrey} and
  {Zolfaghari}}]{brown_2015}
\bibinfo{author}{{Brown}, T.}, \bibinfo{author}{{Breslau}, J.},
  \bibinfo{author}{{Gates}, D.}, \bibinfo{author}{{Pomphrey}, N.},
  \bibinfo{author}{{Zolfaghari}, A.}, \bibinfo{year}{2015}.
\newblock \bibinfo{title}{Engineering optimization of stellarator coils lead to
  improvements in device maintenance}, in: \bibinfo{booktitle}{2015 IEEE 26th
  Symposium on Fusion Engineering (SOFE)}, pp. \bibinfo{pages}{1--6}.
\bibitem[{Canik et~al.(2007)Canik, Anderson, Anderson, Likin, Talmadge and
  Zhai}]{Canik2007}
\bibinfo{author}{Canik, J.M.}, \bibinfo{author}{Anderson, D.T.},
  \bibinfo{author}{Anderson, F.S.B.}, \bibinfo{author}{Likin, K.M.},
  \bibinfo{author}{Talmadge, J.N.}, \bibinfo{author}{Zhai, K.},
  \bibinfo{year}{2007}.
\newblock \bibinfo{title}{Experimental demonstration of improved neoclassical
  transport with quasihelical symmetry}.
\newblock \bibinfo{journal}{Phys. Rev. Lett.} \bibinfo{volume}{98},
  \bibinfo{pages}{085002}.
\newblock \DOIprefix\doi{10.1103/PhysRevLett.98.085002}.
\bibitem[{Conn et~al.(2009)Conn, Scheinberg and
  Vicente}]{ConnScheinbergVicente09}
\bibinfo{author}{Conn, A.R.}, \bibinfo{author}{Scheinberg, K.},
  \bibinfo{author}{Vicente, L.N.}, \bibinfo{year}{2009}.
\newblock \bibinfo{title}{Introduction to derivative-free optimization}.
\newblock \bibinfo{publisher}{SIAM}.
\bibitem[{Dewar and Hudson(1998)}]{Dewar1998}
\bibinfo{author}{Dewar, R.}, \bibinfo{author}{Hudson, S.},
  \bibinfo{year}{1998}.
\newblock \bibinfo{title}{Stellarator symmetry}.
\newblock \bibinfo{journal}{Physica D: Nonlinear Phenomena}
  \bibinfo{volume}{112}, \bibinfo{pages}{275--280}.
\newblock \DOIprefix\doi{https://doi.org/10.1016/S0167-2789(97)00216-9}.
  \bibinfo{note}{proceedings of the Workshop on Time-Reversal Symmetry in
  Dynamical Systems}.
\bibitem[{Drevlak(1998)}]{drevlak_1998}
\bibinfo{author}{Drevlak, M.}, \bibinfo{year}{1998}.
\newblock \bibinfo{title}{Automated optimization of stellarator coils}.
\newblock \bibinfo{journal}{Fusion Technology} \bibinfo{volume}{33},
  \bibinfo{pages}{106--117}.
\newblock \DOIprefix\doi{10.13182/FST98-A21}.
\bibitem[{Drevlak et~al.(2013)Drevlak, Brochard, Helander, Kisslinger,
  Mikhailov, Nührenberg, Nührenberg and Turkin}]{Drevlak_2013}
\bibinfo{author}{Drevlak, M.}, \bibinfo{author}{Brochard, F.},
  \bibinfo{author}{Helander, P.}, \bibinfo{author}{Kisslinger, J.},
  \bibinfo{author}{Mikhailov, M.}, \bibinfo{author}{Nührenberg, C.},
  \bibinfo{author}{Nührenberg, J.}, \bibinfo{author}{Turkin, Y.},
  \bibinfo{year}{2013}.
\newblock \bibinfo{title}{Estell: A quasi-toroidally symmetric stellarator}.
\newblock \bibinfo{journal}{Contributions to Plasma Physics}
  \bibinfo{volume}{53}, \bibinfo{pages}{459--468}.
\newblock \DOIprefix\doi{10.1002/ctpp.201200055}.
\bibitem[{Garabedian(2008)}]{Garabedian_2008}
\bibinfo{author}{Garabedian, P.R.}, \bibinfo{year}{2008}.
\newblock \bibinfo{title}{Three-dimensional analysis of tokamaks and
  stellarators}.
\newblock \bibinfo{journal}{Proceedings of the National Academy of Sciences}
  \bibinfo{volume}{105}, \bibinfo{pages}{13716--13719}.
\newblock \URLprefix \url{https://www.pnas.org/content/105/37/13716},
  \DOIprefix\doi{10.1073/pnas.0806354105}.
\bibitem[{Garren and Boozer(1991)}]{garren1991}
\bibinfo{author}{Garren, D.A.}, \bibinfo{author}{Boozer, A.H.},
  \bibinfo{year}{1991}.
\newblock \bibinfo{title}{Existence of quasihelically symmetric stellarators}.
\newblock \bibinfo{journal}{Physics of Fluids B: Plasma Physics}
  \bibinfo{volume}{3}, \bibinfo{pages}{2822--2834}.
\newblock \URLprefix \url{https://doi.org/10.1063/1.859916},
  \DOIprefix\doi{10.1063/1.859916}.
\bibitem[{Geiger et~al.(2014)Geiger, Beidler, Feng, Maa{\ss}berg, Marushchenko
  and Turkin}]{Geiger_2014}
\bibinfo{author}{Geiger, J.}, \bibinfo{author}{Beidler, C.D.},
  \bibinfo{author}{Feng, Y.}, \bibinfo{author}{Maa{\ss}berg, H.},
  \bibinfo{author}{Marushchenko, N.B.}, \bibinfo{author}{Turkin, Y.},
  \bibinfo{year}{2014}.
\newblock \bibinfo{title}{Physics in the magnetic configuration space of
  {W7-X}}.
\newblock \bibinfo{journal}{Plasma Physics and Controlled Fusion}
  \bibinfo{volume}{57}, \bibinfo{pages}{014004}.
\newblock \DOIprefix\doi{10.1088/0741-3335/57/1/014004}.
\bibitem[{Gunzburger(2003)}]{Gunzburger03}
\bibinfo{author}{Gunzburger, M.D.}, \bibinfo{year}{2003}.
\newblock \bibinfo{title}{Perspectives in Flow Control and Optimization}.
\newblock \bibinfo{publisher}{SIAM}, \bibinfo{address}{Philadelphia}.
\bibitem[{Helander(2014)}]{Helander2014}
\bibinfo{author}{Helander, P.}, \bibinfo{year}{2014}.
\newblock \bibinfo{title}{Theory of plasma confinement in non-axisymmetric
  magnetic fields}.
\newblock \bibinfo{journal}{Reports on Progress in Physics}
  \bibinfo{volume}{77}, \bibinfo{pages}{087001}.
\newblock \DOIprefix\doi{10.1088/0034-4885/77/8/087001}.
\bibitem[{Helander et~al.(2012)Helander, Beidler, Bird, Drevlak, Feng, Hatzky,
  Jenko, Kleiber, Proll, Turkin and Xanthopoulos}]{Helander2012}
\bibinfo{author}{Helander, P.}, \bibinfo{author}{Beidler, C.D.},
  \bibinfo{author}{Bird, T.M.}, \bibinfo{author}{Drevlak, M.},
  \bibinfo{author}{Feng, Y.}, \bibinfo{author}{Hatzky, R.},
  \bibinfo{author}{Jenko, F.}, \bibinfo{author}{Kleiber, R.},
  \bibinfo{author}{Proll, J.H.E.}, \bibinfo{author}{Turkin, Y.},
  \bibinfo{author}{Xanthopoulos, P.}, \bibinfo{year}{2012}.
\newblock \bibinfo{title}{Stellarator and tokamak plasmas: a comparison}.
\newblock \bibinfo{journal}{Plasma Physics and Controlled Fusion}
  \bibinfo{volume}{54}, \bibinfo{pages}{124009}.
\newblock \DOIprefix\doi{10.1088/0741-3335/54/12/124009}.
\bibitem[{Henneberg et~al.(2019)Henneberg, Drevlak, N{\"u}hrenberg, Beidler,
  Turkin, Loizu and Helander}]{Henneberg_2019}
\bibinfo{author}{Henneberg, S.}, \bibinfo{author}{Drevlak, M.},
  \bibinfo{author}{N{\"u}hrenberg, C.}, \bibinfo{author}{Beidler, C.},
  \bibinfo{author}{Turkin, Y.}, \bibinfo{author}{Loizu, J.},
  \bibinfo{author}{Helander, P.}, \bibinfo{year}{2019}.
\newblock \bibinfo{title}{Properties of a new quasi-axisymmetric
  configuration}.
\newblock \bibinfo{journal}{Nuclear Fusion} \bibinfo{volume}{59},
  \bibinfo{pages}{026014}.
\newblock \DOIprefix\doi{10.1088/1741-4326/aaf604}.
\bibitem[{Hirshman et~al.(1986)Hirshman, {van Rij} and Merkel}]{Hirshman1986}
\bibinfo{author}{Hirshman, S.}, \bibinfo{author}{{van Rij}, W.},
  \bibinfo{author}{Merkel, P.}, \bibinfo{year}{1986}.
\newblock \bibinfo{title}{Three-dimensional free boundary calculations using a
  spectral green's function method}.
\newblock \bibinfo{journal}{Computer Physics Communications}
  \bibinfo{volume}{43}, \bibinfo{pages}{143 -- 155}.
\newblock \DOIprefix\doi{https://doi.org/10.1016/0010-4655(86)90058-5}.
\bibitem[{Imbert-Gerard et~al.(2020)Imbert-Gerard, Paul and
  Wright}]{ImbertEtAl20}
\bibinfo{author}{Imbert-Gerard, L.M.}, \bibinfo{author}{Paul, E.J.},
  \bibinfo{author}{Wright, A.M.}, \bibinfo{year}{2020}.
\newblock \bibinfo{title}{An introduction to stellarators: From magnetic fields
  to symmetries and optimization}.
\newblock \href{http://arxiv.org/abs/1908.05360}{{\tt arXiv:1908.05360}}.
\bibitem[{Jackson(1999)}]{jackson_classical_1999}
\bibinfo{author}{Jackson, J.D.}, \bibinfo{year}{1999}.
\newblock \bibinfo{title}{Classical electrodynamics}.
\newblock \bibinfo{edition}{3rd ed.} ed., \bibinfo{publisher}{Wiley},
  \bibinfo{address}{New York, {NY}}.
\newblock \URLprefix \url{http://cdsweb.cern.ch/record/490457}.
\bibitem[{Klinger et~al.(2013)Klinger, Baylard, Beidler, Boscary, Bosch,
  Dinklage, Hartmann, Helander, Ma{\ss}berg, Peacock, Pedersen, Rummel,
  Schauer, Wegener and Wolf}]{Klinger2013}
\bibinfo{author}{Klinger, T.}, \bibinfo{author}{Baylard, C.},
  \bibinfo{author}{Beidler, C.}, \bibinfo{author}{Boscary, J.},
  \bibinfo{author}{Bosch, H.}, \bibinfo{author}{Dinklage, A.},
  \bibinfo{author}{Hartmann, D.}, \bibinfo{author}{Helander, P.},
  \bibinfo{author}{Ma{\ss}berg, H.}, \bibinfo{author}{Peacock, A.},
  \bibinfo{author}{Pedersen, T.}, \bibinfo{author}{Rummel, T.},
  \bibinfo{author}{Schauer, F.}, \bibinfo{author}{Wegener, L.},
  \bibinfo{author}{Wolf, R.}, \bibinfo{year}{2013}.
\newblock \bibinfo{title}{Towards assembly completion and preparation of
  experimental campaigns of {W}endelstein {7-X} in the perspective of a path to
  a stellarator fusion power plant}.
\newblock \bibinfo{journal}{Fusion Engineering and Design}
  \bibinfo{volume}{88}, \bibinfo{pages}{461 -- 465}.
\newblock \DOIprefix\doi{https://doi.org/10.1016/j.fusengdes.2013.02.153}.
  \bibinfo{note}{proceedings of the 27th Symposium On Fusion Technology
  (SOFT-27); Liège, Belgium, September 24-28, 2012}.
\bibitem[{Ku et~al.(2008)Ku, Garabedian, Lyon, Turnbull, Grossman, Mau,
  Zarnstorff and Team}]{ku_2008}
\bibinfo{author}{Ku, L.P.}, \bibinfo{author}{Garabedian, P.R.},
  \bibinfo{author}{Lyon, J.}, \bibinfo{author}{Turnbull, A.},
  \bibinfo{author}{Grossman, A.}, \bibinfo{author}{Mau, T.K.},
  \bibinfo{author}{Zarnstorff, M.}, \bibinfo{author}{Team, A.},
  \bibinfo{year}{2008}.
\newblock \bibinfo{title}{Physics design for {ARIES-CS}}.
\newblock \bibinfo{journal}{Fusion Science and Technology}
  \bibinfo{volume}{54}, \bibinfo{pages}{673--693}.
\newblock \URLprefix \url{https://doi.org/10.13182/FST08-A1899},
  \DOIprefix\doi{10.13182/FST08-A1899}.
\bibitem[{Landreman(2017)}]{Landreman_2017}
\bibinfo{author}{Landreman, M.}, \bibinfo{year}{2017}.
\newblock \bibinfo{title}{An improved current potential method for fast
  computation of stellarator coil shapes}.
\newblock \bibinfo{journal}{Nuclear Fusion} \bibinfo{volume}{57},
  \bibinfo{pages}{046003}.
\newblock \DOIprefix\doi{10.1088/1741-4326/aa57d4}.
\bibitem[{Landreman(2021)}]{Landreman2020}
\bibinfo{author}{Landreman, M.}, \bibinfo{year}{2021}.
\newblock \bibinfo{title}{Figures of merit for stellarators near the magnetic
  axis}.
\newblock \bibinfo{journal}{Journal of Plasma Physics} \bibinfo{volume}{87},
  \bibinfo{pages}{905870112}.
\newblock \DOIprefix\doi{10.1017/S0022377820001658}.
\bibitem[{Landreman and Sengupta(2018)}]{landreman_2018}
\bibinfo{author}{Landreman, M.}, \bibinfo{author}{Sengupta, W.},
  \bibinfo{year}{2018}.
\newblock \bibinfo{title}{Direct construction of optimized stellarator shapes.
  {P}art 1. {T}heory in cylindrical coordinates}.
\newblock \bibinfo{journal}{Journal of Plasma Physics} \bibinfo{volume}{84},
  \bibinfo{pages}{905840616}.
\newblock \DOIprefix\doi{10.1017/S0022377818001289}.
\bibitem[{Landreman and Sengupta(2019)}]{LandremanSengupta2019}
\bibinfo{author}{Landreman, M.}, \bibinfo{author}{Sengupta, W.},
  \bibinfo{year}{2019}.
\newblock \bibinfo{title}{Constructing stellarators with quasisymmetry to high
  order}.
\newblock \bibinfo{journal}{Journal of Plasma Physics} \bibinfo{volume}{85}.
\bibitem[{Landreman et~al.(2019)Landreman, Sengupta and Plunk}]{landreman_2019}
\bibinfo{author}{Landreman, M.}, \bibinfo{author}{Sengupta, W.},
  \bibinfo{author}{Plunk, G.G.}, \bibinfo{year}{2019}.
\newblock \bibinfo{title}{Direct construction of optimized stellarator shapes.
  {P}art 2. {N}umerical quasisymmetric solutions}.
\newblock \bibinfo{journal}{Journal of Plasma Physics} \bibinfo{volume}{85},
  \bibinfo{pages}{905850103}.
\newblock \DOIprefix\doi{10.1017/S0022377818001344}.
\bibitem[{Liu et~al.(2018)Liu, Shimizu, Isobe, Okamura, Nishimura, Suzuki, Xu,
  Zhang, Liu, Huang, Wang, Liu, Tang, Yin, Wan and team}]{liu_2018}
\bibinfo{author}{Liu, H.}, \bibinfo{author}{Shimizu, A.},
  \bibinfo{author}{Isobe, M.}, \bibinfo{author}{Okamura, S.},
  \bibinfo{author}{Nishimura, S.}, \bibinfo{author}{Suzuki, C.},
  \bibinfo{author}{Xu, Y.}, \bibinfo{author}{Zhang, X.}, \bibinfo{author}{Liu,
  B.}, \bibinfo{author}{Huang, J.}, \bibinfo{author}{Wang, X.},
  \bibinfo{author}{Liu, H.}, \bibinfo{author}{Tang, C.}, \bibinfo{author}{Yin,
  D.}, \bibinfo{author}{Wan, Y.}, \bibinfo{author}{team, C.},
  \bibinfo{year}{2018}.
\newblock \bibinfo{title}{Magnetic configuration and modular coil design for
  the chinese first quasi-axisymmetric stellarator}.
\newblock \bibinfo{journal}{Plasma and Fusion Research} \bibinfo{volume}{13},
  \bibinfo{pages}{3405067--3405067}.
\newblock \DOIprefix\doi{10.1585/pfr.13.3405067}.
\bibitem[{McGreivy et~al.(2020)McGreivy, Hudson and Zhu}]{mcgreivy2020}
\bibinfo{author}{McGreivy, N.}, \bibinfo{author}{Hudson, S.R.},
  \bibinfo{author}{Zhu, C.}, \bibinfo{year}{2020}.
\newblock \bibinfo{title}{Optimized finite-build stellarator coils using
  automatic differentiation}.
\newblock \href{http://arxiv.org/abs/2009.00196}{{\tt arXiv:2009.00196}}.
\bibitem[{Merkel(1987)}]{Merkel_1987}
\bibinfo{author}{Merkel, P.}, \bibinfo{year}{1987}.
\newblock \bibinfo{title}{Solution of stellarator boundary value problems with
  external currents}.
\newblock \bibinfo{journal}{Nuclear Fusion} \bibinfo{volume}{27},
  \bibinfo{pages}{867--871}.
\newblock \DOIprefix\doi{10.1088/0029-5515/27/5/018}.
\bibitem[{Mynick et~al.(2010)Mynick, Pomphrey and Xanthopoulos}]{Mynick2010}
\bibinfo{author}{Mynick, H.E.}, \bibinfo{author}{Pomphrey, N.},
  \bibinfo{author}{Xanthopoulos, P.}, \bibinfo{year}{2010}.
\newblock \bibinfo{title}{Optimizing stellarators for turbulent transport}.
\newblock \bibinfo{journal}{Phys. Rev. Lett.} \bibinfo{volume}{105},
  \bibinfo{pages}{095004}.
\newblock \DOIprefix\doi{10.1103/PhysRevLett.105.095004}.
\bibitem[{{Neilson} et~al.(2010){Neilson}, {Gruber}, {Harris}, {Rej}, {Simmons}
  and {Strykowsky}}]{Neilson10}
\bibinfo{author}{{Neilson}, G.H.}, \bibinfo{author}{{Gruber}, C.O.},
  \bibinfo{author}{{Harris}, J.H.}, \bibinfo{author}{{Rej}, D.J.},
  \bibinfo{author}{{Simmons}, R.T.}, \bibinfo{author}{{Strykowsky}, R.L.},
  \bibinfo{year}{2010}.
\newblock \bibinfo{title}{Lessons learned in risk management on {NCSX}}.
\newblock \bibinfo{journal}{IEEE Transactions on Plasma Science}
  \bibinfo{volume}{38}, \bibinfo{pages}{320--327}.
\bibitem[{Nocedal and Wright(2006)}]{NocedalWright06}
\bibinfo{author}{Nocedal, J.}, \bibinfo{author}{Wright, S.J.},
  \bibinfo{year}{2006}.
\newblock \bibinfo{title}{Numerical Optimization}.
\newblock \bibinfo{edition}{second} ed., \bibinfo{publisher}{Springer Verlag},
  \bibinfo{address}{Berlin, Heidelberg, New York}.
\bibitem[{N{\"u}hrenberg and Zille(1988)}]{Nuhrenberg1988}
\bibinfo{author}{N{\"u}hrenberg, J.}, \bibinfo{author}{Zille, R.},
  \bibinfo{year}{1988}.
\newblock \bibinfo{title}{Quasi-helically symmetric toroidal stellarators}.
\newblock \bibinfo{journal}{Physics Letters A} \bibinfo{volume}{129},
  \bibinfo{pages}{113--117}.
\newblock \DOIprefix\doi{https://doi.org/10.1016/0375-9601(88)90080-1}.
\bibitem[{Paul et~al.(2018)Paul, Landreman, Bader and Dorland}]{Paul_2018}
\bibinfo{author}{Paul, E.}, \bibinfo{author}{Landreman, M.},
  \bibinfo{author}{Bader, A.}, \bibinfo{author}{Dorland, W.},
  \bibinfo{year}{2018}.
\newblock \bibinfo{title}{An adjoint method for gradient-based optimization of
  stellarator coil shapes}.
\newblock \bibinfo{journal}{Nuclear Fusion} \bibinfo{volume}{58},
  \bibinfo{pages}{076015}.
\newblock \DOIprefix\doi{10.1088/1741-4326/aac1c7}.
\bibitem[{Reiman et~al.(2001)Reiman, Ku, Monticello, Hirshman, Hudson, Kessel,
  Lazarus, Mikkelsen, Zarnstorff, Berry, Boozer, Brooks, Cooper, Drevlak,
  Fredrickson, Fu, Goldston, Hatcher, Isaev, Jun, Knowlton, Lewandowski, Lin,
  Lyon, Merkel, Mikhailov, Miner, Mynick, Neilson, Nelson, Nührenberg,
  Pomphrey, Redi, Reiersen, Rutherford, Sanchez, Schmidt, Spong, Strickler,
  Subbotin, Valanju and White}]{Reiman2001}
\bibinfo{author}{Reiman, A.}, \bibinfo{author}{Ku, L.},
  \bibinfo{author}{Monticello, D.}, \bibinfo{author}{Hirshman, S.},
  \bibinfo{author}{Hudson, S.}, \bibinfo{author}{Kessel, C.},
  \bibinfo{author}{Lazarus, E.}, \bibinfo{author}{Mikkelsen, D.},
  \bibinfo{author}{Zarnstorff, M.}, \bibinfo{author}{Berry, L.A.},
  \bibinfo{author}{Boozer, A.}, \bibinfo{author}{Brooks, A.},
  \bibinfo{author}{Cooper, W.A.}, \bibinfo{author}{Drevlak, M.},
  \bibinfo{author}{Fredrickson, E.}, \bibinfo{author}{Fu, G.},
  \bibinfo{author}{Goldston, R.}, \bibinfo{author}{Hatcher, R.},
  \bibinfo{author}{Isaev, M.}, \bibinfo{author}{Jun, C.},
  \bibinfo{author}{Knowlton, S.}, \bibinfo{author}{Lewandowski, J.},
  \bibinfo{author}{Lin, Z.}, \bibinfo{author}{Lyon, J.F.},
  \bibinfo{author}{Merkel, P.}, \bibinfo{author}{Mikhailov, M.},
  \bibinfo{author}{Miner, W.}, \bibinfo{author}{Mynick, H.},
  \bibinfo{author}{Neilson, G.}, \bibinfo{author}{Nelson, B.E.},
  \bibinfo{author}{Nührenberg, C.}, \bibinfo{author}{Pomphrey, N.},
  \bibinfo{author}{Redi, M.}, \bibinfo{author}{Reiersen, W.},
  \bibinfo{author}{Rutherford, P.}, \bibinfo{author}{Sanchez, R.},
  \bibinfo{author}{Schmidt, J.}, \bibinfo{author}{Spong, D.},
  \bibinfo{author}{Strickler, D.}, \bibinfo{author}{Subbotin, A.},
  \bibinfo{author}{Valanju, P.}, \bibinfo{author}{White, R.},
  \bibinfo{year}{2001}.
\newblock \bibinfo{title}{Recent advances in the design of quasiaxisymmetric
  stellarator plasma configurations}.
\newblock \bibinfo{journal}{Physics of Plasmas} \bibinfo{volume}{8},
  \bibinfo{pages}{2083--2094}.
\newblock \URLprefix \url{https://doi.org/10.1063/1.1351826},
  \DOIprefix\doi{10.1063/1.1351826}.
\bibitem[{De~los Reyes(2015)}]{Delosreyes15}
\bibinfo{author}{De~los Reyes, J.C.}, \bibinfo{year}{2015}.
\newblock \bibinfo{title}{Numerical PDE-constrained optimization}.
\newblock \bibinfo{publisher}{Springer}.
\bibitem[{Seiwald et~al.(2008)Seiwald, Kasilov, Kernbichler, Kalyuzhnyj, Nemov,
  Tribaldos and Jiménez}]{Seiwald2008}
\bibinfo{author}{Seiwald, B.}, \bibinfo{author}{Kasilov, S.},
  \bibinfo{author}{Kernbichler, W.}, \bibinfo{author}{Kalyuzhnyj, V.},
  \bibinfo{author}{Nemov, V.}, \bibinfo{author}{Tribaldos, V.},
  \bibinfo{author}{Jiménez, J.}, \bibinfo{year}{2008}.
\newblock \bibinfo{title}{Optimization of energy confinement in the 1/$\nu$
  regime for stellarators}.
\newblock \bibinfo{journal}{Journal of Computational Physics}
  \bibinfo{volume}{227}, \bibinfo{pages}{6165 -- 6183}.
\newblock \URLprefix
  \url{http://www.sciencedirect.com/science/article/pii/S0021999108001265},
  \DOIprefix\doi{https://doi.org/10.1016/j.jcp.2008.02.026}.
\bibitem[{Stenson(2019)}]{stenson2019}
\bibinfo{author}{Stenson, E.}, \bibinfo{year}{2019}.
\newblock \bibinfo{title}{Plans for epos: A tabletopsized, superconducting,
  optimized stellarator for matter/antimatter pair plasmas}.
\newblock \bibinfo{journal}{Stellarator News} \bibinfo{volume}{167},
  \bibinfo{pages}{5 -- 8}.
\bibitem[{Stewart(2012)}]{stewart2012essential}
\bibinfo{author}{Stewart, J.}, \bibinfo{year}{2012}.
\newblock \bibinfo{title}{Essential calculus: Early transcendentals}.
\newblock \bibinfo{publisher}{Cengage Learning}.
\bibitem[{Stoneking et~al.(2020)Stoneking, Pedersen, Helander, Chen,
  Hergenhahn, Stenson, Fiksel, von~der Linden, Saitoh, Surko, Danielson,
  Hugenschmidt, Horn-Stanja, Mishchenko, Kennedy, Deller, Card, Nissl, Singer,
  Singer, Koenig, Willingale, Peebles, Edwards and Chin}]{stoneking2020}
\bibinfo{author}{Stoneking, M.}, \bibinfo{author}{Pedersen, T.},
  \bibinfo{author}{Helander, P.}, \bibinfo{author}{Chen, H.},
  \bibinfo{author}{Hergenhahn, U.}, \bibinfo{author}{Stenson, E.},
  \bibinfo{author}{Fiksel, G.}, \bibinfo{author}{von~der Linden, J.},
  \bibinfo{author}{Saitoh, H.}, \bibinfo{author}{Surko, C.},
  \bibinfo{author}{Danielson, J.}, \bibinfo{author}{Hugenschmidt, C.},
  \bibinfo{author}{Horn-Stanja, J.}, \bibinfo{author}{Mishchenko, A.},
  \bibinfo{author}{Kennedy, D.}, \bibinfo{author}{Deller, A.},
  \bibinfo{author}{Card, A.}, \bibinfo{author}{Nissl, S.},
  \bibinfo{author}{Singer, M.}, \bibinfo{author}{Singer, M.},
  \bibinfo{author}{Koenig, S.}, \bibinfo{author}{Willingale, L.},
  \bibinfo{author}{Peebles, J.}, \bibinfo{author}{Edwards, M.},
  \bibinfo{author}{Chin, K.}, \bibinfo{year}{2020}.
\newblock \bibinfo{title}{A new frontier in laboratory physics: magnetized
  electron-positron plasmas}.
\newblock \bibinfo{journal}{Journal of Plasma Physics} ,
  \bibinfo{pages}{Accepted for publication}.
\bibitem[{Strickler et~al.(2002)Strickler, Berry and Hirshman}]{strickler_2002}
\bibinfo{author}{Strickler, D.J.}, \bibinfo{author}{Berry, L.A.},
  \bibinfo{author}{Hirshman, S.P.}, \bibinfo{year}{2002}.
\newblock \bibinfo{title}{Designing coils for compact stellarators}.
\newblock \bibinfo{journal}{Fusion Science and Technology}
  \bibinfo{volume}{41}, \bibinfo{pages}{107--115}.
\newblock \DOIprefix\doi{10.13182/FST02-A206}.
\bibitem[{Strickler et~al.(2004)Strickler, Hirshman, Spong, Cole, Lyon, Nelson,
  Williamson and Ware}]{strickler_2004}
\bibinfo{author}{Strickler, D.J.}, \bibinfo{author}{Hirshman, S.P.},
  \bibinfo{author}{Spong, D.A.}, \bibinfo{author}{Cole, M.J.},
  \bibinfo{author}{Lyon, J.F.}, \bibinfo{author}{Nelson, B.E.},
  \bibinfo{author}{Williamson, D.E.}, \bibinfo{author}{Ware, A.S.},
  \bibinfo{year}{2004}.
\newblock \bibinfo{title}{Development of a robust quasi-poloidal compact
  stellarator}.
\newblock \bibinfo{journal}{Fusion Science and Technology}
  \bibinfo{volume}{45}, \bibinfo{pages}{15--26}.
\newblock \DOIprefix\doi{10.13182/FST04-A421}.
\bibitem[{{Strykowsky} et~al.(2009){Strykowsky}, {Brown}, {Chrzanowski},
  {Cole}, {Heitzenroeder}, {Neilson}, {Rej} and {Viol}}]{Strykowsky09}
\bibinfo{author}{{Strykowsky}, R.L.}, \bibinfo{author}{{Brown}, T.},
  \bibinfo{author}{{Chrzanowski}, J.}, \bibinfo{author}{{Cole}, M.},
  \bibinfo{author}{{Heitzenroeder}, P.}, \bibinfo{author}{{Neilson}, G.H.},
  \bibinfo{author}{{Rej}, D.}, \bibinfo{author}{{Viol}, M.},
  \bibinfo{year}{2009}.
\newblock \bibinfo{title}{Engineering cost schedule lessons learned on {NCSX}},
  in: \bibinfo{booktitle}{2009 23rd IEEE/NPSS Symposium on Fusion Engineering},
  pp. \bibinfo{pages}{1--4}.
\bibitem[{Trefethen and Weideman(2014)}]{trefethen_trapezoidal}
\bibinfo{author}{Trefethen, L.N.}, \bibinfo{author}{Weideman, J.A.C.},
  \bibinfo{year}{2014}.
\newblock \bibinfo{title}{The exponentially convergent trapezoidal rule}.
\newblock \bibinfo{journal}{SIAM Review} \bibinfo{volume}{56},
  \bibinfo{pages}{385--458}.
\newblock \DOIprefix\doi{10.1137/130932132}.
\bibitem[{Wechsung(2020)}]{pyplasmaopt-zenodo}
\bibinfo{author}{Wechsung, F.}, \bibinfo{year}{2020}.
\newblock \bibinfo{title}{florianwechsung/pyplasmaopt: Release accompanying the
  submission of "single-stage gradient-based stellarator coil design:
  Optimization for near-axis quasi-symmetry"}.
\newblock \URLprefix \url{https://zenodo.org/record/4058482},
  \DOIprefix\doi{10.5281/ZENODO.4058482}.
\bibitem[{Wechsung et~al.(2021)Wechsung, Giuliani, Landreman, Cerfon and
  Stadler}]{WechsungGiulianiLandremanEtAl21}
\bibinfo{author}{Wechsung, F.}, \bibinfo{author}{Giuliani, A.},
  \bibinfo{author}{Landreman, M.}, \bibinfo{author}{Cerfon, A.},
  \bibinfo{author}{Stadler, G.}, \bibinfo{year}{2021}.
\newblock \bibinfo{title}{Single-stage gradient-based stellarator coil design:
  stochastic optimization}.
\newblock \href{http://arxiv.org/abs/2106.12137}{{\tt arXiv:2106.12137}}.
  \bibinfo{note}{{Nuclear Fusion (revised, Aug 2021)}}.
\bibitem[{Weideman and Reddy(2000)}]{weideman2000matlab}
\bibinfo{author}{Weideman, J.A.}, \bibinfo{author}{Reddy, S.C.},
  \bibinfo{year}{2000}.
\newblock \bibinfo{title}{A {MATLAB} differentiation matrix suite}.
\newblock \bibinfo{journal}{ACM Transactions on Mathematical Software (TOMS)}
  \bibinfo{volume}{26}, \bibinfo{pages}{465--519}.
\bibitem[{Wobig(1987)}]{Wobig1987}
\bibinfo{author}{Wobig, H.}, \bibinfo{year}{1987}.
\newblock \bibinfo{title}{Magnetic surfaces and localized perturbations in the
  {W}endelstein {VII-A} stellarator}.
\newblock \bibinfo{journal}{Zeitschrift für Naturforschung A}
  \bibinfo{volume}{42}, \bibinfo{pages}{1054 -- 1066}.
\newblock \DOIprefix\doi{https://doi.org/10.1515/zna-1987-1003}.
\bibitem[{Zarnstorff et~al.(2001)Zarnstorff, Berry, Brooks, Fredrickson, Fu,
  Hirshman, Hudson, Ku, Lazarus, Mikkelsen, Monticello, Neilson, Pomphrey,
  Reiman, Spong, Strickler, Boozer, Cooper, Goldston, Hatcher, Isaev, Kessel,
  Lewandowski, Lyon, Merkel, Mynick, Nelson, Nuehrenberg, Redi, Reiersen,
  Rutherford, Sanchez, Schmidt and White}]{Zarnstorff2001}
\bibinfo{author}{Zarnstorff, M.C.}, \bibinfo{author}{Berry, L.A.},
  \bibinfo{author}{Brooks, A.}, \bibinfo{author}{Fredrickson, E.},
  \bibinfo{author}{Fu, G.Y.}, \bibinfo{author}{Hirshman, S.},
  \bibinfo{author}{Hudson, S.}, \bibinfo{author}{Ku, L.P.},
  \bibinfo{author}{Lazarus, E.}, \bibinfo{author}{Mikkelsen, D.},
  \bibinfo{author}{Monticello, D.}, \bibinfo{author}{Neilson, G.H.},
  \bibinfo{author}{Pomphrey, N.}, \bibinfo{author}{Reiman, A.},
  \bibinfo{author}{Spong, D.}, \bibinfo{author}{Strickler, D.},
  \bibinfo{author}{Boozer, A.}, \bibinfo{author}{Cooper, W.A.},
  \bibinfo{author}{Goldston, R.}, \bibinfo{author}{Hatcher, R.},
  \bibinfo{author}{Isaev, M.}, \bibinfo{author}{Kessel, C.},
  \bibinfo{author}{Lewandowski, J.}, \bibinfo{author}{Lyon, J.F.},
  \bibinfo{author}{Merkel, P.}, \bibinfo{author}{Mynick, H.},
  \bibinfo{author}{Nelson, B.E.}, \bibinfo{author}{Nuehrenberg, C.},
  \bibinfo{author}{Redi, M.}, \bibinfo{author}{Reiersen, W.},
  \bibinfo{author}{Rutherford, P.}, \bibinfo{author}{Sanchez, R.},
  \bibinfo{author}{Schmidt, J.}, \bibinfo{author}{White, R.B.},
  \bibinfo{year}{2001}.
\newblock \bibinfo{title}{Physics of the compact advanced stellarator {NCSX}}.
\newblock \bibinfo{journal}{Plasma Physics and Controlled Fusion}
  \bibinfo{volume}{43}, \bibinfo{pages}{A237--A249}.
\newblock \DOIprefix\doi{10.1088/0741-3335/43/12a/318}.
\bibitem[{Zhu et~al.(2017)Zhu, Hudson, Song and Wan}]{Zhu2017}
\bibinfo{author}{Zhu, C.}, \bibinfo{author}{Hudson, S.R.},
  \bibinfo{author}{Song, Y.}, \bibinfo{author}{Wan, Y.}, \bibinfo{year}{2017}.
\newblock \bibinfo{title}{New method to design stellarator coils without the
  winding surface}.
\newblock \bibinfo{journal}{Nuclear Fusion} \bibinfo{volume}{58},
  \bibinfo{pages}{016008}.

\end{thebibliography}
